\documentclass[10pt]{article}

\usepackage{graphicx}
\usepackage{color}
\usepackage{subfig}
\usepackage{amsmath,amsthm,amssymb}
\usepackage{setspace}
\usepackage{mathtools}
\usepackage{epsfig}
\usepackage{epstopdf}
\usepackage{float}
\usepackage{multirow}
\usepackage{verbatim}
\usepackage{pdflscape}
\usepackage[table,xcdraw]{xcolor}

\usepackage{geometry} 

\title{\Large{\bf A Novel Reinitialization Scheme for Conservative Level Set Method}}
\author{S. Parameswaran, J. C. Mandal}
\date{}

\begin{document}
\maketitle



\begin{abstract}
Existing artificial compression based reinitialization scheme for
conservative level set method has a few drawbacks, like distortion of
fluid-fluid interface, unphysical patch formation away from the
interface and lack of mass conservation. In this paper, a novel reinitialization approach has been
presented which circumvents these limitations by reformulating the
reinitialization equation. With the reformulated procedure, the
present approach is able to reinitialize the level set function
without causing any unwanted movement of the interface contour. The
unphysical patch formation away from the interface is also resolved
here by avoiding the use of ill-conditioned contour normal vectors. As
a result of this measure, there is a significant improvement in the
mass conservation property. In addition, the simplified form of the
new reinitialization equation enables one to choose a much larger time
step during the reinitialization iteration. Moreover, the new
formulation also helps in reducing the amount of numerical computations
per time step, leading to an overall reduction in the computational
efforts. In order to evaluate the performance of the present
formulation, a set of test problems involving reinitialization of
stationary level set functions is carried out first. Then, the
efficacy of the proposed reinitialization scheme is demonstrated using
a set of standard two-dimensional scalar advection based test problems
and incompressible two-phase flow problems. Finally, the ability to
deal with complex mesh types is demonstrated by solving a test problem
on an unstructured mesh consisting of finite volume cells having
triangular and quadrilateral shapes.
\end{abstract}

{\bf Keywords: }{\it
Multiphase flow; Finite volume method; Conservative
level set method; Reformulation of reinitialization;
Artificial compressibility method; Unstructured mesh.
}


\section*{Highlights}
\begin{itemize}
\item Developed new reinitialization procedure by reformulating the artificial compression approach.

\item Problems of interface movement and unphysical patch formation during reinitialization are completely resolved.

\item Applicable to a wide variety of meshes including unstructured hybrid meshes and computationally efficient.

\item Numerical results show superior mass conservation and good agreement with analytical and experimental results.

\end{itemize}

\section{Introduction}
\label{sec:intro}
Numerical simulation of incompressible two-phase flows poses great
challenges due to the presence of the fluid-fluid interface. Popular
contact capturing methods, such as, Volume of Fluid (VOF) method and
Level Set (LS) method, use an additional interface advection equation
along with the incompressible Navier-Stokes (NS) equations. Any
inaccuracy in solving the interface advection equation will
significantly affect the quality of the numerical solution. Issues of
the formation of jetsam/ floatsam in VOF method, violation of mass
conservation in LS method are a few examples associated with errors in
the computation of the interface advection equation. In order to
overcome the mass conservation error in the classical LS method, a
variant of the LS method, known as Conservative Level Set (CLS)
method, is proposed by Olsson and Kreiss~\cite{Olsson2005} and
Olsson~et~al~\cite{Olsson2007}. The enhanced mass conservation 
property is achieved here by replacing the signed distance function
used in the classical level set method with a hyperbolic tangent type
level set function. This level set function is then advected using a
scalar conservation law. In order to recover from the excessive
numerical dissipation error, an artificial compression based
reinitialization procedure is also formulated for the level set
function. With the improved mass conservation property, the CLS method
shows promising capabilities and provides a good alternative to
the classical LS method in solving incompressible two-phase
flow problems. However, in practice, the reinitialization procedure
for level set function in CLS method often runs into various numerical
difficulties, leading to undesirable results.

Primarily, two major issues with the reinitialization procedure are
reported in the literature~\cite{Shukla2010, Mccaslin2014,
  Desjardins2008}. The first one is the undesired movement of the
interface contour during reinitialization. The problem gets further
aggravated with the frequent use of reinitialization. This issue
arises particularly in two and three dimensions, where, the interface
curvature gets involved in the computations. It is demonstrated
in reference~\cite{Shukla2010} that the degree of movement of
interface contour depends upon the strength of the interface
curvature. Also, a set of numerical experiments presented in
reference~\cite{Mccaslin2014} verifies that the frequent use of
reinitialization results in substantial movement of the interface,
leading to inaccuracies in the numerical solution. Several attempts to
resolve this issue can be found in literature~\cite{Yohei2012,
  Mccaslin2014}. The efforts are focused mainly on localizing the
reinitialization process only to a selected region of the level set
field. Excessive reinitialization at less dissipated regions is thus
avoided. In reference~\cite{Yohei2012}, the reinitialization is
localized by defining a local coefficient based on the degree of
sharpness of the level set function. Whereas, in
reference~\cite{Mccaslin2014}, a metric which depends on the local
flow conditions and the numerical diffusion errors is used. It is,
therefore, clear that these methods introduce additional complexity
and computational efforts in obtaining the local coefficients.

The second issue of the original CLS method is the formation of
unphysical fluid patches away from the interface due to the
ill-conditioned behaviour of the interface contour normal
vectors. The contour normal vectors decide the direction of the
compression and diffusion fluxes during the reinitialization
process. Far away from the fluid-fluid interface these contour normal
vectors are expected to be zero, resulting in no
reinitialization. However, due to their ill-conditioned nature, even
far away from the interface the contour normal vectors may not
necessarily be zero. This results in undesirable compression
at far away region, which leads to formation of unphysical fluid
patches there. Several cures for this problem can be found in
literature~\cite{Desjardins2008, Shukla2010, Zhao2014,
  Waclawczyk2015, Chiodi2017, Tabar2018}. Most of the efforts are
targeted towards replacing the ill-conditioned contour normal vectors
with some alternatives. In the Accurate Conservative Level Set (ACLS)
method, proposed in reference~\cite{Desjardins2008}, the contour
normal vectors are computed from an auxiliary signed distance
function. The auxiliary signed distance function is constructed here
from the level set function using a fast marching method. In the
method proposed by Shukla~et~al.~\cite{Shukla2010}, a modified form of
the reinitialization equation is used. Here, the level set function is
replaced with a smooth function constructed from the level set
function itself by employing a mapping procedure. A technique, by
combining the reinitialization schemes of both the classical and
conservative level set method, named as Improved Conservative Level
Set (ICLS) Method, is reported in reference~\cite{Zhao2014}. In
reference~\cite{Waclawczyk2015} and its improved version in
reference~\cite{Chiodi2017}, reformulations of the original
reinitialization equation are presented which take care of the
spurious movement of interface contour as well as the ill-conditioned
behaviour of the contour normal vectors. Recently in
reference~\cite{Tabar2018}, the ill-conditioned unit contour normal
vectors are replaced with another normal vectors, such that, their
magnitudes start to diminish away from the interface. This ensures
that the reinitialization process is activated only near the interface
regions. Though improvements in the contour normal vectors partially
circumvent the issue of formation of unphysical fluid patches, they
involve evaluation of more complicated terms adding to the overall
computational cost.

In the present work, a much simpler technique to reinitialize the
level set function is presented. Here, the existing artificial
compression based reinitialization equation is revised by isolating
and removing terms that have potential to move the interface
contours. The remaining terms in the modified equation are then
reformulated considerably, such that, the usage of the contour normal
vectors is completely avoided. With the new reformulation, issues
such as, distortion of the interface contour and the unphysical patch
formation away from the interface are resolved. As a consequence, the area conservation property is improved significantly. In addition, absence of a
viscous dissipation like (second derivative) term in the new
reinitialization equation enables one to choose a much larger time
step during the reinitialization iteration. The simplified
terms also help in significantly reducing the numerical computations
per reinitialization time step, aiding an overall reduction in the
computational efforts. In order to demonstrate the efficacy of the
proposed reinitialization scheme, a set of standard two-dimensional
test problems involving reinitialization of stationary level set
functions (henceforth, we name it as in-place reinitialization
problems), advection of level set function under predefined velocity
fields and a few standard incompressible two-phase flow problems are
solved. Finally, in order to demonstrate the ability to deal with 
complex mesh types, an incompressible two-phase flow problem is solved
on an unstructured mesh consisting of finite volume cells having
triangular and quadrilateral shapes.

Rest of the paper is organized as follows. The original CLS method and
its reinitialization scheme is briefly described in
section~2. The limitations of the existing artificial compression
based reinitialization approach and its new reformulation are also
discussed in the same section. In section~3, the mathematical
formulation of incompressible two-phase flows is briefly
described. The numerical discretization of the governing equations and
the new reinitialization equation are also described in
section~3. Several numerical test problems are solved in
section~4. Finally the conclusions are given in section~5.

\section{Conservative Level set Method}
The fluid-fluid interface in conservative level set method is
represented in the form of an iso-contour of a hyperbolic tangent type
level set function, defined as,
\begin{equation}
\psi({\bf x}, t) = \frac{1}{1 +  \exp \left(\frac{-\phi({\bf x}, t)}{\varepsilon}\right)} \equiv
\frac{1}{2} \left( \tanh \left(\frac{\phi({\bf x}, t)}{2 \varepsilon}
  \right) + 1 \right) 
\label{eq:level_set_function}
\end{equation}
where, $\phi({\bf x}, t)$ is the standard signed distance function
defined in terms of the minimum distance $d({\bf x}, t)$ from the
interface, as,
$$
\phi({\bf x}, t) = 
\begin{cases}
-d({\bf x}, t), &\text{ inside the first fluid} \\
0,  &\text{ at the fluid-fluid interface} \\
+d({\bf x}, t), &\text{ inside the second fluid} 
\end{cases}
$$
The function $\psi$ takes a value $0$ at regions occupied by the
first fluid and $1$ at the second fluid. Within a thin transition
region between the two fluids, $\psi$ varies smoothly from $0$ to
$1$. Width of the transition region is dictated by the parameter
$\varepsilon$. The contour corresponds to $\psi({\bf x},0) = 0.5$
represents the actual fluid-fluid interface. The geometric parameters
associated with the interface, such as interface normal vector (${\bf
  n}$) and interface curvature (${\kappa}$), are obtained from the
level set function as,
\begin{equation}
{\bf n} = \frac{\nabla \psi}{\lvert \nabla \psi \rvert}
\label{eq:normal}
\end{equation}
\begin{equation}
\kappa = - \nabla \cdot {\bf n}
\label{eq:curvature}
\end{equation}
Finally, movement of the fluid-fluid interface is achieved by
advecting the level set function according to the flow field, as,
\begin{equation}
\label{eq:ls-advec}
\frac{\partial \psi}{\partial t} + \nabla \cdot \left({\bf u}
  \psi\right) = 0
\end{equation}
where, ${\bf u} = u \hat{\text{i}} + v \hat{\text{j}}$, is the
divergence-free velocity field. 

\subsection{Artificial Compression based Reinitialization Procedure}
\label{sec:artificial_compression}
It is well known that the level set function suffers from excessive
dissipation due to numerical errors~\cite{Olsson2005}. This leads the
level set function to deviate from its original hyperbolic tangent
type profile. An artificial compression based reinitialization is
developed in reference~\cite{Olsson2007} in order to re-establish the
pre-specified thickness and the profile of the level set function. The
discretized level set advection equation together with the
reinitialization should satisfy the following three
requirements~\cite{Olsson2005}. Firstly, the method should ensure
discrete conservation of mass while advecting the level set
function. Secondly, the method should not introduce any spurious
oscillations. Finally, the initial properties of the level set
function should be maintained throughout the simulation.

The equation for reinitializing the level set function can be written
as per~\cite{Olsson2007} as,
\begin{equation}
\frac{\partial \psi}{\partial \tau_r} + \nabla \cdot \left(\psi (1 -
  \psi) {\bf n}_0 \right) - \nabla \cdot \left(\varepsilon
  \left(\nabla \psi \cdot {\bf n}_0 \right)~{\bf n}_0\right) = 0
\label{eq:reinit_htf}
\end{equation}
where, ${\bf n}_0 = \displaystyle \frac{\nabla \psi_0}{\lvert
  \nabla \psi_0 \rvert}$ is the interface contour normal vector
defined before the reinitialization starts. The variable $\tau_r$ is a
time like variable and $\psi_0$ is the level set function defined at
$\tau_r = 0$. In equation~(\ref{eq:reinit_htf}), the second and third
terms are responsible for the interface compression and diffusion
respectively, which balances each other once the
equation~(\ref{eq:reinit_htf}) converges in time~$\tau_r$.

In practice, the reinitialization formulation suffers from various
deficiencies. As described in the introduction, the above
reinitialization procedure may involve error in the ${\bf n}_0$
computation. It may also be noted, the reinitialization equation moves
the interface contour based on the local curvature. In order to
identify terms that have potential to move the interface contour
during the reinitialization, equation~(\ref{eq:reinit_htf}) is
rewritten in non-conservative form by expanding the the compressive
and diffusive terms. After rewriting equation~(\ref{eq:reinit_htf}) in
non-conservative form, a curvature dependent velocity like term, ${\bf
  v} = \varepsilon \kappa_0 \bigg({\bf n}_0 - \displaystyle
\frac{\nabla \phi}{\lvert \nabla \phi \rvert^2}\bigg)$, is isolated
which is found to be responsible for undesired movement of the interface
(details shown in reference \cite{Parameswaran2020}).

\subsection{Reformulation of the Reinitialization Equation}
\label{sec:reformulation}
\begin{figure}[H]
\centering
\includegraphics[scale=1.0]{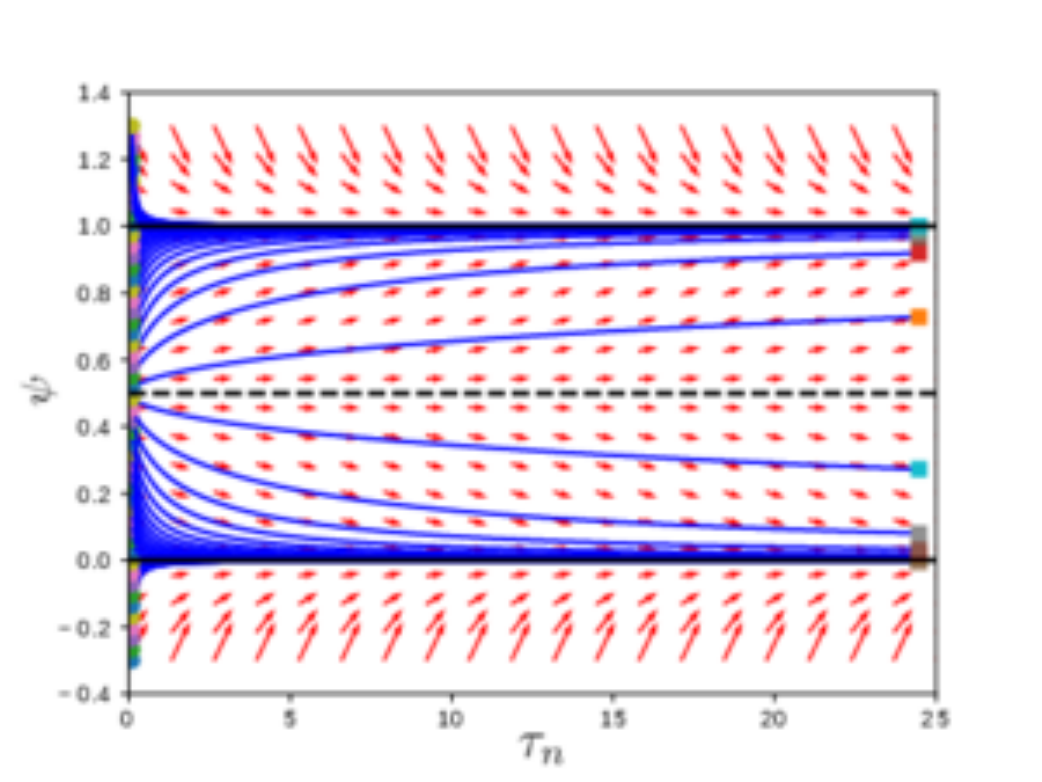}
\caption{Phase plot for the new reinitialization equation. The arrows
  indicate the magnitude and direction of change of $\psi$ with
  respect to $\tau_n$. The blue lines show the trajectories of
  different initial values of $\psi$ as $\tau_n$ progresses. The
  initial and final values of $\psi$ are represented using circles and
  squares respectively.}
\label{fig:phase_plot}
\end{figure}

As described above, the curvature dependent advection term moves the
interface; thus it is undesired in a reinitialization
procedure. Therefore, in the new formulation of the reinitialization
equation, we remove this curvature dependent advection term from the
reinitialization equation. It may also be noted, the reinitialization
equation is sensitive towards numerical errors arising from the
ill-conditioned behaviour of the contour normal vectors. In order to
overcome the difficulty to deal with the these terms, in the new
formulation we seek for terms that are easy to compute and are less
sensitive to numerical errors arising form the ill-conditioned contour
normal vectors. After some manipulations (details shown in reference
\cite{Parameswaran2020}), the final reformulated reinitialization
equation can be written as,
\begin{equation} 
\frac{\partial \psi}{\partial \tau_n} = - \psi (1 - \psi) (1 - 2 \psi)
+ \varepsilon (1 - 2 \psi) \lvert \nabla \psi \rvert
\label{eq:reinit_new}
\end{equation} 
It can be easily shown that the level set function, $\psi$, as given
in equation~(\ref{eq:level_set_function}), trivially satisfies the
steady state form of equation~(\ref{eq:reinit_new}). Furthermore, the
overall behaviour of equation~(\ref{eq:reinit_new}) can be described
by considering each terms in the RHS individually. In the absence of
the second term in RHS, equation~(\ref{eq:reinit_new}) behaves like an
ordinary differential equation with $\psi = 1$ and $\psi = 0$ as two
stable equilibrium points and $\psi=0.5$ as an unstable equilibrium
point. Figure~\ref{fig:phase_plot} shows the phase plot of
equation~(\ref{eq:reinit_new}) with only first term in the RHS. From
the phase plot, it is clear that the first term results in sharpening
the level set function profile. Moreover, the first term also helps in
stabilizing the overshoot ($\psi > 1.0$) and undershoot ($\psi < 0.0$)
issues arising in case of the use of non-TVD numerical schemes for the
original level set advection equation. Nature of the second term in
the RHS of equation~(\ref{eq:reinit_new}) is to balance the first
term. Since the $\varepsilon$ and $\lvert \nabla \psi \rvert$ are
always positive quantities, the sign of the second term depends on the
sign of $(1 - 2\psi)$. That is, the sign of the second term is
positive when $\psi < 0.5$ and negative when $\psi > 0.5$. At $\psi =
0.5$, the second term is zero. In other words, the second term drives
the level set function towards a flat profile with $\psi = 0.5$
everywhere, thus balancing the first term. With the above mentioned
sharpening and balancing actions, the equation~(\ref{eq:reinit_new})
reinitializes the level set function.

\section{Mathematical Formulation of Incompressible Two-Phase Flows}
\subsection{Governing Equations}
A dual time-stepping based artificial compressibility approach is
followed here for modelling incompressible two-phase flows. The
governing system of equations describing the unsteady incompressible
viscous two-phase flow can be written as,
\begin{equation}
\frac{\partial \bf U}{\partial \tau} + I^t \frac{\partial \bf
	U}{\partial t} + \left[\frac{\partial {\bf (F -
    F_v)}}{\partial
	x} + \frac{\partial {\bf (G - G_v)}}{\partial y}\right] =
    {\bf F_g} + {\bf F_s}
\label{eq:gov_eq_two_phase}
\end{equation}
where,
\begin{center}
	${\bf U} =\left\{\begin{array}{c}
	p/\beta\\
	\rho u\\
	\rho v\\
	\psi
	\end{array}\right\}$; ${\bf F} = \left\{\begin{array}{c}
	u\\
	\rho u^{2}+p\\
	\rho uv\\
	u \psi
	\end{array}\right\}$; ${\bf G} = \left\{\begin{array}{c}
	v\\
	\rho uv\\
	\rho v^{2}+p\\
	v \psi
	\end{array}\right\}$; ${\bf F_v} = \left\{\begin{array}{c}
	0 \\
	2\mu \frac{\partial u}{\partial x} \\
	\mu (\frac{\partial u}{\partial y} + \frac{\partial
    v}{\partial x})\\
	0
	\end{array}\right\}$; ${\bf G_v} = \left\{\begin{array}{c}
	0 \\
	\mu (\frac{\partial u}{\partial y} + \frac{\partial
    v}{\partial x})\\
	2\mu \frac{\partial v}{\partial y}\\
	0
	\end{array}\right\}$; ${\bf F_g} = \left\{\begin{array}{c}
	0 \\
	-\rho g_x \\
	-\rho g_y \\
	0
	\end{array}\right\}$; \hspace{0.5cm} ${\bf F_s} =
  \left\{\begin{array}{c}
	0 \\
	-\sigma \kappa \frac{\partial \psi}{\partial x}\\ \\
	-\sigma \kappa \frac{\partial \psi}{\partial y}\\
	0
	\end{array}\right\}$; \hspace{0.5cm} $I^t =
    \left[\begin{array}{cccc}
	0 & 0 & 0 & 0\\
	0 & 1 & 0 & 0\\
	0 & 0 & 1 & 0\\
	0 & 0 & 0 & 1\\
	\end{array}\right]$.
\end{center}
where, the vector ${\bf U}$ in equation~(\ref{eq:gov_eq_two_phase})
denotes the vector of conservative variables and the vectors $({\bf
  F}, {\bf G})$ and $({\bf F_v}, {\bf G_v})$ denote the convective and
viscous flux vectors respectively. The vectors ${\bf F_g}$ and ${\bf
  F_s}$ denote the source terms containing gravitational and surface
tension forces respectively. Here, the surface tension term is
modelled using a continuum surface forces (CSF) method, proposed by
Brackbill~et~al.~\cite{Brackbill1992}. The parameter $\sigma$ denotes
the surface tension coefficient per unit length of interface and $g_x$
and $g_y$ denote the $x$ and $y$ components of the acceleration due to
gravity. The variable $p$, $\rho$ and $\mu$ denote the pressure,
density and dynamic viscosity of the fluid respectively. The variables
$\tau$ and $t$ appearing in equation~(\ref{eq:gov_eq_two_phase})
denote the pseudo time and the real time respectively. The parameter
$\beta$ denotes the artificial compressibility parameter, which is
usually taken as constant for a given test problem. The artificial
compressibility term added in the continuity equation is similar to
the one introduced by Chorin in~\cite{Chorin1967}. Once
equation~(\ref{eq:gov_eq_two_phase}) converges to a pseudo-steady
state, it recovers the set of unsteady incompressible two-phase flow
equations. It can be noticed that the level set advection, described
by equation(\ref{eq:ls-advec}), is combined here with the system of
equations~(\ref{eq:gov_eq_two_phase}), and solved simultaneously along
with the Navier-Stokes equations. The density and viscosity used in
equation~(\ref{eq:gov_eq_two_phase}) are defined in terms of the level
set function, as,
\begin{eqnarray}
\label{eq:den_lsm}
\rho = \rho(\psi) = \rho_{\text{2}}\psi + (1 - \psi) \rho_{\text{1}}
\\
\mu  = \mu(\psi) = \mu_{\text{2}}\psi + (1 - \psi) \mu_{\text{1}}
\label{eq:visc_lsm}
\end{eqnarray}
where the subscripts ``1'' and ``2'' indicate the properties
corresponds to the first and the second fluids respectively.

\subsection{Numerical Discretization of Governing Equations}
\label{sec:fv_formulation}
A finite volume approach is followed here for solving the governing
system of equations~(\ref{eq:gov_eq_two_phase}). In order to proceed
with finite volume discretization, the governing system of
equations~(\ref{eq:gov_eq_two_phase}) is first integrated over a
control volume. The computational domain is then discretized
into a finite number of non-overlapping finite volume cells. The final
space discretized form of equations~(\ref{eq:gov_eq_two_phase}) for
an $i^{\text{th}}$ finite volume cell can be written as,
\begin{equation}
	\Omega_{i} \frac{\partial {\overline{ \bf U}}}{\partial \tau} + I^t
	\Omega_{i} \frac{\partial {\overline{ \bf U}}}{\partial t} + R\left({
		\overline{\bf U}}\right) = 0
	\label{eq:fv_disc_eq}
\end{equation}
where,
\begin{equation*}
	R\left({ \overline{\bf U}}\right) = \sum_{m=1}^{M} \left( {\bf F}~n^m_x + {\bf
		G}~n^m_y  \right)_m \Gamma_m -  \sum_{m=1}^{M} \left( {\bf F_v}~n^m_x
	+ {\bf G_v}~n^m_y  \right)_m \Gamma_m - \Omega_{i} {\overline{\bf
			F}_g} - \Omega_{i} {\overline{\bf F}_s},
\end{equation*}
$\Omega_i$ is the area, $\Gamma_m$ and ${\bf n}^m = (n^m_x,
n^m_y)$ are the length and edge normals of the $m^{\text{th}}$ edge
respectively and $M$ is the total number of edges of the finite volume
cell~$i$. The vectors ${\overline{\bf U}}$, ${\overline{\bf F}_g}$ and
${\overline{\bf F}_s}$ represent the cell averaged values of ${\bf 
	U}$, ${\bf F_g}$ and ${\bf F_s}$ respectively. The source
term vector, ${\bf \overline{F}_g}$, appearing in
equation~(\ref{eq:fv_disc_eq}) is computed by multiplying the cell
averaged value of density and the acceleration due to
gravity. Evaluation of the surface tension vector, ${\overline{\bf
		F}_s}$, involves the computation
of gradient, contour normal and curvature of the level set
function. The contour normal and curvature are computed from the
gradient of the level set function as per equation~(\ref{eq:normal})
and equation~(\ref{eq:curvature}) respectively, where the gradient
vector is evaluated using central least square method as explained
in~\cite{Parameswaran2019-2}. The convective flux vector $\left( {\bf
	F}~n^m_x + {\bf G}~n^m_y  \right)$ and the viscous flux vector
$\left({\bf F_v}~n^m_x  + {\bf G_v}~n^m_y \right)$ in
equation~(\ref{eq:fv_disc_eq}) are computed at the edges of each cell
using a Roe-type Riemann solver, developed in~\cite{Parameswaran2019},
and a Green-Gauss integral approach over a Coirier's diamond path,
described in~\cite{Parameswaran2019-2}, respectively. The real time
derivatives appearing in equation~(\ref{eq:fv_disc_eq}) are computed
using a three point implicit backward differencing procedure. Finally,
an explicit three stage Strong Stability Preserving Runge-Kutta
(SSP-RK) method, described in~\cite{Gottlieb2005}, is used for
iterating in pseudo-time. The time step required for the pseudo-time
iteration is computed by considering the convective, viscous and
surface tension effects. For faster convergence, a local time stepping
approach is adopted here, in which, each cell is updated using its own
$\Delta \tau_i$. The local time step $\Delta \tau_i$ is computed as,
\begin{equation}
	\label{eq:pseudo_dt}
	\Delta \tau_i = \text{min}(\Delta \tau^{\text{conv}}_i, \Delta \tau^{\text{visc}}_i, \Delta \tau^{\text{surf}}_i)
\end{equation}
where, $\Delta \tau^{\text{conv}}_i, \Delta \tau^{\text{visc}}_i$ and
$\Delta \tau^{\text{surf}}_i$ are the maximum allowed time steps due to
convective flux, viscous flux and surface tension force
respectively. These time steps are evaluated as,
\begin{equation*}
	\Delta \tau^{\text{conv}}_i = \frac{\nu ~
		\Omega_{i}}{\sum_{m=1}^{M}\left(\lvert u_n \rvert +
		\sqrt{\left(u_n \right)^2 + \frac{\beta}{\rho_i}} \right)_m
		\Gamma_m } \text{;} 
	\Delta \tau^{\text{visc}}_i = \frac{\nu ~
		\Omega_{i}^2}{\left(\frac{8}{3}\right)\frac{\mu_i}{\rho_i}
		\sum_{m=1}^{M}\left(\Gamma_m \right)^2} \text{ and } 
	\Delta \tau^{\text{surf}}_i = \nu ~ \sqrt{\frac{(\rho_1 + \rho_2)
			h^3}{4 \pi \sigma}}
\end{equation*}
where, $\nu$ is the Courant number, $h$ is the average cell size and
$u_n$ is the velocity component along the edge normal direction. For
stability reasons, the Courant number, $\nu$, is always taken less
than unity. The $\Delta \tau_i$ computed using
equation~(\ref{eq:pseudo_dt}) is further restricted based on the real
time step~\cite{Gaitonde1998}, as, $\Delta \tau_i \leq
\frac{2}{3}\Delta t$. Detailed descriptions of the numerical methods
used for solving incompressible two-phase flows are excluded from here
due to brevity reasons. One can refer~\cite{Parameswaran2019} and
\cite{Parameswaran2019-2} for more details.

\subsection{Numerical Discretization of Reinitialization Equation}
The RHS of equation~(\ref{eq:reinit_new}) consists of two terms. In
order to evaluate the first term at a given cell, only the cell center
values of $\psi$ is sufficient. However, evaluation of the second term
involves computation of the gradient of $\psi$. Here, the gradient of
$\psi$ at the cell centers are evaluated using a central least square
approach discussed in~\ref{sec:central_least_square}. Finally,
the time integration of equation~(\ref{eq:reinit_new}) is carried out
using a three stage Strong Stability Preserving Runge-Kutta (SSP-RK-3)
method described in section~\ref{sec:ssprk-3}.

\subsubsection{Time integration for the reinitialization equation}
\label{sec:ssprk-3}
According to the SSP-RK-3 approach described in~\cite{Gottlieb2005},
the cell averaged value of the unknown function in
equation~(\ref{eq:reinit_new}) is updated as,
\begin{equation}
\begin{array}{r@{}l@{}l@{}l@{}l@{}l@{}l}
	\overline{\psi_i}^{1} &{}=&{}&{}&{}\overline{\psi_i}^{n} &{}+&{} \displaystyle \frac{\Delta
      \tau_n}{\Omega_i}~L\left({\overline{\psi_i}}^{n}\right)\\
	\overline{\psi_i}^{2} &{}= \displaystyle
    \frac{3}{4}&{}\overline{\psi_i}^{n} &{}+ \displaystyle
    \frac{1}{4} &{}\overline{\psi_i}^{1} &{}+
    \displaystyle \frac{1}{4} &{}\displaystyle \frac{\Delta
      \tau_n}{\Omega_i}~L\left({\overline{\psi_i}}^{1}\right)\\
	\overline{\psi_i}^{n+1} &{}= \displaystyle
    \frac{1}{3}&{}\overline{\psi_i}^{n} &{}+ \displaystyle \frac{2}{3}&{}\overline{\psi_i}^{2} &{}+
    \displaystyle \frac{2}{3} &{}\displaystyle \frac{\Delta \tau_n}{\Omega_i}~L\left({\overline{\psi_i}}^{2}\right)
\end{array}
\label{eq:ssprk}
\end{equation}
where, $\overline{\psi_i}^n$ and $\overline{\psi_i}^{n+1}$ are the cell
averaged level set function defined at $n^{\text{th}}$ and
$(n+1)^{\text{th}}$ time levels respectively, $\overline{\psi_i}^1$ and
$\overline{\psi_i}^2$ are the intermediate values of $\psi$ and
$L\left({\overline{\psi_i}}^{*}\right) = - \overline{\psi_i}^{*}(1
- \overline{\psi_i}^{*})(1 - 2\overline{\psi_i}^{*}) + \varepsilon
\lvert \nabla \overline{\psi_i}^{*} \rvert (1 -
2\overline{\psi_i}^{*})$. For the explicit time integration of
equation~(\ref{eq:reinit_new}), the time step is restricted based on
the nature of the reinitialization equation. In order to find out the
allowable time step, equation~(\ref{eq:reinit_new}) is rewritten in a
Hamilton-Jacobi form with a velocity like variable, ${\bf \mathcal{S}}
= \varepsilon(1 - 2\psi) \left( \frac{1 - \lvert \nabla \phi
  \rvert}{\lvert \nabla \phi \rvert} \right){\bf n}$. Since the
solution variable $\psi$ is updated according to the sharpening
velocity vector ${\bf \mathcal{S}}$, a stable explicit time
integration scheme for equation~(\ref{eq:reinit_new}) is possible
only with a restricted time step,
\begin{equation}
\Delta \tau_n \leq 2h^d
\label{eq:reinit_time_step}
\end{equation}
It may be noticed that the time step, $\Delta \tau_n$, given in
equation~(\ref{eq:reinit_time_step}) is larger by a factor of
$4/h$ in comparison with the allowable time step for
the artificial compression based reinitialization
procedure~\cite{Olsson2005}. Presence of a viscous dissipation term in
the artificial compression based approach restricts the
reinitialization time step to a smaller value~\cite{Olsson2005}.

\section{Numerical Experiments}
\label{sec:num_exp}
Performance of the new reinitialization procedure is evaluated using
three types of test problems. In order to illustrate the movement of
the interface contour during the reinitialization, a set of test
problems involving reinitialization of stationary level set function,
is carried out first in section~\ref{sec:stationary_pblms}. These
problems are named here as in-place reinitialization
problems. Followed to the in-place reinitialization problems, a set of
scalar advection based test problems are considered in
section~\ref{sec:scalar_aadv}, where, the area and shape errors during
the level set advection are quantified. In the
sections~\ref{sec:db}~-~\ref{sec:br}, several incompressible two-phase
flow test problems are presented. These problems are arranged
according to their increasing levels of complexities, starting from an
inviscid flow problem to problems involving viscous and surface
tension forces. All test problems are solved on simple Cartesian type
meshes. However, in order to demonstrate the the ability to deal with
complex mesh types, the last test problem is also solved on an
unstructured mesh consisting of finite volume cells having triangular
and quadrilateral shapes.

\subsection{In-Place Reinitialization Problems}
\label{sec:stationary_pblms}
As illustrated in section~2.1, the reinitialization scheme often
results in moving the interface contour according to the sign and
strength of the interface curvature. In order to demonstrate this, a
set of test problems involving reinitialization of stationary level
set function, similar to the one reported in~\cite{Chiodi2017}, is
carried out here. In order to perform in-place reinitialization tests,
a level set function, corresponds to some given geometry, is
constructed first. For the present study, three standard shapes,
namely, a circle with diameter of 4 units, an ellipse with 4 units and
2 units of major and minor axes respectively and a square with size 3
units, are chosen. The geometric shapes are placed at the center of a
computational domain of square shaped region bounded between $-5 \leq
x \leq 5$ and $-5 \leq y \leq 5$. The computational domain is
discretized using a $200 \times 200$ Cartesian mesh. The level set
function corresponds to the given shape is then taken as the initial
condition for the reinitialization equation and carried out large number
of pseudo-time iterations. Under ideal situations, the
reinitialization process should not result in movement of the
interface contour. However, due to errors in the reinitialization
scheme, this need not be satisfied always. Moreover, unlike other test
problems, numerical errors associated with the advection of level set
function are not present here. Therefore, these tests help in
isolating the errors associated with only the reinitialization
process.

\begin{figure}[H]
\centering
\subfloat[CLS-Olsson (no.iter = 0)]
{\includegraphics[scale=0.47]{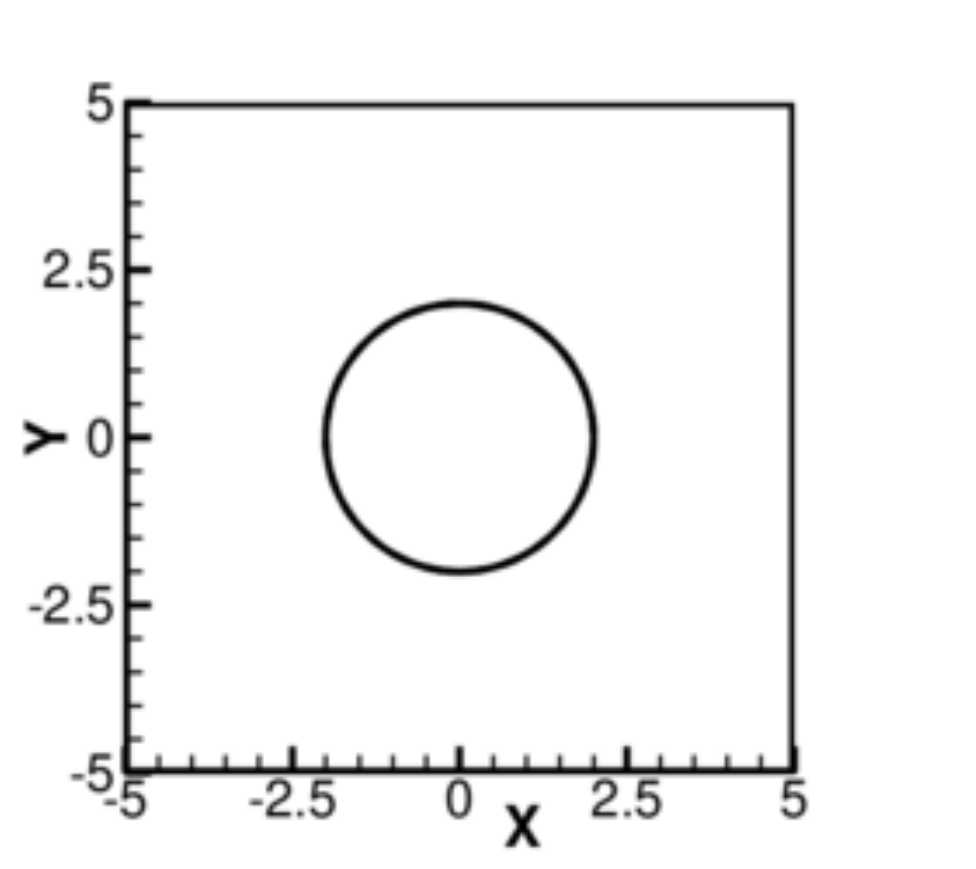}}
\subfloat[CLS-Olsson (no.iter = 10)]
{\includegraphics[scale=0.47]{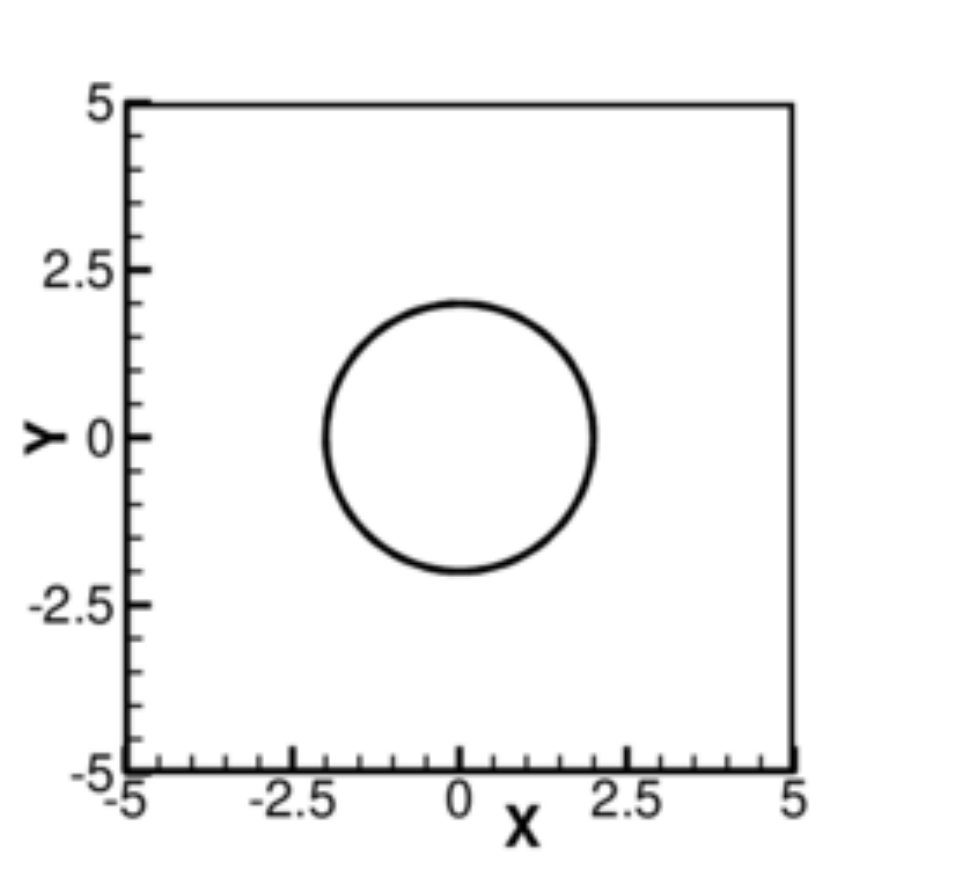}}
\subfloat[CLS-Olsson (no.iter = 100)]
{\includegraphics[scale=0.47]{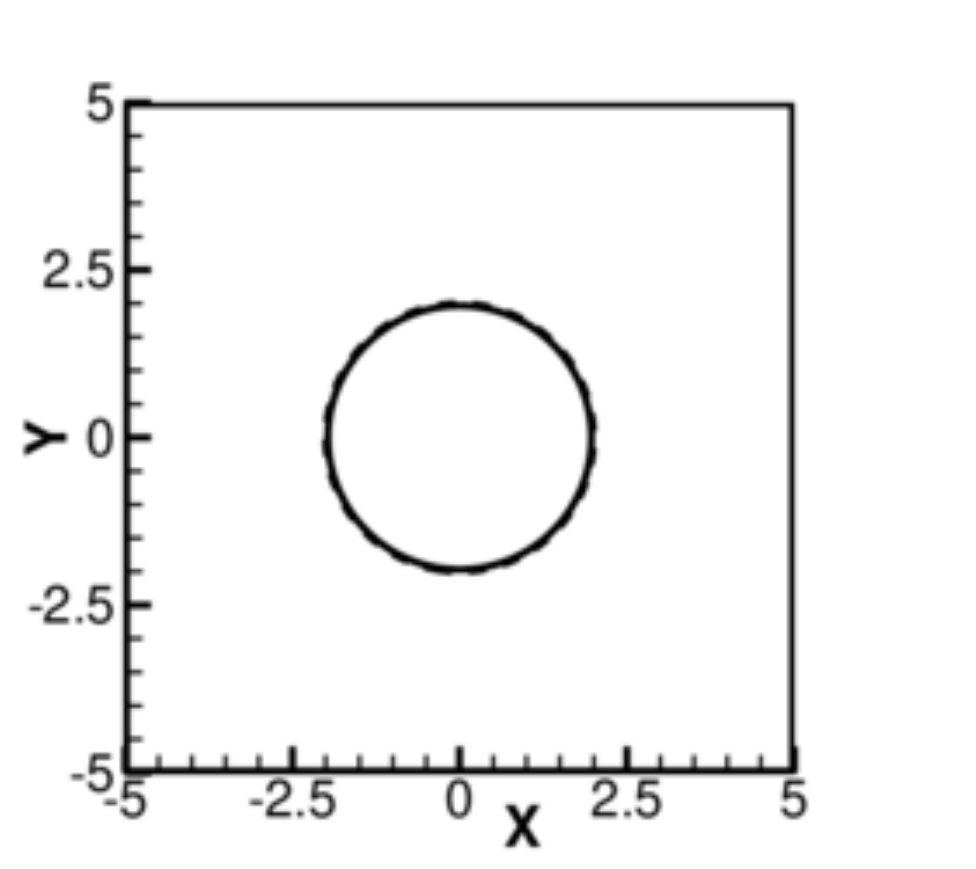}}
\subfloat[CLS-Olsson (no.iter = 250)]
{\includegraphics[scale=0.47]{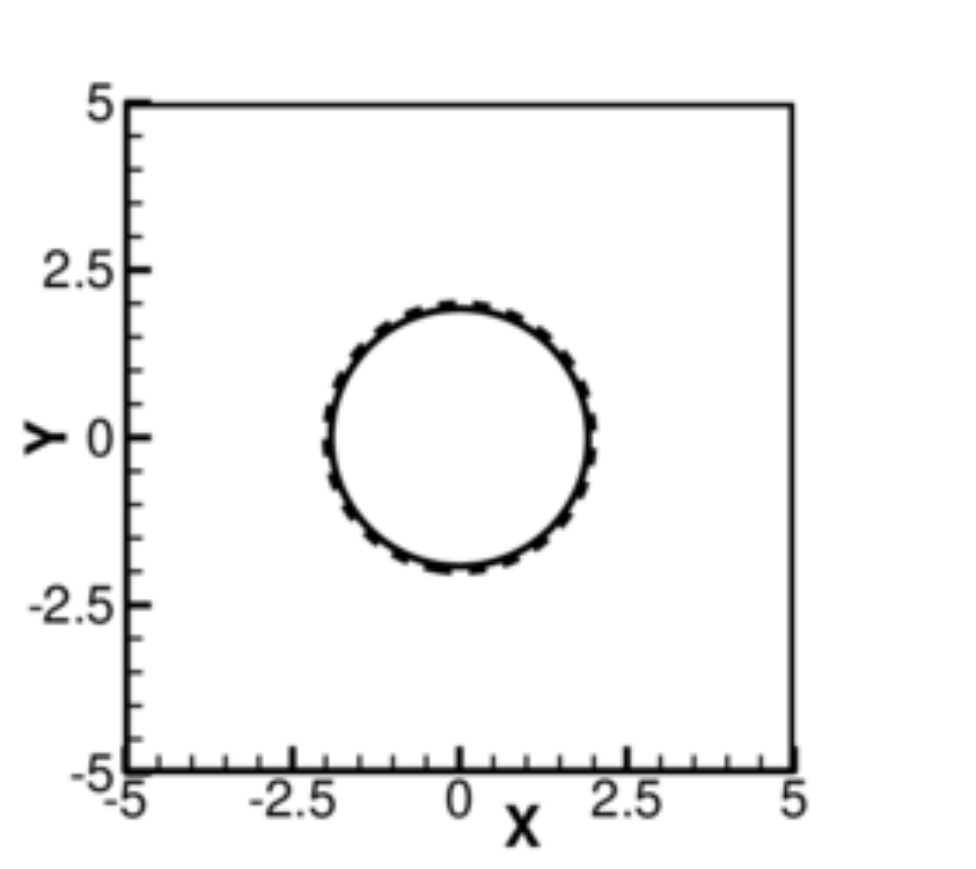}}

\subfloat[New (no.iter = 0)]
{\includegraphics[scale=0.47]{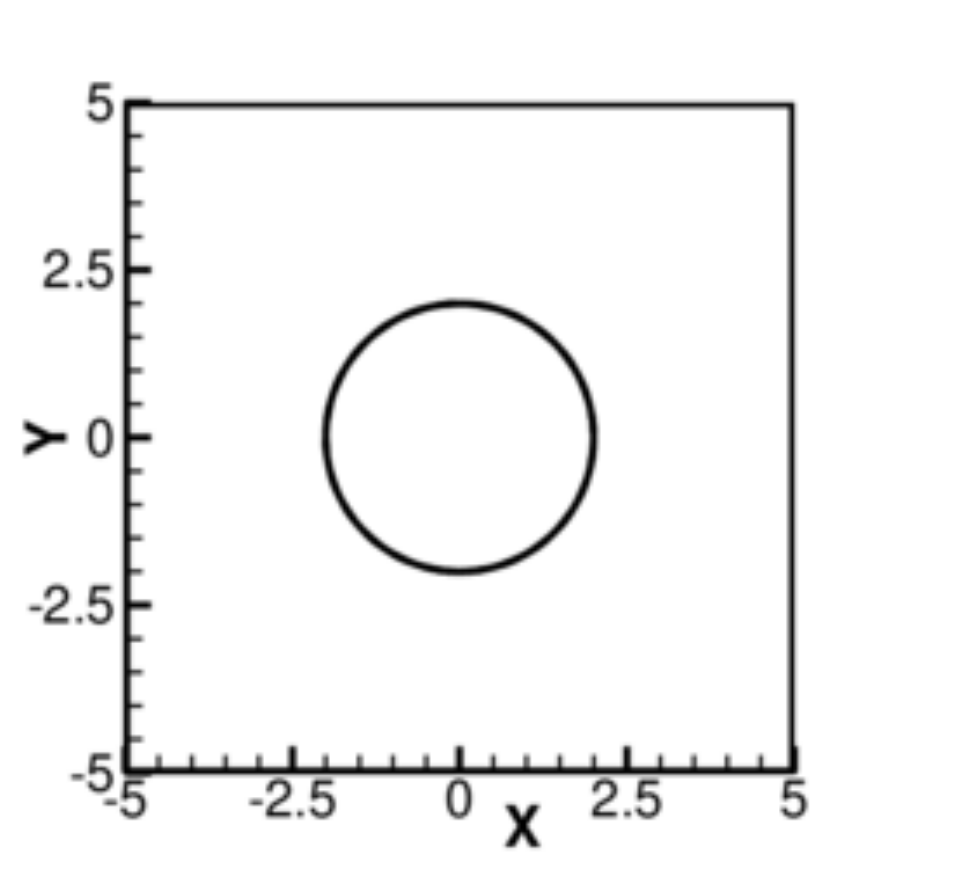}}
\subfloat[New (no.iter = 10)]
{\includegraphics[scale=0.47]{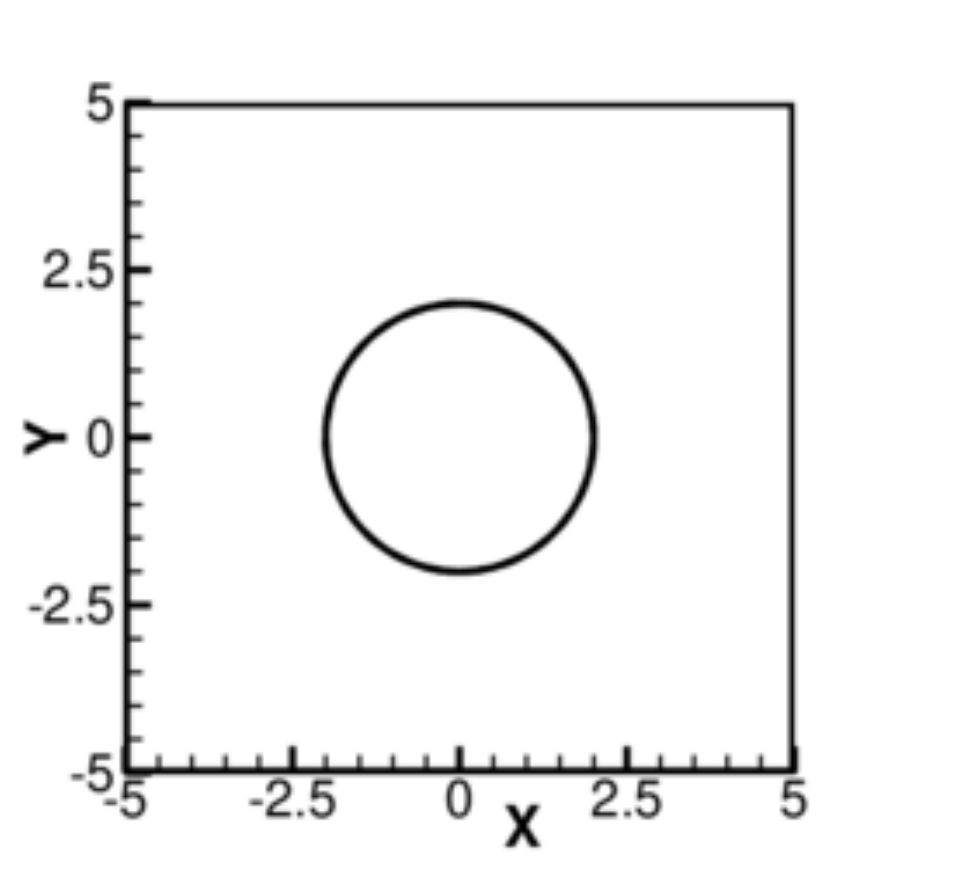}}
\subfloat[New (no.iter = 100)]
{\includegraphics[scale=0.47]{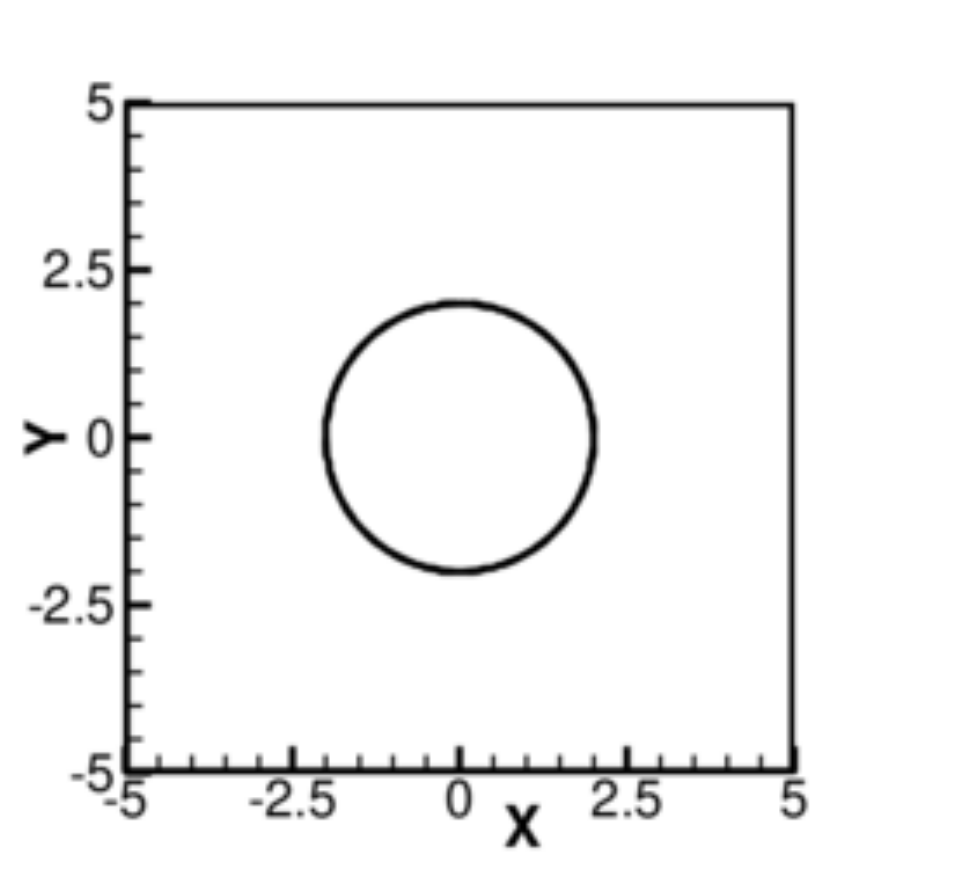}}
\subfloat[New (no.iter = 250)]
{\includegraphics[scale=0.47]{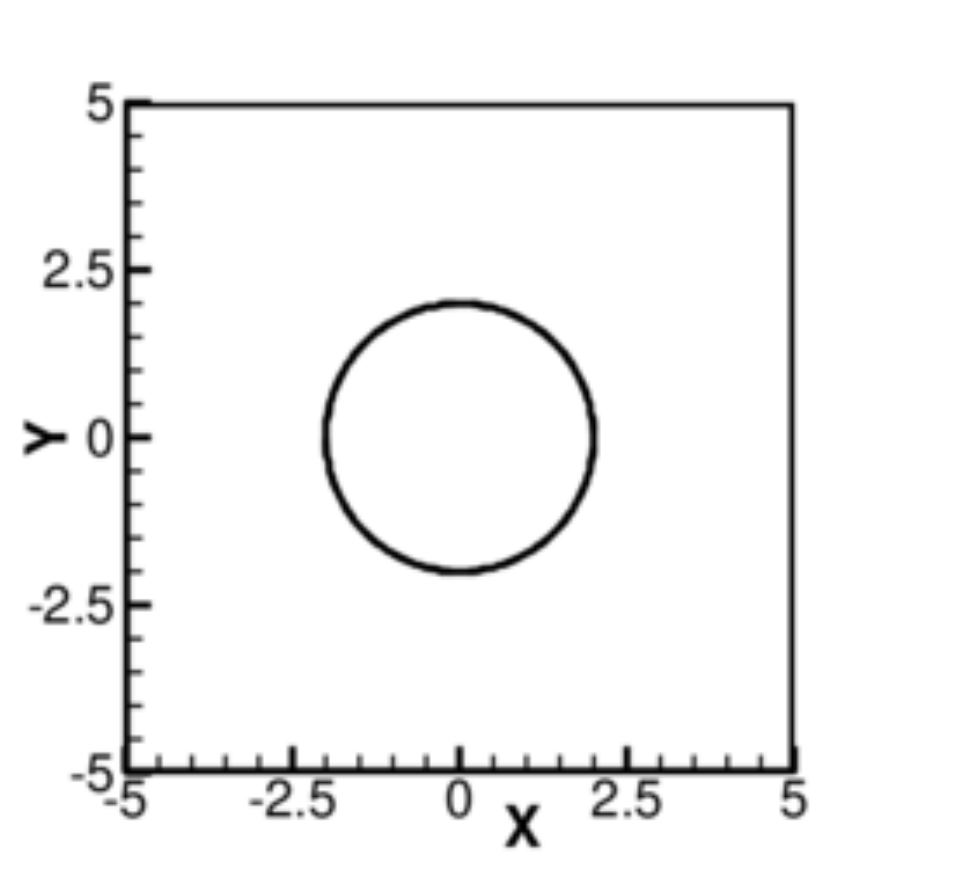}}
\caption{In-place reinitialization of a circular interface using the
  original CLS approach (a) to (d) and the new approach (e) to
  (h). The dashed black curve denotes the initial interface and the
  solid black curves denote the interfaces at the respective
  pseudo-time iterations.}
\label{fig:ipr_circle}
\end{figure}

\begin{figure}[H]
\centering
\subfloat[CLS-Olsson (no.iter = 0)]
{\includegraphics[scale=0.47]{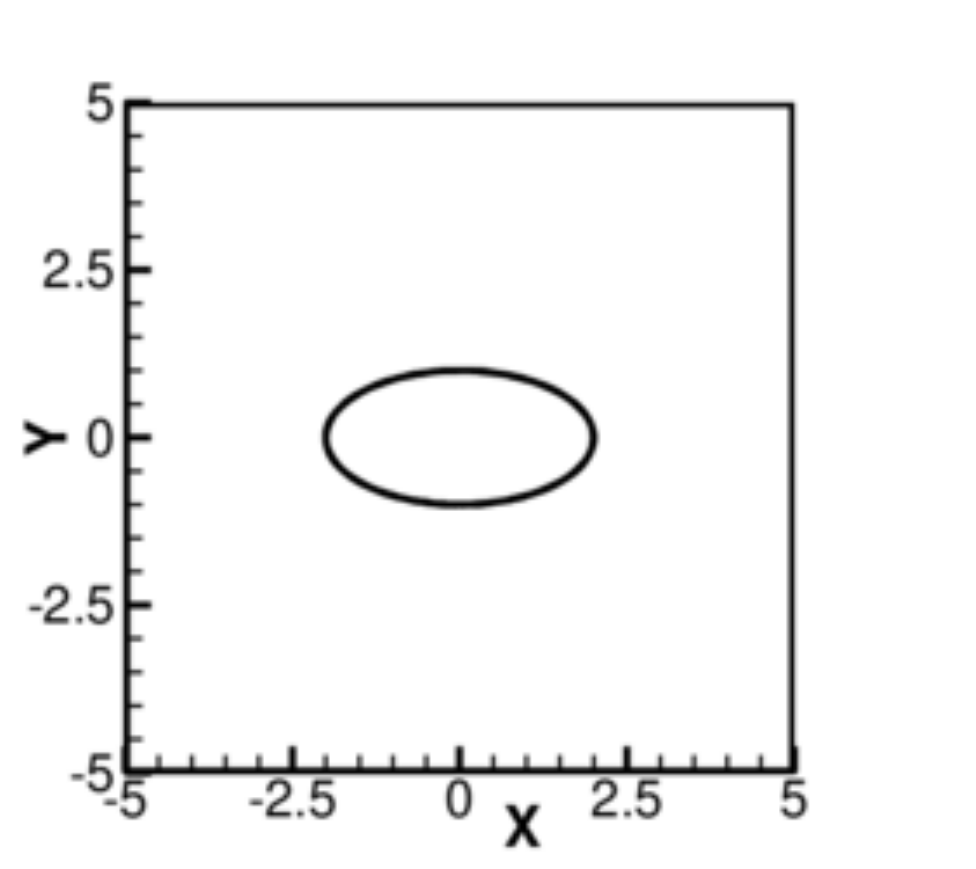}}
\subfloat[CLS-Olsson (no.iter = 10)]
{\includegraphics[scale=0.47]{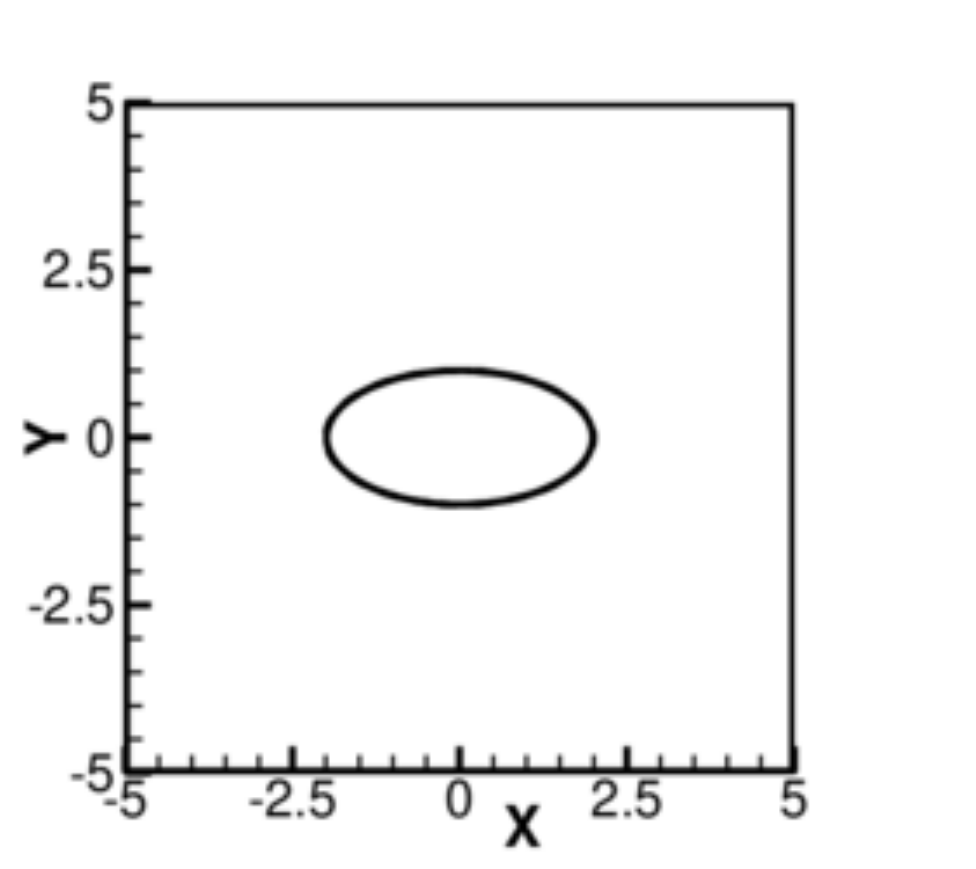}}
\subfloat[CLS-Olsson (no.iter = 100)]
{\includegraphics[scale=0.47]{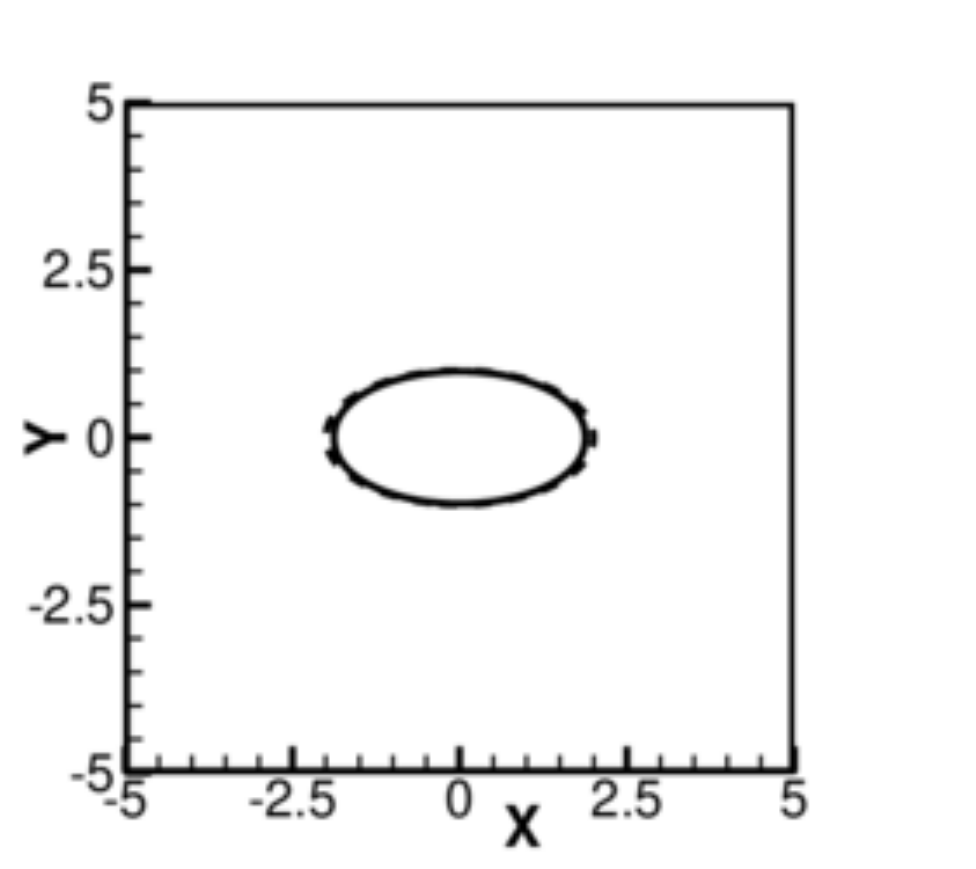}}
\subfloat[CLS-Olsson (no.iter = 250)]
{\includegraphics[scale=0.47]{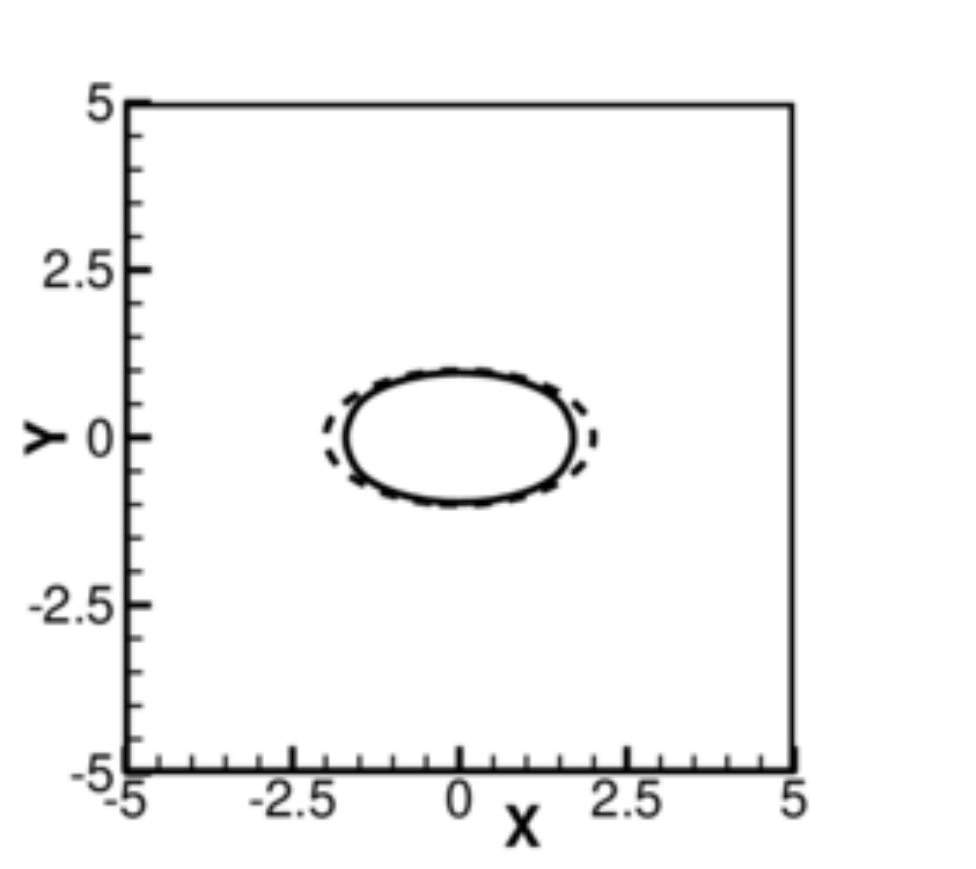}}

\subfloat[New (no.iter = 0)]
{\includegraphics[scale=0.47]{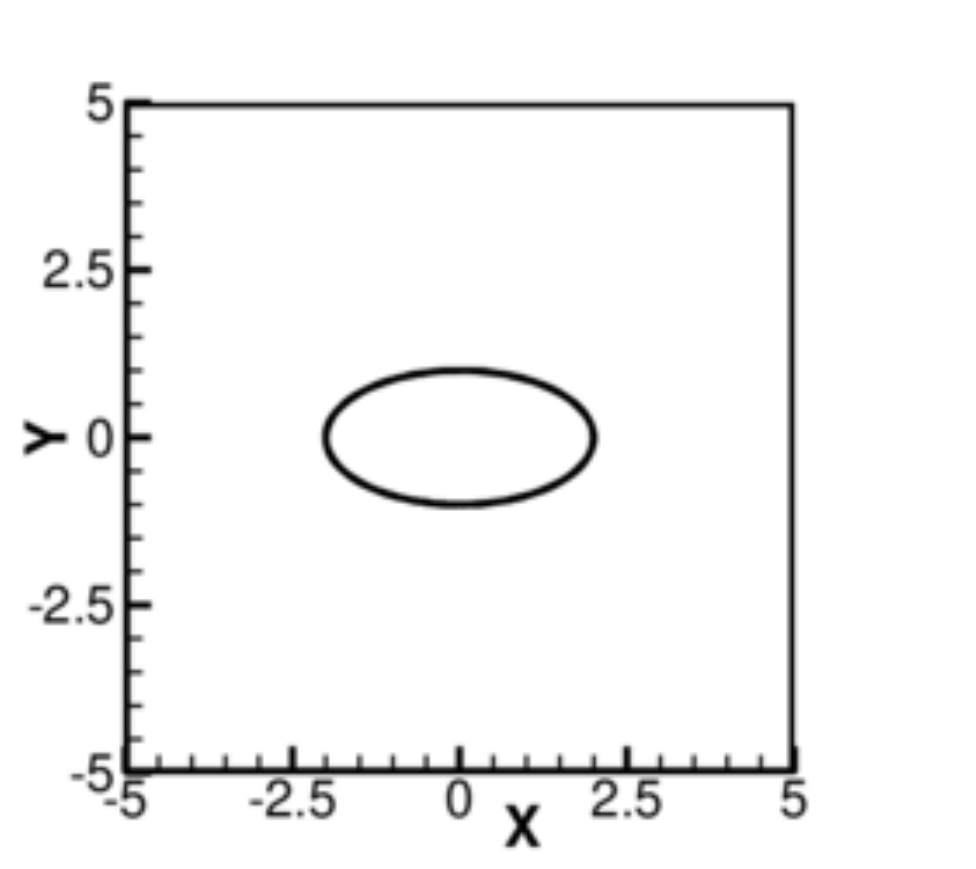}}
\subfloat[New (no.iter = 10)]
{\includegraphics[scale=0.47]{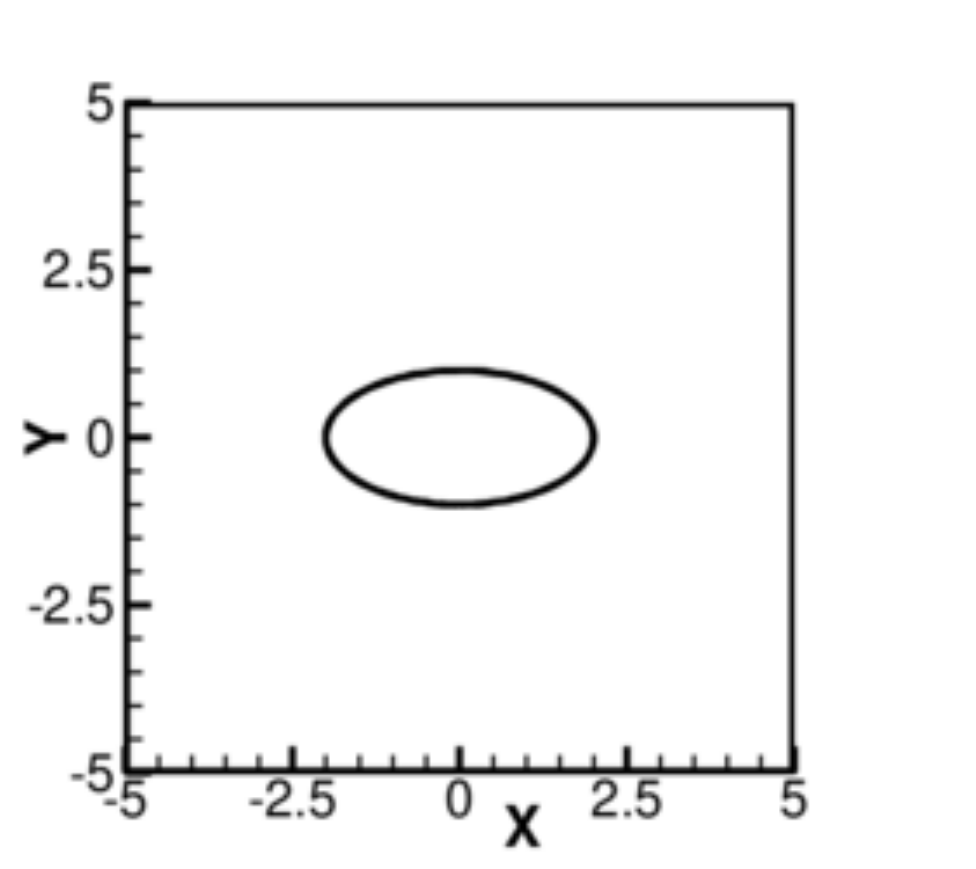}}
\subfloat[New (no.iter = 100)]
{\includegraphics[scale=0.47]{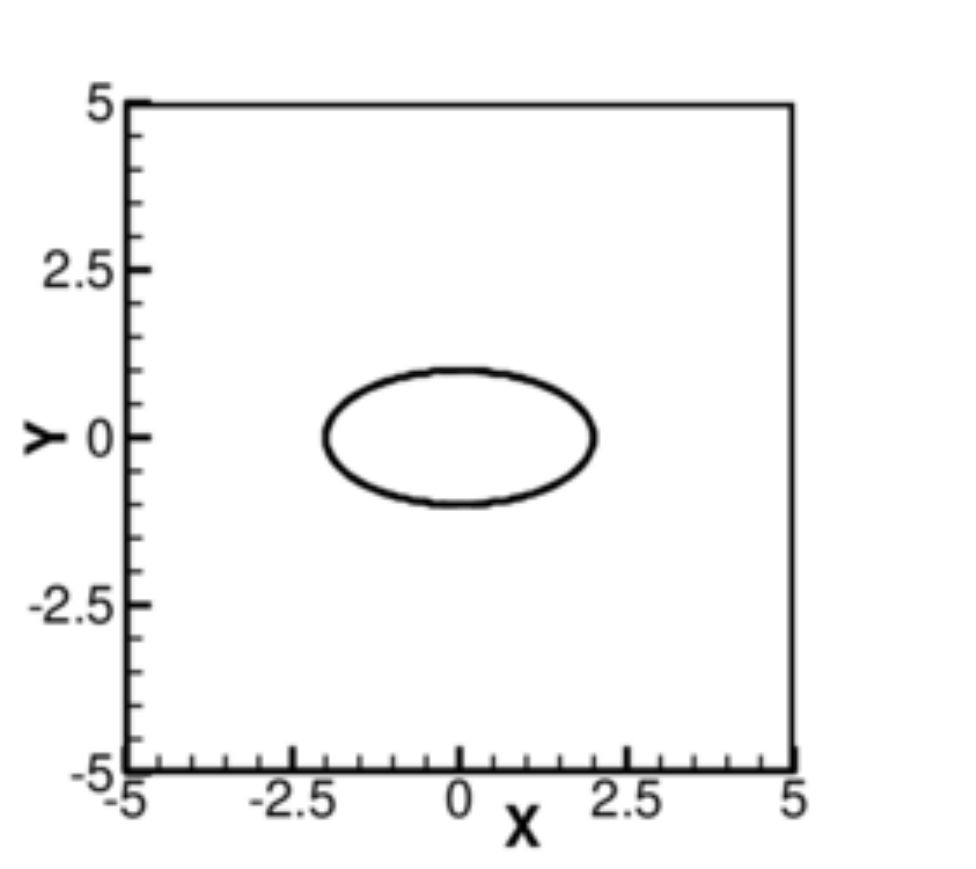}}
\subfloat[New (no.iter = 250)]
{\includegraphics[scale=0.47]{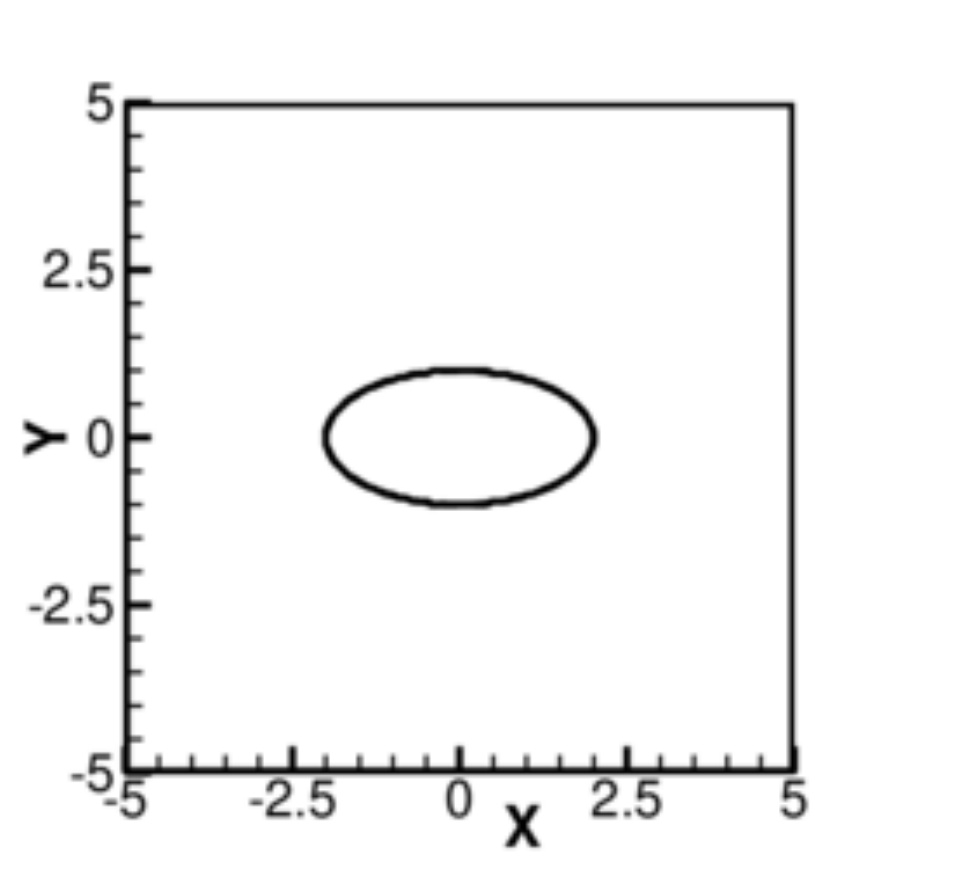}}
\caption{In-place reinitialization of an elliptical interface using
  the original CLS approach (a) to (d) and the new approach (e)
  to (h). The dashed black curve denotes the initial interface and the
  solid black curves denote the interfaces at the respective
  pseudo-time iterations.}
\label{fig:ipr_ellipse}
\end{figure}

\begin{figure}[H]
\centering
\subfloat[CLS-Olsson (no.iter = 0)]
{\includegraphics[scale=0.47]{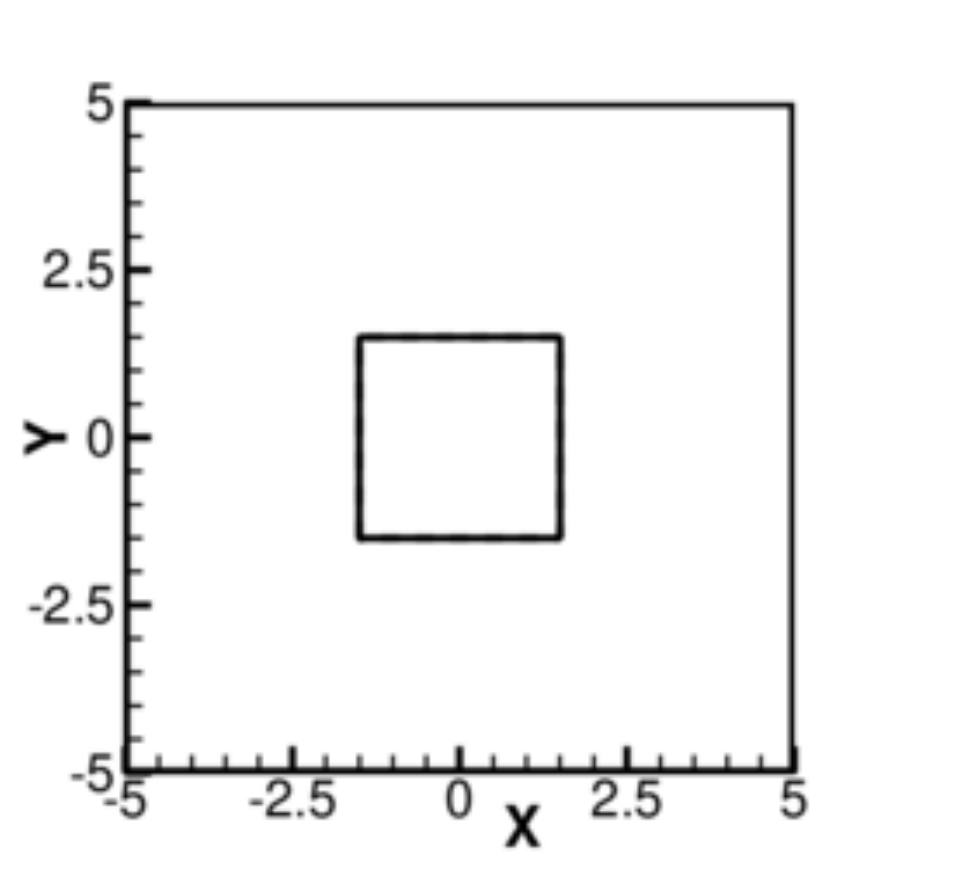}}
\subfloat[CLS-Olsson (no.iter = 10)]
{\includegraphics[scale=0.47]{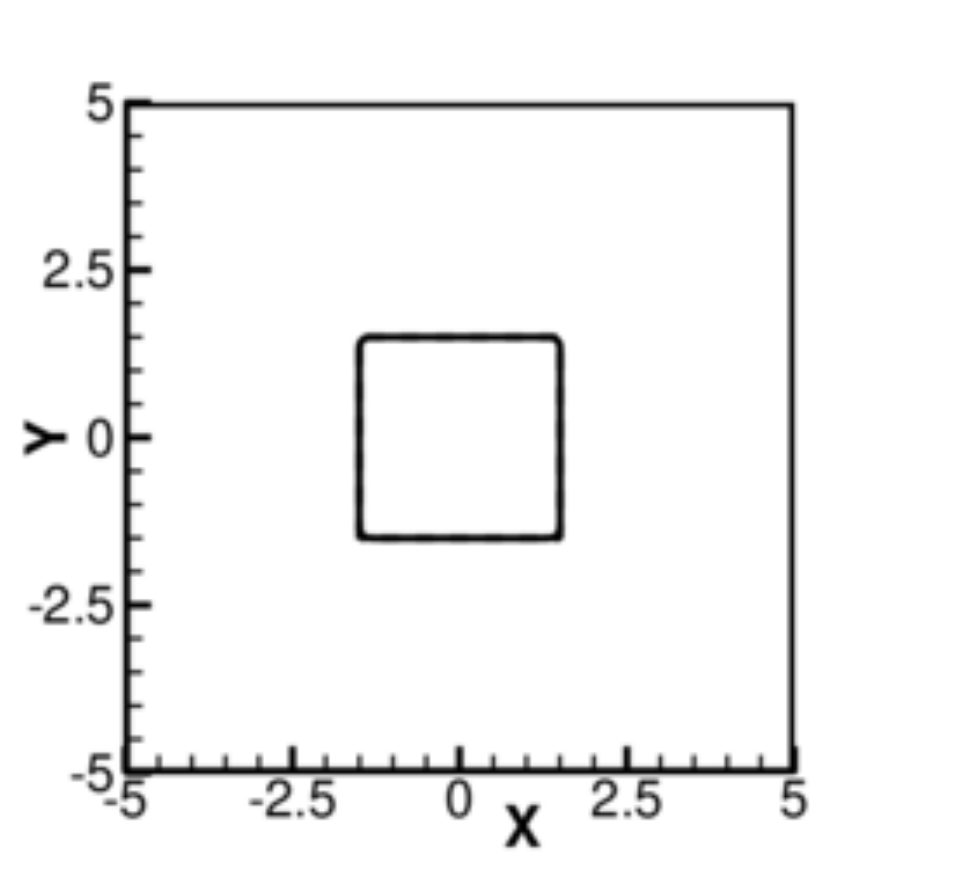}}
\subfloat[CLS-Olsson (no.iter = 100)]
{\includegraphics[scale=0.47]{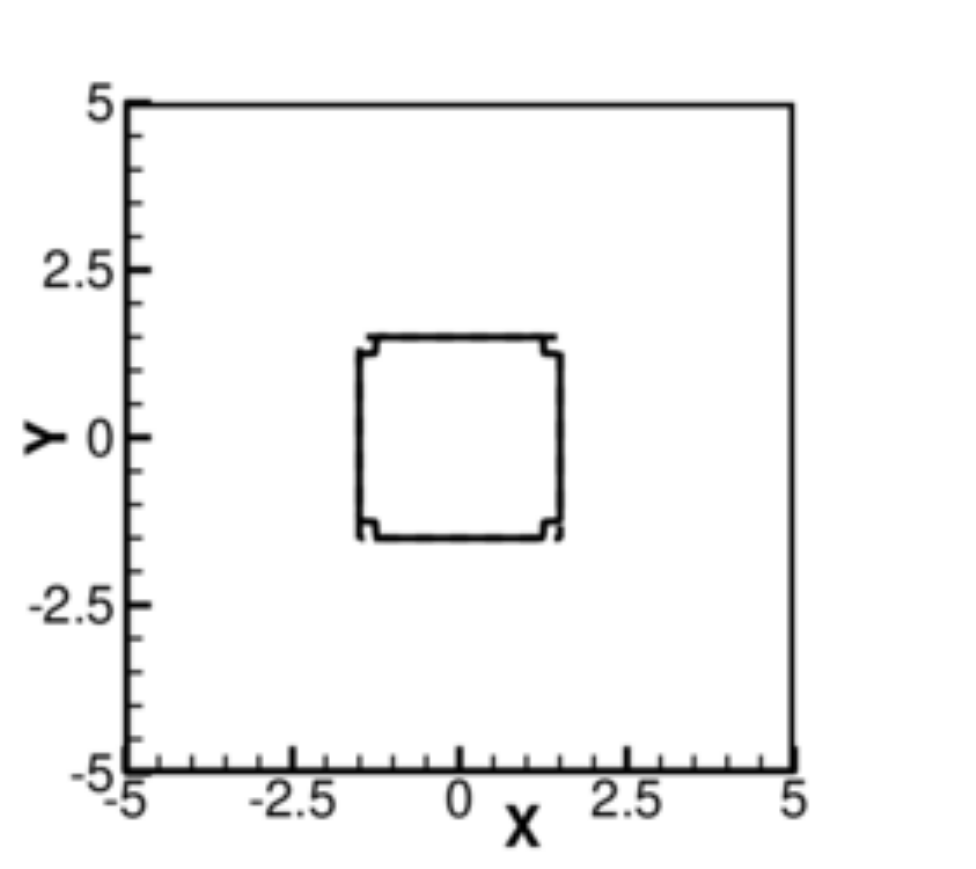}}
\subfloat[CLS-Olsson (no.iter = 250)]
{\includegraphics[scale=0.47]{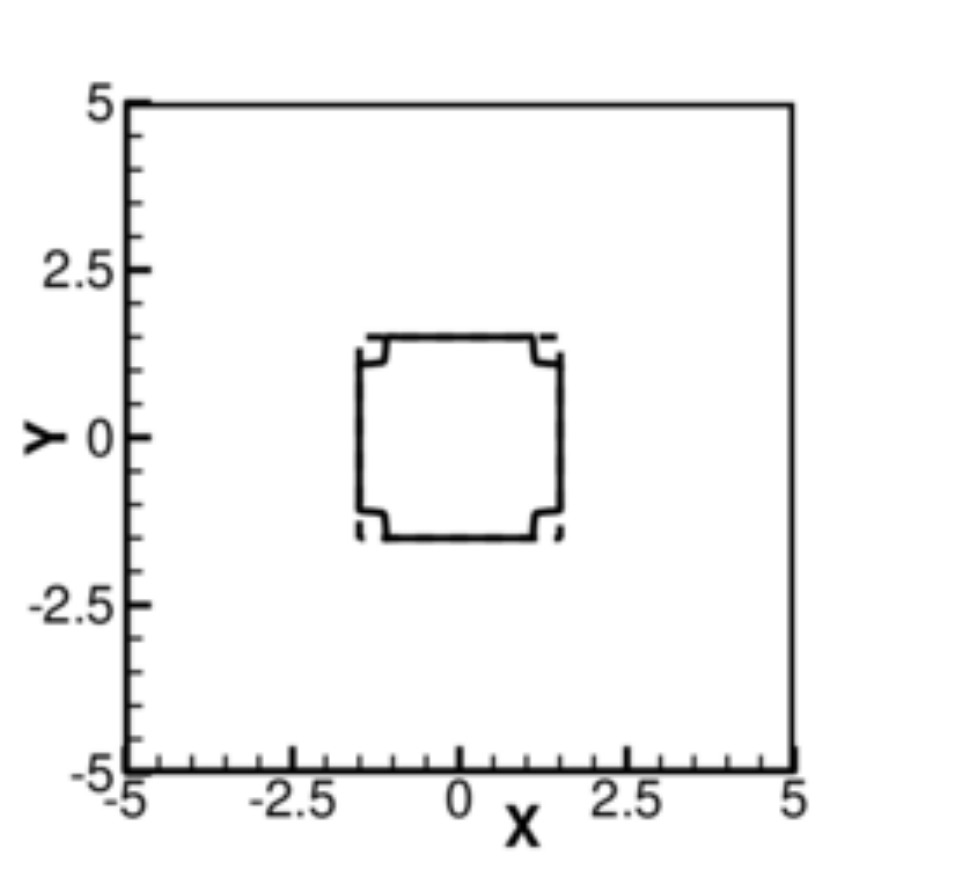}}

\subfloat[New (no.iter = 0)]
{\includegraphics[scale=0.47]{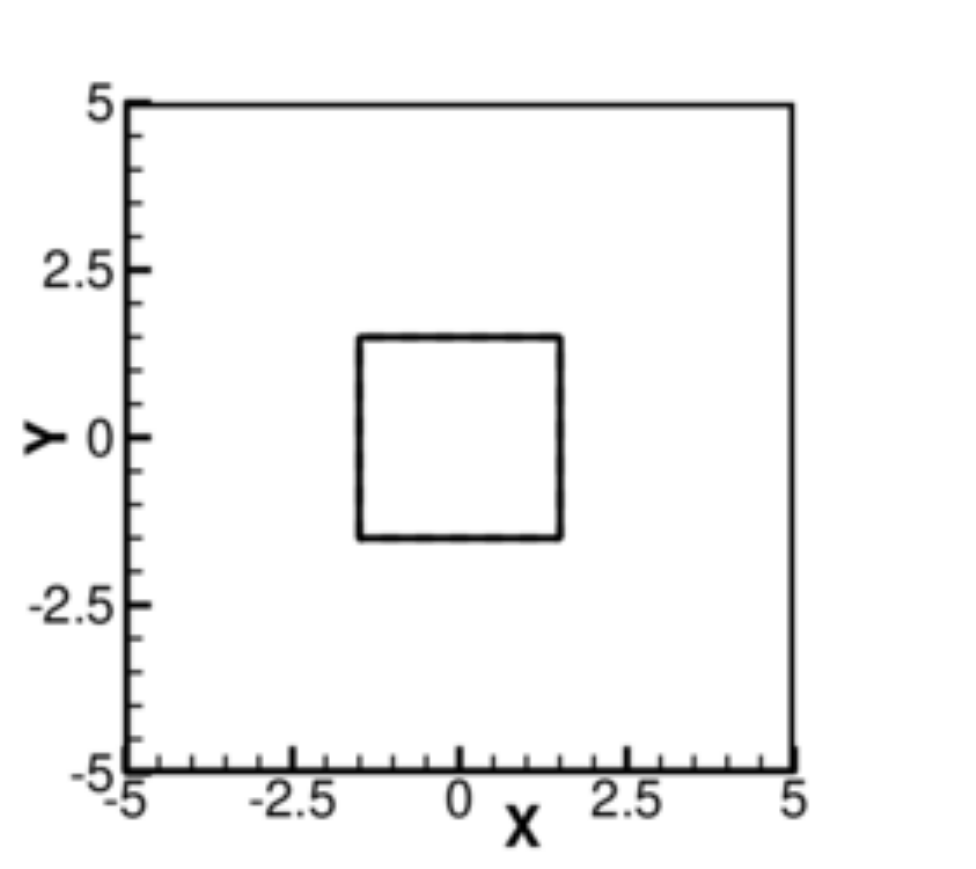}}
\subfloat[New (no.iter = 10)]
{\includegraphics[scale=0.47]{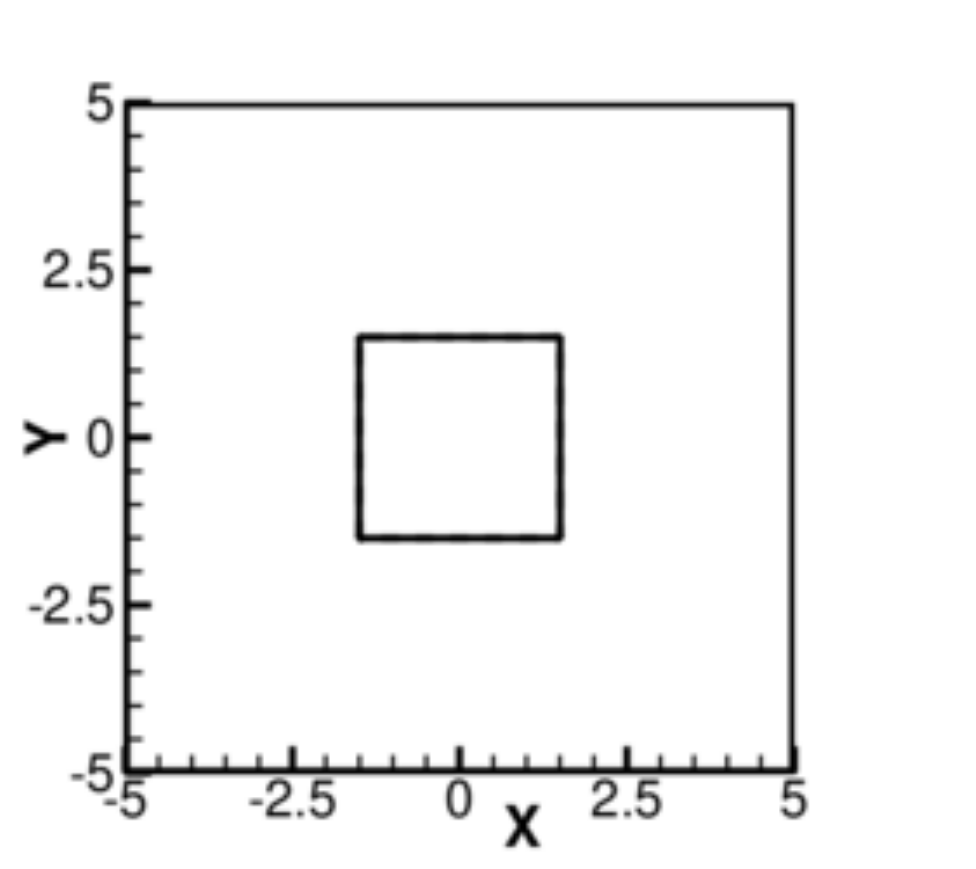}}
\subfloat[New (no.iter = 100)]
{\includegraphics[scale=0.47]{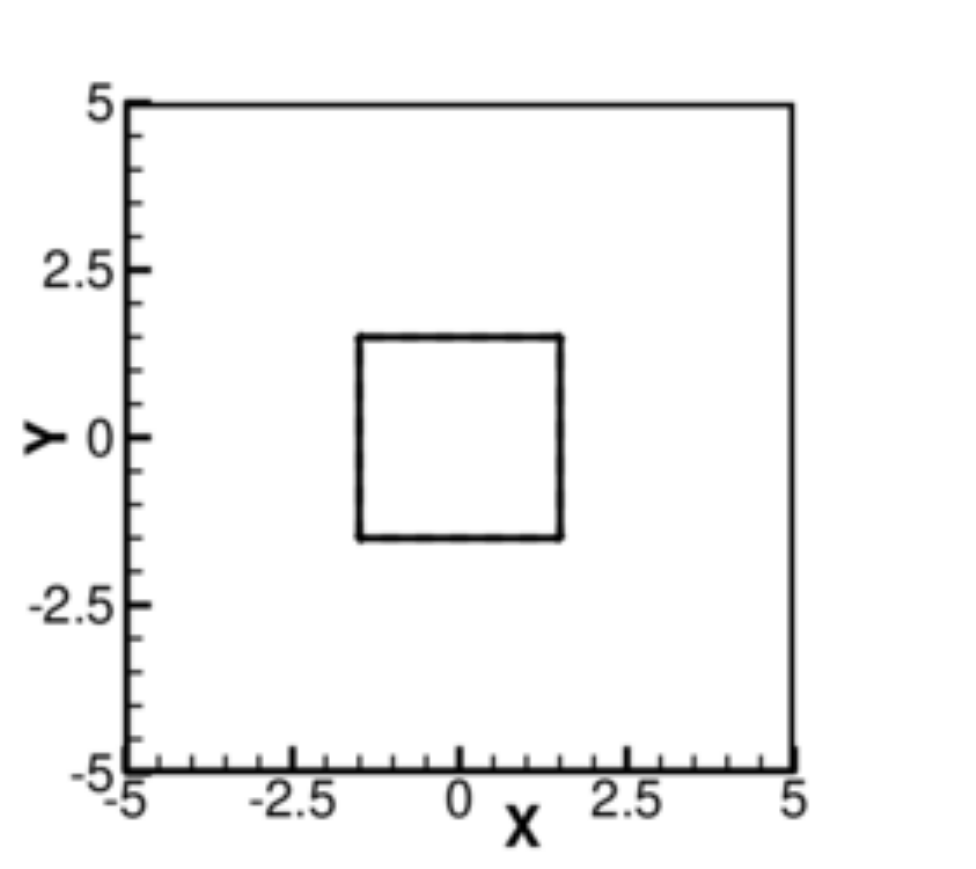}}
\subfloat[New (no.iter = 250)]
{\includegraphics[scale=0.47]{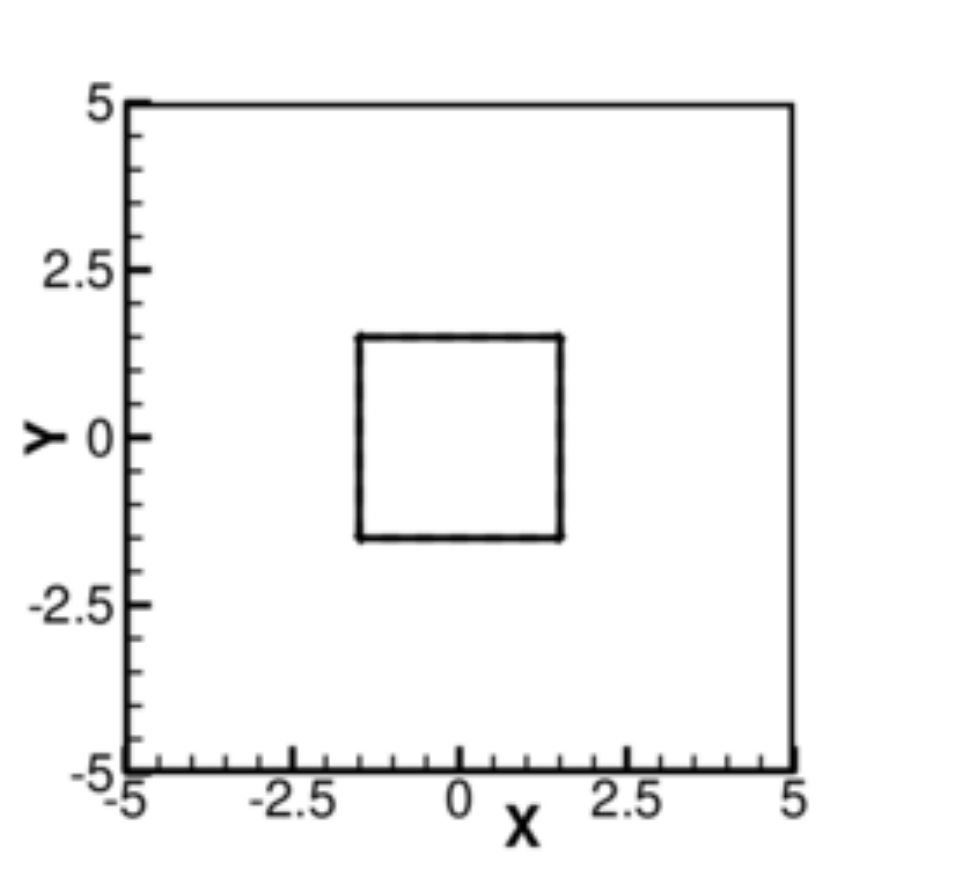}}
\caption{In-place reinitialization of a square shaped interface using
  original CLS approach (a) to (d) and the the new approach (e)
  to (h). The dashed black curve denotes the initial interface and the
  solid black curves denote the interfaces at the respective
  pseudo-time iterations.}
\label{fig:ipr_square}
\end{figure}

In the present study, the original CLS reinitialization
algorithm, described in~\cite{Olsson2007}, and the newly proposed
reinitialization algorithm are compared. The deformation of the
interface contour in both the cases are monitored constantly during
the pseudo-time
iterations. Figure~\ref{fig:ipr_circle},~\ref{fig:ipr_ellipse}~and~\ref{fig:ipr_square}
show the interface contours during the in-place reinitialization
compared with the initial contours in case of the circle, ellipse and
square shapes respectively. The solid black curve denotes the
interface contour during reinitialization and the dashed black curve
denotes the initial interface contour. One can notice that, for all
the three shapes, up to 10 number of reinitialization iterations no
significant changes in the interface contours are visible. However, as
the number of reinitialization iterations increases, the original CLS
approach leads to interface contour deformations. Especially, more
deformations can be observed at regions having higher
curvature. Whereas, there are no visible deformations of the interface
contours even after 250 iterations in case of the new reinitialization approach. This
observation is in well agreement with the discussion given in
section~\ref{sec:artificial_compression}~and~\ref{sec:reformulation}.

\subsection{Reinitialization of Scalar Advection Problems}
\label{sec:scalar_aadv}
In order to further study the performance of the new reinitialization
formulation, a set of standard two-dimensional scalar advection based
test problems are considered next. In the scalar advection problems,
the initial interface is placed at $(0.25,~0.5)$ on a unit square
domain and advected upon a predefined velocity field. Simulations are
carried out till the initial interface completes one full
rotation. The area confined by the interface and the $L^1$ and $L^2$
error norms are monitored during the simulation. The percentage area
error, at any given time $t$, is computed as,
\begin{equation}
\text{Area Error (\%)} = \Bigg(\frac{A^{t} - A^0}{A^0}\Bigg) \times 100
\label{eq:area_error}
\end{equation}
where, $A$ is the area enclosed by the 0.5 contour of the level set
function and the superscript $t$ and $0$ represent the data computed
at time $t$ and at the initial time, $t=0$, respectively. Similarly,
the $L^1$ and $L^2$ error norms are defined as,
\begin{equation}
L^1 = \left( \frac{1}{Nx \times Ny} \right)
\sum_{i=1}^{Nx}\sum_{j=1}^{Ny}\lvert \psi^{2\pi}_{ij} - \psi^0_{ij}\rvert
\end{equation}
and
\begin{equation}
L^2 = \left( \frac{1}{Nx \times Ny} \right)
\sqrt{\sum_{i=1}^{Nx}\sum_{j=1}^{Ny}\Big(\psi^{2\pi}_{ij} - \psi^0_{ij}\Big)^2}
\end{equation}
where, $\psi^{2\pi}_{ij}$ and $\psi^0_{ij}$ are the discretized level
set functions defined at time level $t = 2\pi$ and $t = 0$ respectively,
and $Nx$ and $Ny$ are the total number of cells along $x$ and $y$
directions respectively. In this section, the test problems are solved
using the reinitialization schemes reported in~\cite{Olsson2007,
  Desjardins2008, Waclawczyk2015, Chiodi2017} along with the new
scheme. For better clarity, the schemes are denoted here as
CLS-Olsson, ACLS-Desjardins, CLS-Wac{\l}awczyk and CLS-Chiodi for
references \cite{Olsson2007}, \cite{Desjardins2008},
\cite{Waclawczyk2015} and \cite{Chiodi2017} respectively.

\subsubsection{Reinitialization of circular disc rotation problem}
\begin{figure}[H]
\centering
\subfloat[CLS-Olsson]
{\includegraphics[scale=1.8]{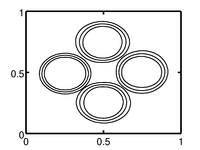}}
\subfloat[ACLS-Desjardins]
{\includegraphics[scale=1.8]{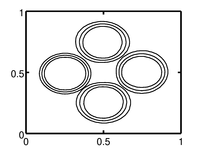}}
\subfloat[CLS-Wac{\l}awczyk]
{\includegraphics[scale=1.8]{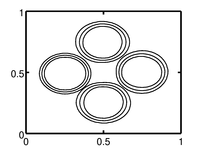}}

\subfloat[CLS-Chiodi]
{\includegraphics[scale=1.8]{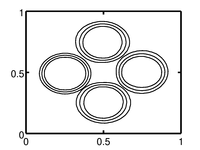}}
\subfloat[New]
{\includegraphics[scale=1.8]{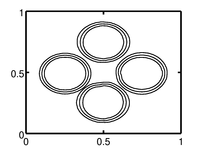}}
\caption{Comparison of reinitialization schemes for the
  reinitialization of circular disc rotation problem solved on $100
  \times 100$ grid. Each subfigure shows the 0.05, 0.5 and 0.95 level
  contours of the level set functions at time levels of 0, $\pi$/4,
  $\pi$/2 and $3\pi$/4 (left, top, right and bottom respectively).}
\label{fig:OC_contours}
\end{figure}

\begin{figure}[H]
\centering
\subfloat[$L^1$ Error]
{\includegraphics[scale=0.22]{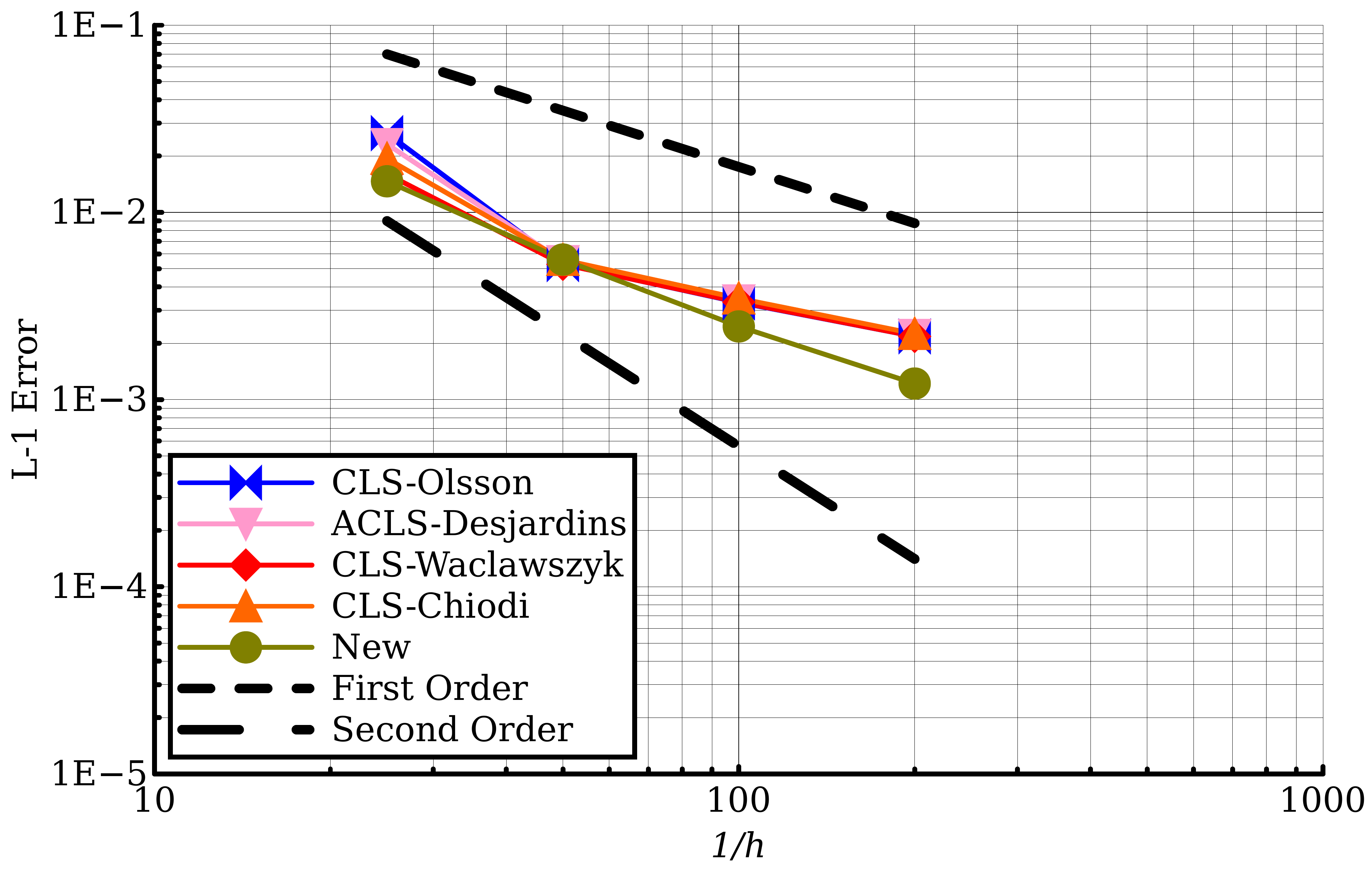}} \hspace{0.5cm}
\subfloat[$L^2$ Error]
{\includegraphics[scale=0.22]{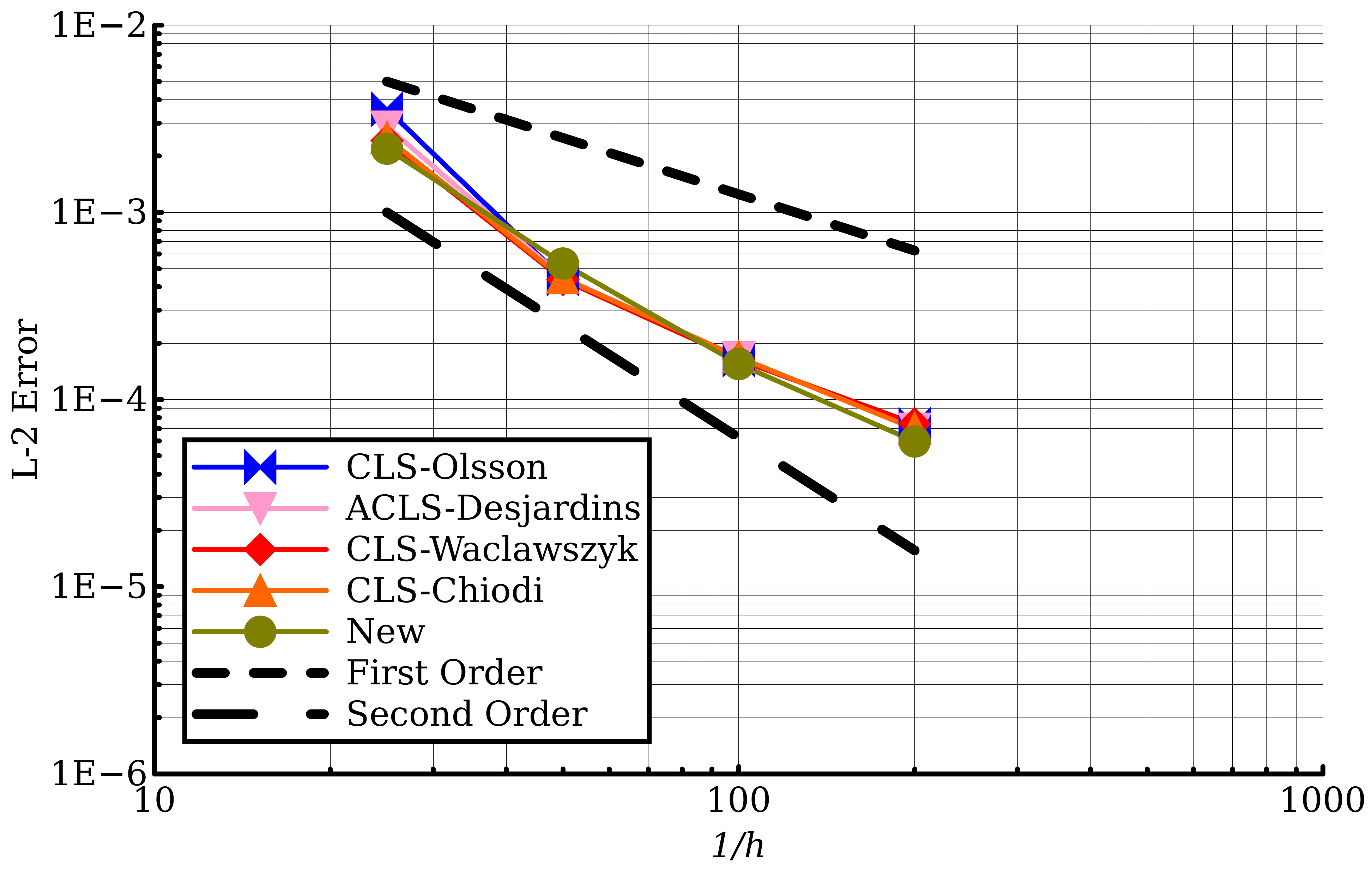}}
\caption{Convergence of the $L^1$ and $L^2$ error norms for the
  reinitialization of circular disc rotation problem solved on four
  ($25 \times 25$, $50 \times 50$, $100 \times 100$ and $200 \times
  200$) grid levels.}
\label{fig:OC_order_error}
\end{figure}

Rotation of a circular disc, similar to the test reported in
\cite{Olsson2005}, is considered first, where, a circular disc of
radius 0.15 units is advected upon a velocity field $u = (y-0.5)$ and
$v = (0.5-x)$. Test problem is solved on four levels of Cartesian
meshes starting from $25 \times 25$ up to $200 \times 200$. The level
set contours correspond to $\psi = 0.05, \psi = 0.5$ and $\psi =
0.95$ for the $100 \times 100$ case at time levels $t = 0$,
$t = \pi/4$, $t = \pi/2$ and $t = 3\pi/4$ for all five
reinitialization schemes are plotted in
Figure~\ref{fig:OC_contours}. Further,
Table~\ref{tab:OC_area_error}~and~\ref{tab:OC_l1_l2} show the
percentage area error and error norms computed after the disc
completes one full rotation ($ie.,$ at $t = 2 \pi$). Moreover, the
error norms are plotted against the mesh size in
Figure~\ref{fig:OC_order_error} along with reference slopes for the
first and second order rate of convergence. Looking at the above
figures and tables, it can be noticed that the new reinitialization
shows the least error among all.

\begin{table}[H]
\begin{center}
\caption{Percentage area errors computed after completion of one full rotation of different reinitialization schemes for the reinitialization of circular disc rotation problem.}
\label{tab:OC_area_error}
\begin{tabular}{|c|c|c|c|c|c|}
\hline
{\color[HTML]{000000} }                                & \multicolumn{5}{c|}{{\color[HTML]{000000} \text{Reinitialization Schemes}}}                                                                                                                                                    \\ \cline{2-6} 
\multirow{-2}{*}{{\color[HTML]{000000} \text{Mesh}}} & {\color[HTML]{000000} \text{CLS-Olsson}} & {\color[HTML]{000000} \text{ACLS-Desjardins}} & {\color[HTML]{000000} \text{CLS-Wac{\l}awszyk}} & {\color[HTML]{000000} \text{CLS-Chiodi}} & {\color[HTML]{000000} \text{New}} \\ \hline
{\color[HTML]{000000} $25 \times 25$}                  & {\color[HTML]{000000} -5.48500}            & {\color[HTML]{000000} -5.04690}                 & {\color[HTML]{000000} -4.38460}                & {\color[HTML]{000000} -4.41610}            & {\color[HTML]{000000} -1.95050}     \\ \hline
{\color[HTML]{000000} $50 \times 50$}                  & {\color[HTML]{000000} -2.21890}            & {\color[HTML]{000000} -2.12190}                 & {\color[HTML]{000000} -2.15960}                & {\color[HTML]{000000} -2.03820}            & {\color[HTML]{000000} -0.15707}     \\ \hline
{\color[HTML]{000000} $100 \times 100$}                & {\color[HTML]{000000} -1.99710}            & {\color[HTML]{000000} -1.94980}                 & {\color[HTML]{000000} -1.91780}                & {\color[HTML]{000000} -1.74080}            & {\color[HTML]{000000} -0.02158}     \\ \hline
{\color[HTML]{000000} $200 \times 200$}                & {\color[HTML]{000000} -1.71600}            & {\color[HTML]{000000} -1.68580}                 & {\color[HTML]{000000} -1.71300}                & {\color[HTML]{000000} -1.65980}            & {\color[HTML]{000000} -0.00865}     \\ \hline
\end{tabular}
\end{center}
\end{table}

\begin{table}[H]
\begin{center}
\caption{The $L^1$ and $L^2$ error norms of different reinitialization schemes for the reinitialization of circular disc rotation problem.}
\label{tab:OC_l1_l2}
\begin{tabular}{|c|c|c|c|c|c|c|}
\hline
{\color[HTML]{000000} }                                     & {\color[HTML]{000000} }                                & \multicolumn{5}{c|}{{\color[HTML]{000000} \text{Reinitialization Schemes}}}                                                                                                                                                    \\ \cline{3-7} 
\multirow{-2}{*}{{\color[HTML]{000000} \text{Errors}}}    &
\multirow{-2}{*}{{\color[HTML]{000000} \text{Mesh}}} &
         {\color[HTML]{000000} \text{CLS-Olsson}} &
         {\color[HTML]{000000} \text{ACLS-Desjardins}} &
         {\color[HTML]{000000} \text{CLS-Wac{\l}awszyk}} &
         {\color[HTML]{000000} \text{CLS-Chiodi}} &
         {\color[HTML]{000000} \text{New}} \\ \hline 
\hline
{\color[HTML]{000000} }                              & {\color[HTML]{000000} $25 \times 25$}         & {\color[HTML]{000000} 2.6488E-02} & {\color[HTML]{000000} 2.3154E-02}      & {\color[HTML]{000000} 2.1835E-02}     & {\color[HTML]{000000} 1.9378E-02} & {\color[HTML]{000000} 1.4665E-02} \\ \cline{2-7} 
{\color[HTML]{000000} }                              & {\color[HTML]{000000} $50 \times 50$}         & {\color[HTML]{000000} 5.2915E-03} & {\color[HTML]{000000} 5.4820E-03}      & {\color[HTML]{000000} 5.2766E-03}     & {\color[HTML]{000000} 5.5800E-03} & {\color[HTML]{000000} 5.5977E-03} \\ \cline{2-7} 
{\color[HTML]{000000} }                              & {\color[HTML]{000000} $100 \times 100$}       & {\color[HTML]{000000} 3.3020E-03} & {\color[HTML]{000000} 3.3816E-03}      & {\color[HTML]{000000} 3.3176E-03}     & {\color[HTML]{000000} 3.4732E-03} & {\color[HTML]{000000} 2.4589E-03} \\ \cline{2-7} 
\multirow{-4}{*}{{\color[HTML]{000000} $L^1$ Error}} & {\color[HTML]{000000} $200 \times 200$}       & {\color[HTML]{000000} 2.1747E-03} & {\color[HTML]{000000} 2.2122E-03}      & {\color[HTML]{000000} 2.1747E-03}     & {\color[HTML]{000000} 2.2454E-03} & {\color[HTML]{000000} 1.2175E-03} \\ \hline \hline
{\color[HTML]{000000} }                              & {\color[HTML]{000000} $25 \times 25$}         & {\color[HTML]{000000} 3.5611E-03} & {\color[HTML]{000000} 2.8689E-03}      & {\color[HTML]{000000} 2.4137E-03}     & {\color[HTML]{000000} 2.4929E-03} & {\color[HTML]{000000} 2.4929E-03} \\ \cline{2-7} 
{\color[HTML]{000000} }                              & {\color[HTML]{000000} $50 \times 50$}         & {\color[HTML]{000000} 4.4320E-04} & {\color[HTML]{000000} 4.4905E-04}      & {\color[HTML]{000000} 4.3691E-04}     & {\color[HTML]{000000} 4.4555E-04} & {\color[HTML]{000000} 4.4555E-04} \\ \cline{2-7} 
{\color[HTML]{000000} }                              & {\color[HTML]{000000} $100 \times 100$}       & {\color[HTML]{000000} 1.6393E-04} & {\color[HTML]{000000} 1.6817E-04}      & {\color[HTML]{000000} 1.6307E-04}     & {\color[HTML]{000000} 1.6827E-04} & {\color[HTML]{000000} 1.6827E-04} \\ \cline{2-7} 
\multirow{-4}{*}{{\color[HTML]{000000} $L^2$ Error}} & {\color[HTML]{000000} $200 \times 200$}       & {\color[HTML]{000000} 7.3160E-05} & {\color[HTML]{000000} 6.9776E-05}      & {\color[HTML]{000000} 7.4308E-05}     & {\color[HTML]{000000} 6.9736E-05} & {\color[HTML]{000000} 6.9736E-05} \\ \hline
\end{tabular}
\end{center}
\end{table}

\subsubsection{Reinitialization of Zalesak's disc rotation problem}
\begin{figure}[H]
\centering
\subfloat[CLS-Olsson]
{\includegraphics[scale=1.8]{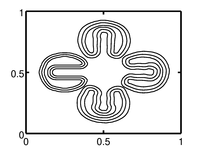}}
\subfloat[ACLS-Desjardins]
{\includegraphics[scale=1.8]{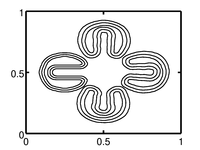}}
\subfloat[CLS-Wac{\l}awczyk]
{\includegraphics[scale=1.8]{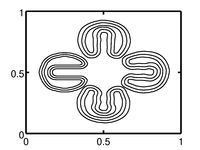}}

\subfloat[CLS-Chiodi]
{\includegraphics[scale=1.8]{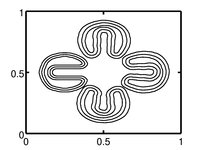}}
\subfloat[New]
{\includegraphics[scale=1.8]{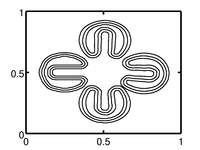}}
\caption{Comparison of reinitialization schemes for the
  reinitialization of Zalesak's disc rotation problem solved on $100
  \times 100$ grid. Each subfigure shows the 0.05, 0.5 and 0.95 level
  contours of the level set functions at time levels of 0.0, $\pi$/4,
  $\pi$/2 and $3\pi$/4 (left, top, right and bottom respectively).}
\label{fig:ZD_contours}
\end{figure}

\begin{figure}[H]
\centering
\subfloat[$L^1$ Error]
{\includegraphics[scale=0.22]{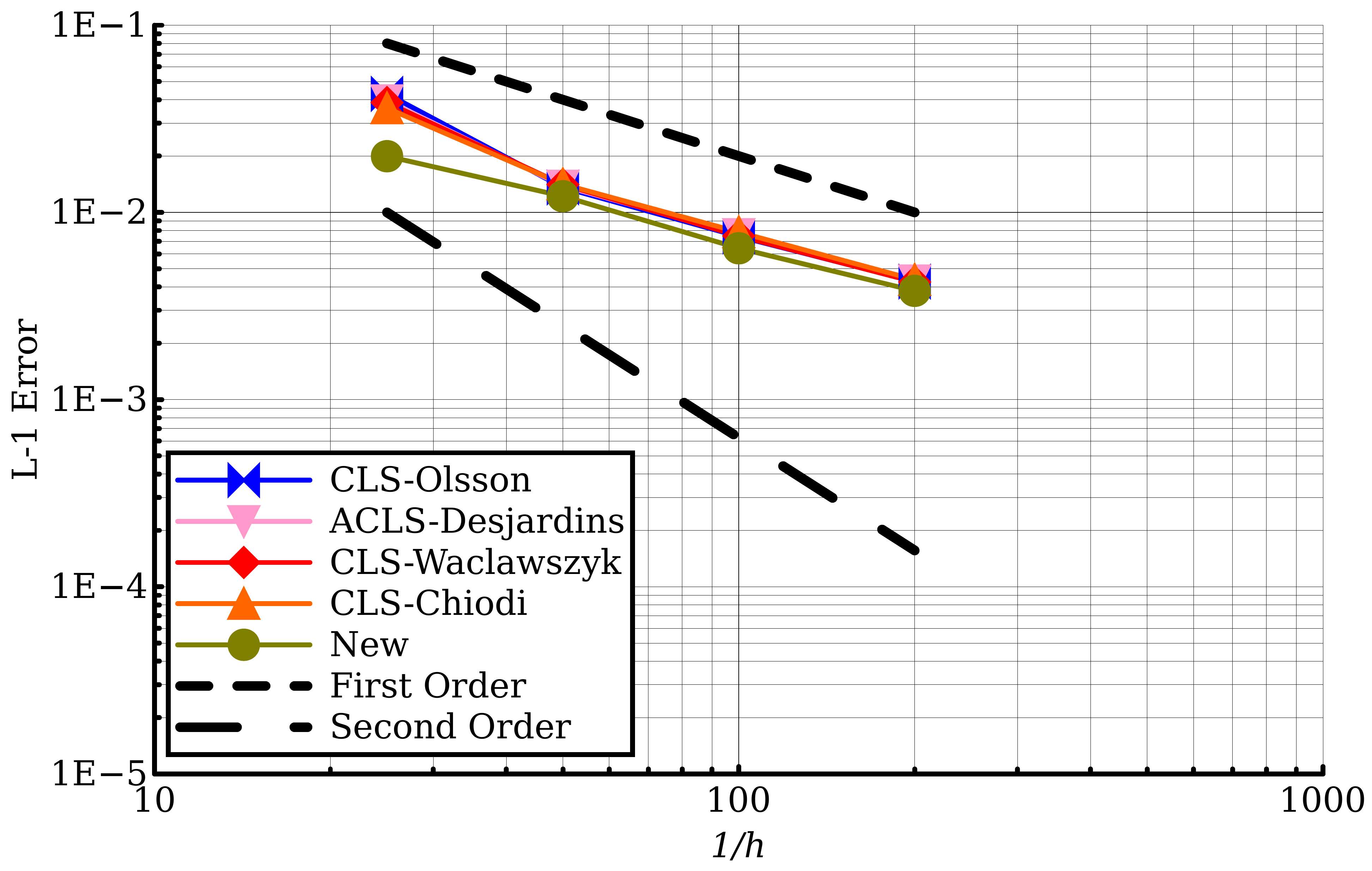}} \hspace{0.5cm}
\subfloat[$L^2$ Error]
{\includegraphics[scale=0.22]{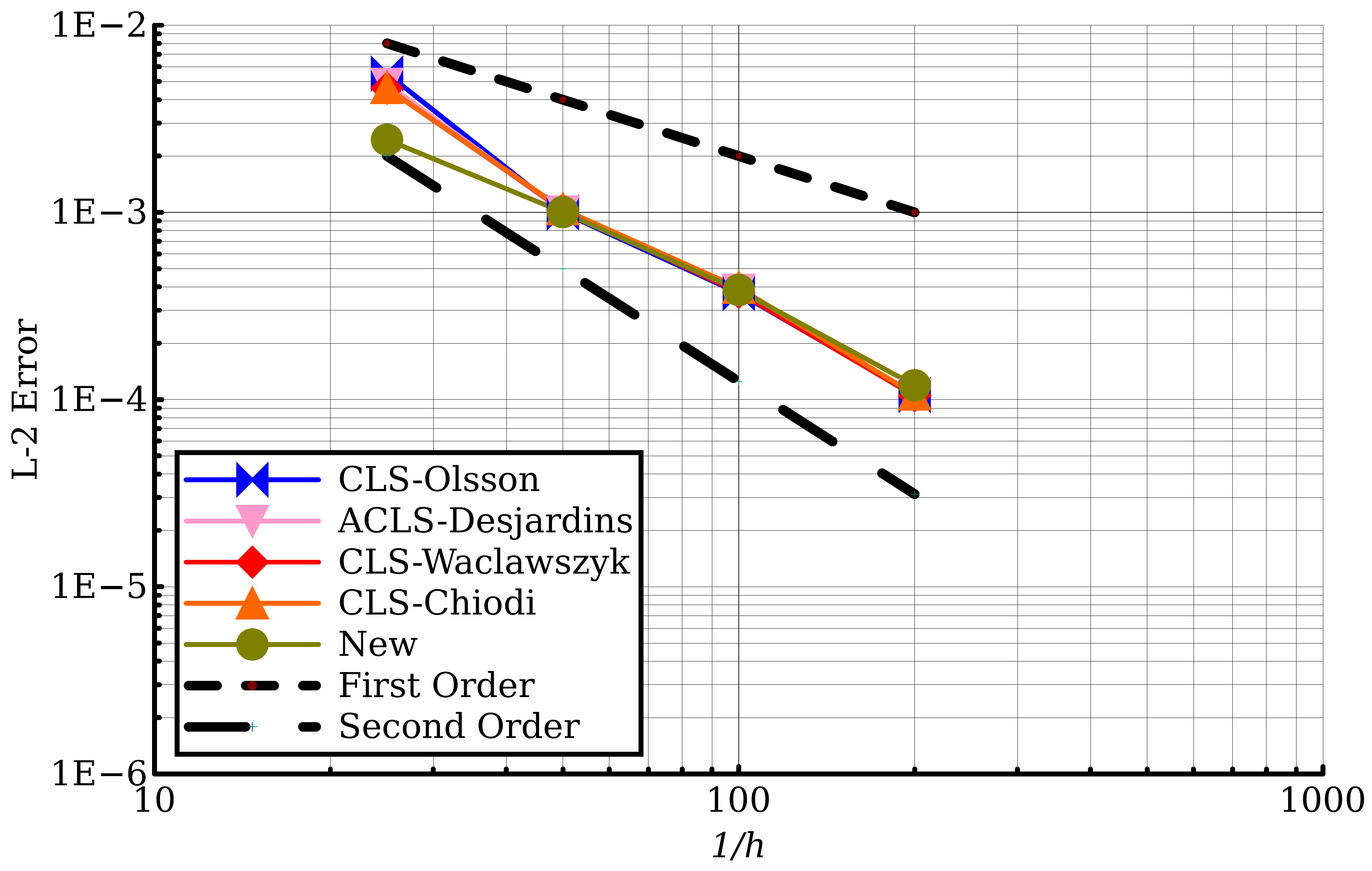}}
\caption{Convergence of the $L^1$ and $L^2$ error norms for the
  reinitialization of Zalesak's disc rotation problem solved on four
  ($25 \times 25$, $50 \times 50$, $100 \times 100$ and $200 \times
  200$) grid levels.}
\label{fig:ZD_order_error}
\end{figure}

The second problem considered is the advection of Zalesak's disc,
similar to the test reported in \cite{Desjardins2008}. The Zalesak's
disc is of radius 0.15 units, and notch length and width of 0.3 units
and 0.1 units respectively. Due to the presence of sharp corners, this
test problem shows more numerical errors compared to the previous
problem. Similar to the rotation of circular disc problem, this
problem also is solved on four levels of Cartesian meshes starting
from $25 \times 25$ up to $200 \times 200$. The level set contours
correspond to $\psi = 0.05, \psi = 0.5$ and $\psi = 0.95$ for the
$100 \times 100$ case at time levels $t = 0$, $t = \pi/4$, $t = \pi/2$
and $t = 3\pi/4$ for all five reinitialization schemes are plotted in
Figure~\ref{fig:ZD_contours}. Further,
Table~\ref{tab:ZD_area_error}~and~\ref{tab:ZD_l1_l2} show the
percentage area error and error norms computed after the disc
completes one full rotation ($ie.,$ at $t = 2 \pi$). Moreover, the
error norms are plotted against the mesh size in
Figure~\ref{fig:ZD_order_error} along with reference slopes for the
first and second order rate of convergence. Similar to the previous
problem, here also it can be seen that the new reinitialization scheme
is having the least error among all.

\begin{table}[H]
\begin{center}
\caption{Percentage area errors after completion of one full rotation of different reinitialization schemes for the reinitialization of Zalesak's disc rotation problem.}
\label{tab:ZD_area_error}
\begin{tabular}{|c|c|c|c|c|c|}
\hline
{\color[HTML]{000000} }                                & \multicolumn{5}{c|}{{\color[HTML]{000000} \text{Reinitialization Schemes}}}                                                                                                                                                    \\ \cline{2-6} 
\multirow{-2}{*}{{\color[HTML]{000000} \text{Mesh}}} & {\color[HTML]{000000} \text{CLS-Olsson}} & {\color[HTML]{000000} \text{ACLS-Desjardins}} & {\color[HTML]{000000} \text{CLS-Wac{\l}awszyk}} & {\color[HTML]{000000} \text{CLS-Chiodi}} & {\color[HTML]{000000} \text{New}} \\ \hline
{\color[HTML]{000000} $25 \times 25$}                  & {\color[HTML]{000000} -9.49000}            & {\color[HTML]{000000} -9.51580}                 & {\color[HTML]{000000} -8.99110}                & {\color[HTML]{000000} -9.37870}            & {\color[HTML]{000000} -7.10470}     \\ \hline
{\color[HTML]{000000} $50 \times 50$}                  & {\color[HTML]{000000} -5.33530}            & {\color[HTML]{000000} -5.22570}                 & {\color[HTML]{000000} -5.19310}                & {\color[HTML]{000000} -4.68600}            & {\color[HTML]{000000} -3.70040}     \\ \hline
{\color[HTML]{000000} $100 \times 100$}                & {\color[HTML]{000000} -3.09830}            & {\color[HTML]{000000} -3.45670}                 & {\color[HTML]{000000} -3.29800}                & {\color[HTML]{000000} -2.00520}            & {\color[HTML]{000000} -1.28230}     \\ \hline
{\color[HTML]{000000} $200 \times 200$}                & {\color[HTML]{000000} -2.38510}            & {\color[HTML]{000000} -2.33920}                 & {\color[HTML]{000000} -2.38360}                & {\color[HTML]{000000} -2.33040}            & {\color[HTML]{000000} -0.90177}     \\ \hline
\end{tabular}
\end{center}
\end{table}

\begin{table}[H]
\begin{center}
\caption{The $L^1$ and $L^2$ error norms of different reinitialization schemes for the reinitialization of Zalesak's disc rotation problem.}
\label{tab:ZD_l1_l2}
\begin{tabular}{|c|c|c|c|c|c|c|}
\hline
{\color[HTML]{000000} }                                     & {\color[HTML]{000000} }                                & \multicolumn{5}{c|}{{\color[HTML]{000000} \text{Reinitialization Schemes}}}                                                                                                                                                    \\ \cline{3-7} 
\multirow{-2}{*}{{\color[HTML]{000000} \text{Errors}}}    &
\multirow{-2}{*}{{\color[HTML]{000000} \text{Mesh}}} &
         {\color[HTML]{000000} \text{CLS-Olsson}} &
         {\color[HTML]{000000} \text{ACLS-Desjardins}} &
         {\color[HTML]{000000} \text{CLS-Wac{\l}awszyk}} &
         {\color[HTML]{000000} \text{CLS-Chiodi}} &
         {\color[HTML]{000000} \text{New}} \\ \hline 
\hline
{\color[HTML]{000000} }                              & {\color[HTML]{000000} $25 \times 25$}         & {\color[HTML]{000000} 4.2972E-02} & {\color[HTML]{000000} 3.9579E-02}      & {\color[HTML]{000000} 4.1547E-02}     & {\color[HTML]{000000} 3.6041E-02} & {\color[HTML]{000000} 1.9959E-02} \\ \cline{2-7} 
{\color[HTML]{000000} }                              & {\color[HTML]{000000} $50 \times 50$}         & {\color[HTML]{000000} 1.3590E-02} & {\color[HTML]{000000} 1.3860E-02}      & {\color[HTML]{000000} 1.4083E-02}     & {\color[HTML]{000000} 1.4145E-02} & {\color[HTML]{000000} 1.2202E-02} \\ \cline{2-7} 
{\color[HTML]{000000} }                              & {\color[HTML]{000000} $100 \times 100$}       & {\color[HTML]{000000} 7.4338E-03} & {\color[HTML]{000000} 7.6362E-03}      & {\color[HTML]{000000} 7.4889E-03}     & {\color[HTML]{000000} 7.8407E-03} & {\color[HTML]{000000} 6.4325E-03} \\ \cline{2-7} 
\multirow{-4}{*}{{\color[HTML]{000000} $L^1$ Error}} & {\color[HTML]{000000} $200 \times 200$}       & {\color[HTML]{000000} 4.2656E-03} & {\color[HTML]{000000} 4.3188E-03}      & {\color[HTML]{000000} 4.2485E-03}     & {\color[HTML]{000000} 4.3666E-03} & {\color[HTML]{000000} 3.8105E-03} \\ \hline \hline
{\color[HTML]{000000} }                              & {\color[HTML]{000000} $25 \times 25$}         & {\color[HTML]{000000} 5.5200E-03} & {\color[HTML]{000000} 4.8633E-03}      & {\color[HTML]{000000} 4.6080E-03}     & {\color[HTML]{000000} 4.6220E-03} & {\color[HTML]{000000} 2.4447E-03} \\ \cline{2-7} 
{\color[HTML]{000000} }                              & {\color[HTML]{000000} $50 \times 50$}         & {\color[HTML]{000000} 1.0000E-03} & {\color[HTML]{000000} 1.0103E-03}      & {\color[HTML]{000000} 1.0334E-03}     & {\color[HTML]{000000} 1.0402E-03} & {\color[HTML]{000000} 1.0036E-03} \\ \cline{2-7} 
{\color[HTML]{000000} }                              & {\color[HTML]{000000} $100 \times 100$}       & {\color[HTML]{000000} 3.7050E-04} & {\color[HTML]{000000} 3.8588E-04}      & {\color[HTML]{000000} 3.7405E-04}     & {\color[HTML]{000000} 3.9550E-04} & {\color[HTML]{000000} 3.8000E-04} \\ \cline{2-7} 
\multirow{-4}{*}{{\color[HTML]{000000} $L^2$ Error}} & {\color[HTML]{000000} $200 \times 200$}       & {\color[HTML]{000000} 1.0556E-04} & {\color[HTML]{000000} 1.0664E-04}      & {\color[HTML]{000000} 1.0473E-04}     & {\color[HTML]{000000} 1.0647E-04} & {\color[HTML]{000000} 1.2000E-04} \\ \hline
\end{tabular}
\end{center}
\end{table}

\subsubsection{Reinitialization of circular disc deformation problem}
\begin{figure}[H]
\centering
\subfloat[CLS-Olsson]
{\includegraphics[scale=1.8]{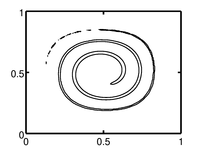}}
\subfloat[ACLS-Desjardins]
{\includegraphics[scale=1.8]{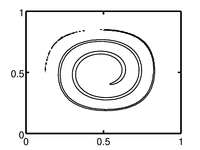}}
\subfloat[CLS-Wac{\l}awczyk]
{\includegraphics[scale=1.8]{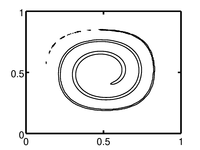}}

\subfloat[CLS-Chiodi]
{\includegraphics[scale=1.8]{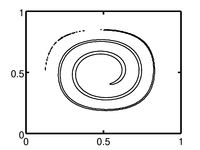}}
\subfloat[New]
{\includegraphics[scale=1.8]{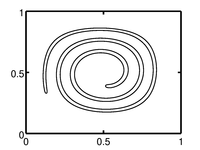}}
\caption{Comparison of reinitialization schemes for the
  reinitialization of circular disc deformation problem solved on $200
  \times 200$ grid. Each subfigure shows the 0.5 level contours of
  the level set function at time levels of 4~s.}
\label{fig:SW_contours}
\end{figure}

The last scalar advection test problem considered is the
advection of circular disc subjected to a shear velocity field, $u =
\sin^2(\pi x) \sin(2\pi y)$, $v = -\sin^2(\pi y) \sin(2\pi
x)$. Figure~\ref{fig:SW_contours} shows a qualitative comparison of the
interface contour for all five reinitialization scheme at $t = 4$~s
solved a $200 \times 200$ Cartesian mesh. From
Figure~\ref{fig:SW_contours}, one can clearly see that the tail of the
interface is fully resolved without breaking in case of the new
reinitialization scheme.

\subsection{Broken Dam Problem}
\label{sec:db}
In previous sections, in-place reinitialization problems and scalar advection
based test problems are solved. In order to compare its performance on
more realistic problems, an inviscid broken dam problem, reported in
\cite{Yang2014, Parameswaran2019} is considered next. Here, an
initial water column of height 0.9~m and width 0.45~m is kept at zero
velocity field and subjected to a hydrostatic pressure distribution
inside a computational domain bounded between $0 \leq x \leq 2$~m and
$0 \leq y \leq 1$~m. The density of water and air are taken as
998.2~kg/m$^3$ and 1.2040~kg/m$^3$ respectively. The computational
domain is discretized into $100 \times 50$ Cartesian mesh. The
artificial compressibility parameter, $\beta$, is taken as 10000. All
four boundaries are set to free-slip boundary condition. As time
progresses, due to the presence of gravitational force, the water
column collapses. In order to accurately capture the interface
movement, a small real-time step of 0.005~s is chosen. For stability
reasons, a smaller Courant number of 0.1 is chosen for the computation
of pseudo-time step.

\begin{figure}[H]
\centering
\subfloat[CLS-Olsson ($t = 0s$)]
{\includegraphics[scale=0.8]{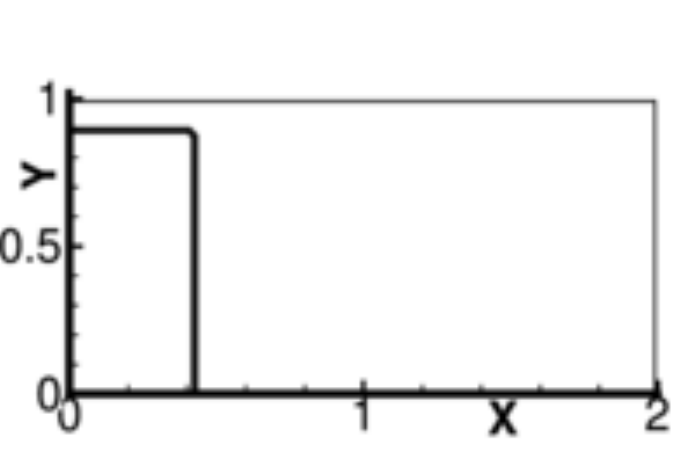}} \hspace{2.5cm}
\subfloat[New ($t = 0s$)]
{\includegraphics[scale=0.8]{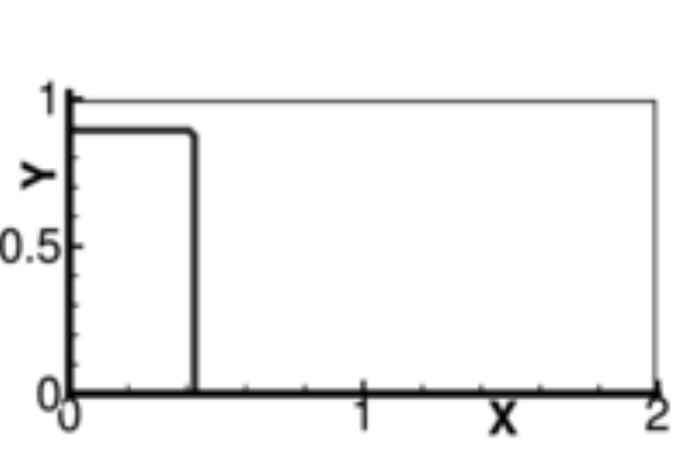}}

\subfloat[CLS-Olsson ($t = 0.5s$)]
{\includegraphics[scale=0.8]{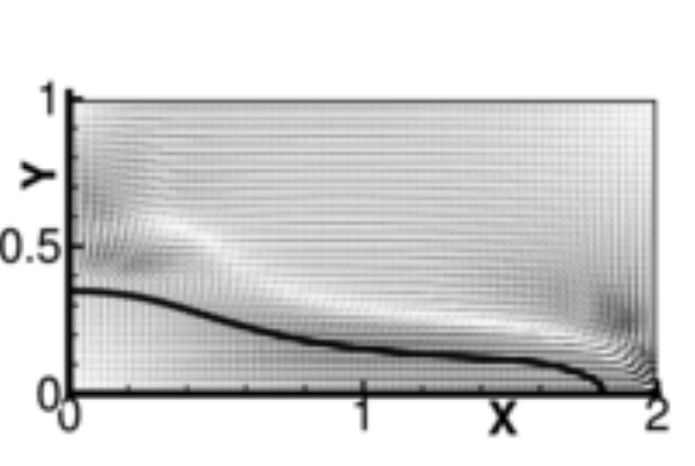}} \hspace{2.5cm}
\subfloat[New ($t = 0.5s$)]
{\includegraphics[scale=0.8]{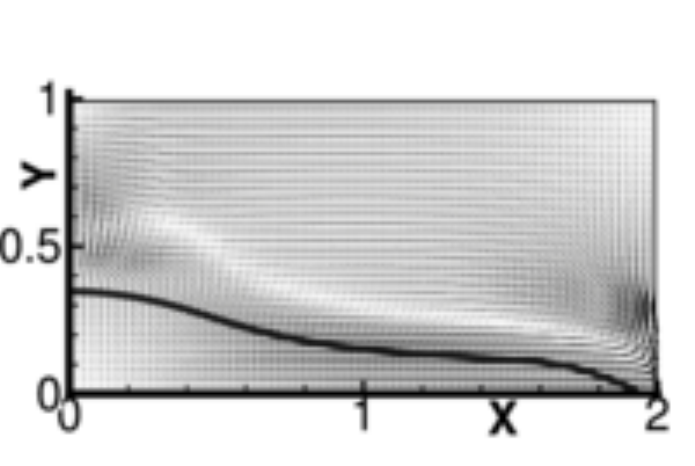}}

\subfloat[CLS-Olsson ($t = 0.75s$)]
{\includegraphics[scale=0.8]{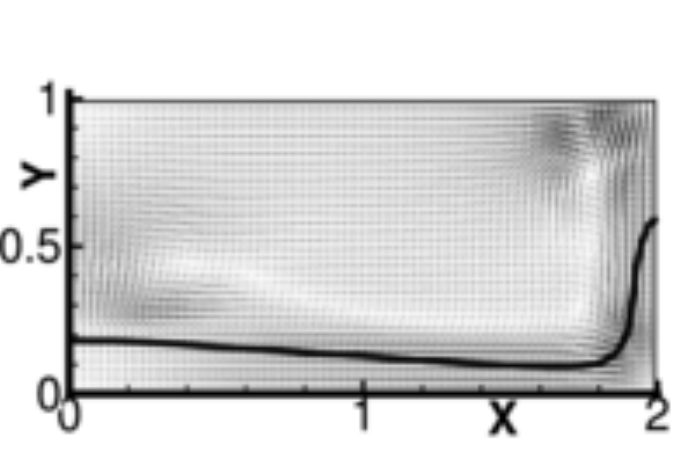}} \hspace{2.5cm}
\subfloat[New ($t = 0.75s$)]
{\includegraphics[scale=0.8]{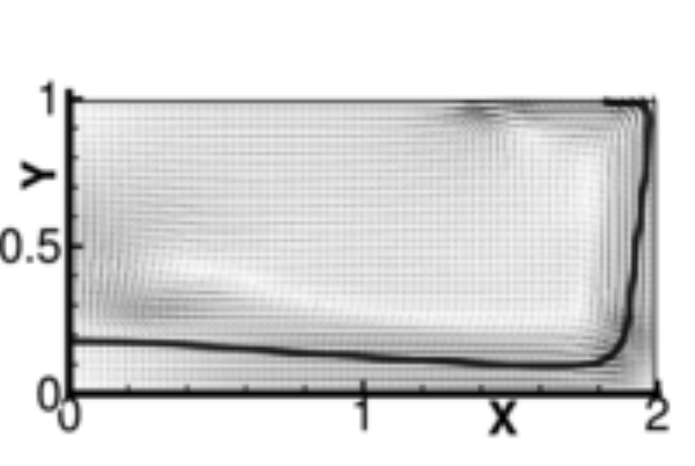}}

\subfloat[CLS-Olsson ($t = 1.0s$)]
{\includegraphics[scale=0.8]{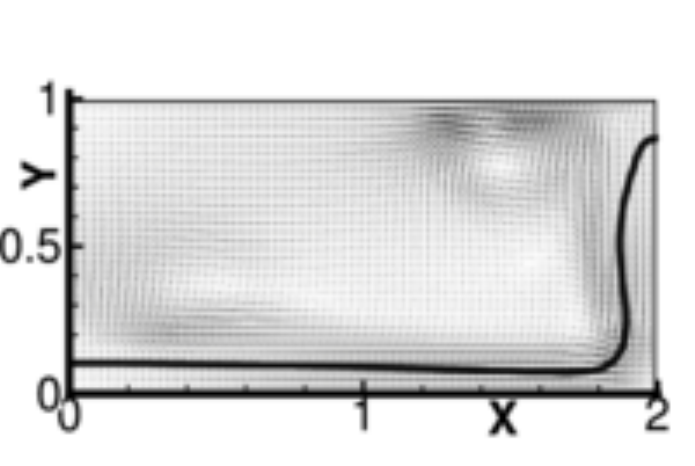}} \hspace{2.5cm}
\subfloat[New ($t = 1.0s$)]
{\includegraphics[scale=0.8]{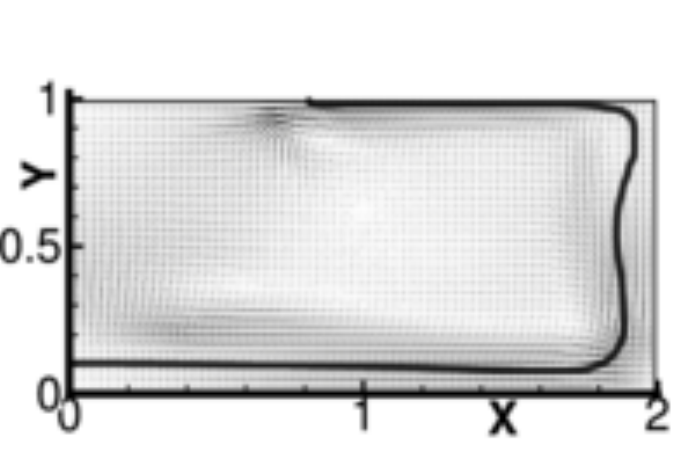}}

\subfloat[CLS-Olsson ($t = 1.5s$)]
{\includegraphics[scale=0.8]{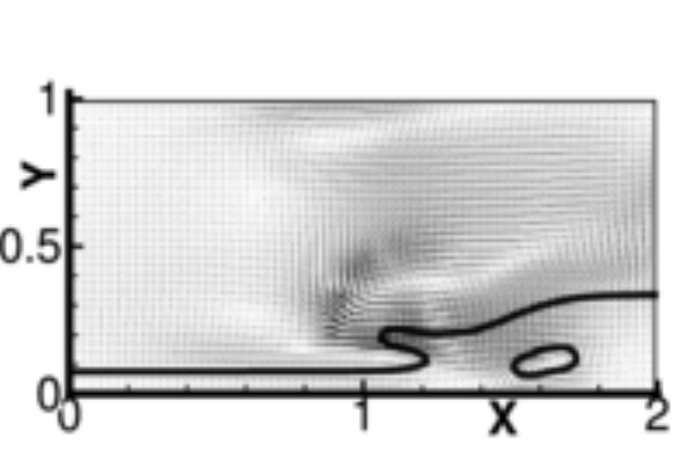}} \hspace{2.5cm}
\subfloat[New ($t = 1.5s$)]
{\includegraphics[scale=0.8]{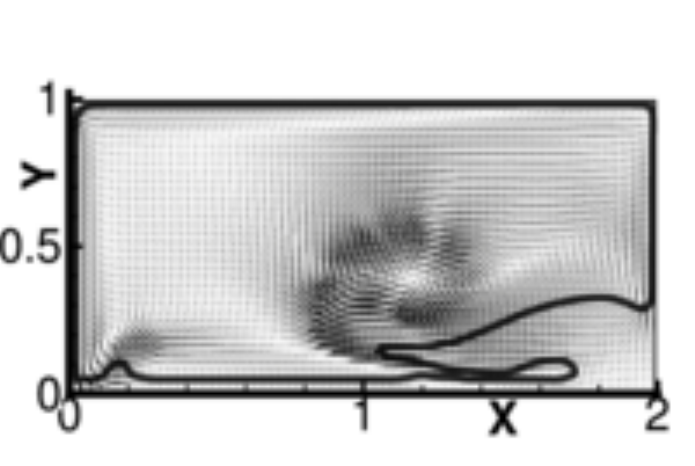}}
\caption{Interface profiles at different time levels, starting from $t
  = 0.0$~s up to $t = 1.5$~s, for the broken dam problem. Subfigures
  (a), (c), (e), (g) and (i) correspond to the CLS-Olsson scheme
  and (b), (d), (f), (h) and (j) correspond to the new
  reinitialization scheme.}
\label{fig:DB_Interfaces}
\end{figure}

\begin{figure}[H]
\centering
\subfloat[$s$ versus $T_s$]
{\includegraphics[scale=1.0]{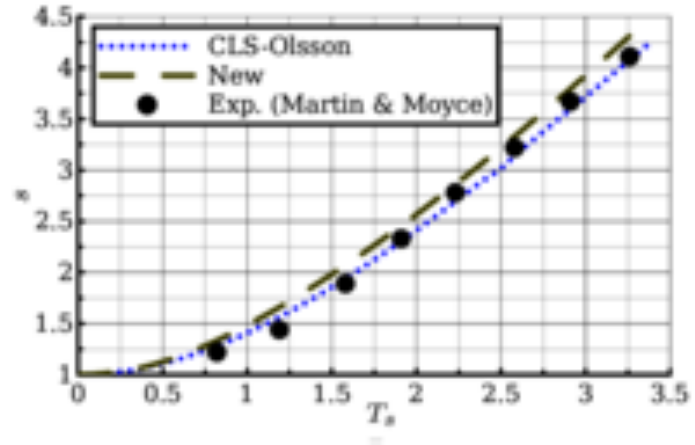}} \hspace{2.5cm}
\subfloat[$h$ versus $T_h$]
{\includegraphics[scale=1.]{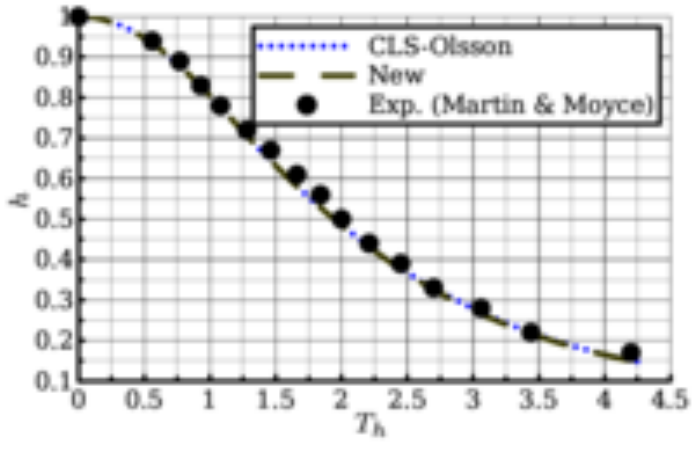}}
\caption{The non-dimensional surge front and water column height
  plotted with respect to the corresponding non-dimensional time
  scales for the broken dam problem.}
\label{fig:BD_exp_comparison}
\end{figure}

\begin{figure}[H]
\centering
\includegraphics[scale=1.2]{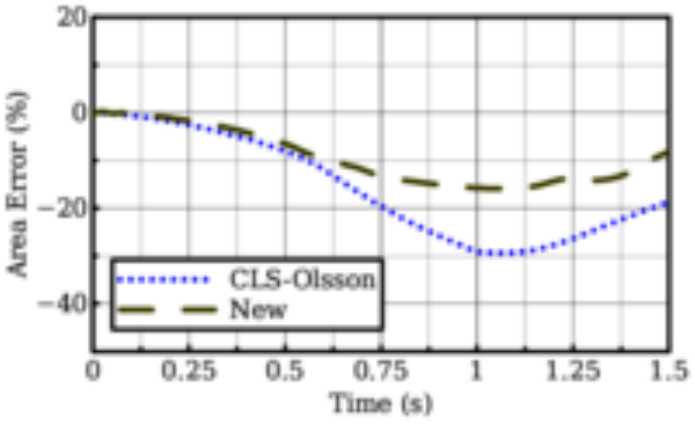}
\caption{The percentage area error plotted with respect to time for
  the broken dam problem.}
\label{fig:DB_area_error}
\end{figure}

Figure~\ref{fig:DB_Interfaces} shows the snapshots comparing the air-water
interface computed using the CLS-Olsson and the new reinitialization
procedure. One can see that the surge front, in case of the new
reinitialization scheme, touches the top wall, then reaches the left
wall and, finally, falls to the bottom pool of water. Since the
free-slip wall boundary conditions do not offer any frictional
resistance, such a behaviour is expected. However, in case of
CLS-Olsson, the thin surge front is spoiled due to inaccuracies
arising form the reinitialization scheme. One can see that, the surge
front does not even touch the top wall. In order to compare the
numerical results with experimental data reported
in~\cite{Martin1952}, the non-dimensional surge front, $s$, and
non-dimensional water column height, $h$, are plotted with respect to
the corresponding non-dimensional time scales $T_s = t\sqrt{2g/a}$ and
$T_h = t\sqrt{g/a}$ respectively in
Figure~\ref{fig:BD_exp_comparison}. The surge front and water column
heights are non-dimensionalized with respect to their initial
sizes. Looking at Figure~\ref{fig:BD_exp_comparison}, one can see an
excellent match of both the numerical results with the experimental
data reported by Martin and Moyce in~\cite{Martin1952}. Finally,
Figure~\ref{fig:DB_area_error} shows the percentage area errors,
computed using equation~(\ref{eq:area_error}), with respect to time
for both the cases. Clearly, the area loss is high for the CLS-Olsson
case, especially during $t = 1$~s, when the surge front becomes thin.

\subsection{Rayleigh-Taylor Instability Problem}
\label{sec:rti}
The second test problem considered is a Rayleigh-Taylor Instability
problem similar to the one reported in~\cite{Puckett1997,
  Parameswaran2019}. Unlike the previous problem, viscosity plays an
important role here. In this problem, a heavier fluid of density
1.225~kg/m$^3$ is placed on top of a lighter fluid of density
0.1694~kg/m$^3$ inside the computational domain bounded between $0
\leq x \leq 1$ and $0 \leq y \leq 4$. The dynamic viscosity for both
the fluids are taken to be the same, $\mu_1 = \mu_2 = 3.1304952 \times
10^{-3}$~kg/m~s. The two fluids are initially separated by an
interface, defined as, $y = 2.0 + 0.05\cos(2 \pi x)$. The problem is
solved on a Cartesian mesh of $32 \times 128$ finite volume cells. The
top and bottom boundaries are set to no-slip boundary condition,
whereas, the left and right boundaries are set as symmetric boundary
condition. The initial velocity field is set to be zero and the
pressure field is set based on gravity. The artificial compressibility
parameter, $\beta$, is taken as 1000 and the real-time step is taken
as 0.01~s. For stability reasons, the Courant number is chosen as 0.9
for the computation of the pseudo-time step.

\begin{figure}[H]
\centering
\subfloat[($t = 0.8s$)]
{\includegraphics[scale=0.6]{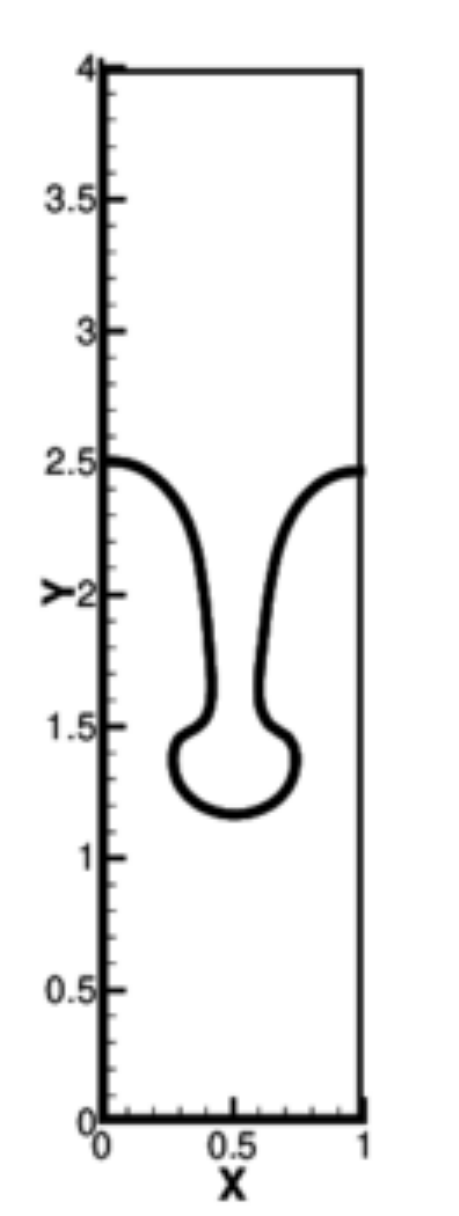}} \hspace{0.8cm}
\subfloat[($t = 0.9s$)]
{\includegraphics[scale=0.6]{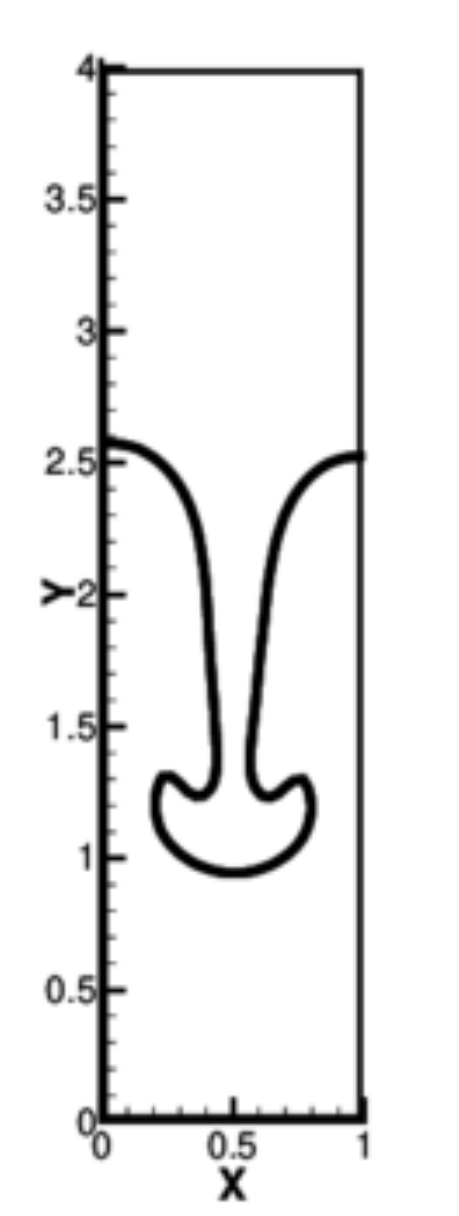}} \hspace{0.8cm}
\subfloat[($t = 1.0s$)]
{\includegraphics[scale=0.6]{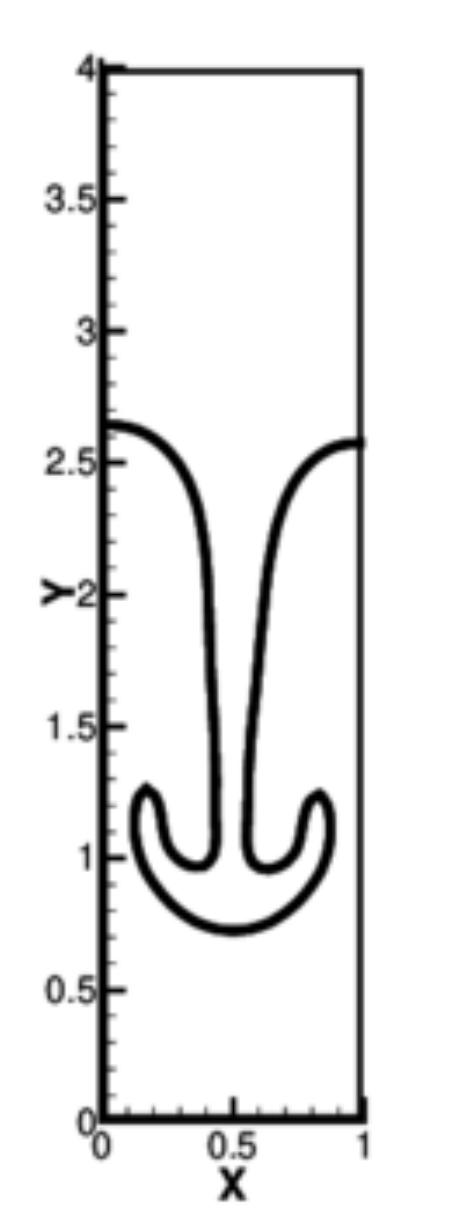}} \hspace{0.8cm}
\subfloat[($t = 1.2s$)]
{\includegraphics[scale=0.6]{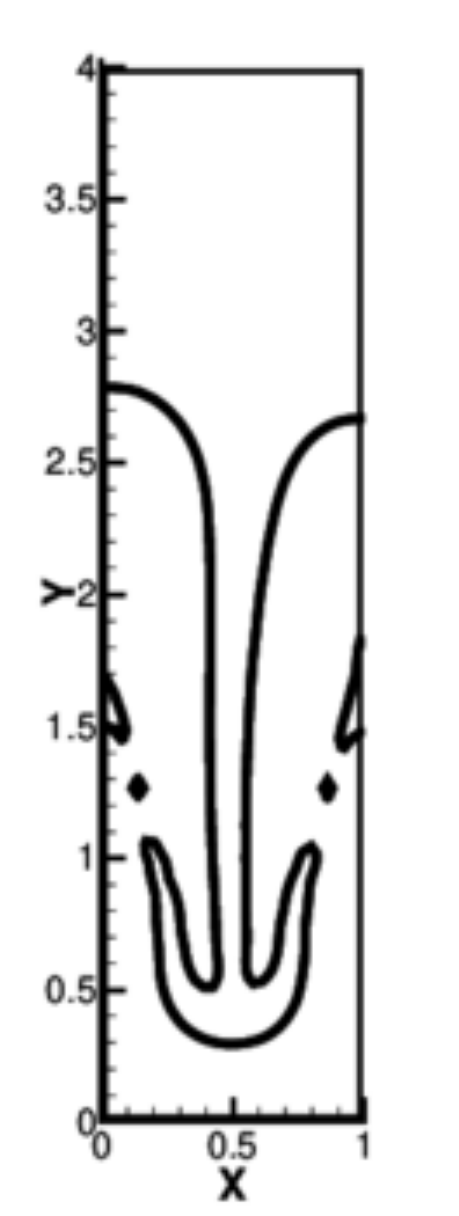}} \hspace{0.8cm}
\subfloat[($t = 1.4s$)]
{\includegraphics[scale=0.6]{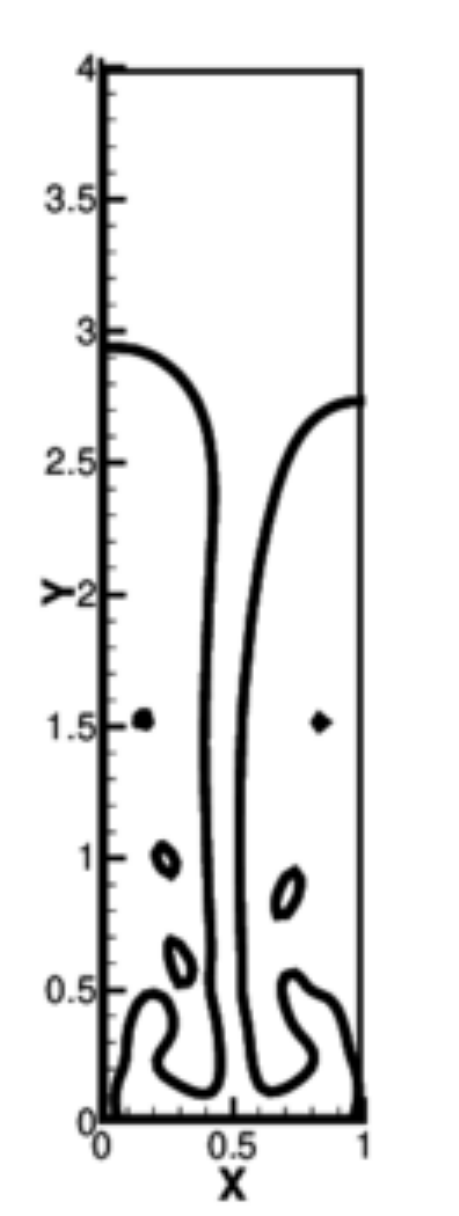}}

\subfloat[($t = 0.8s$)]
{\includegraphics[scale=0.6]{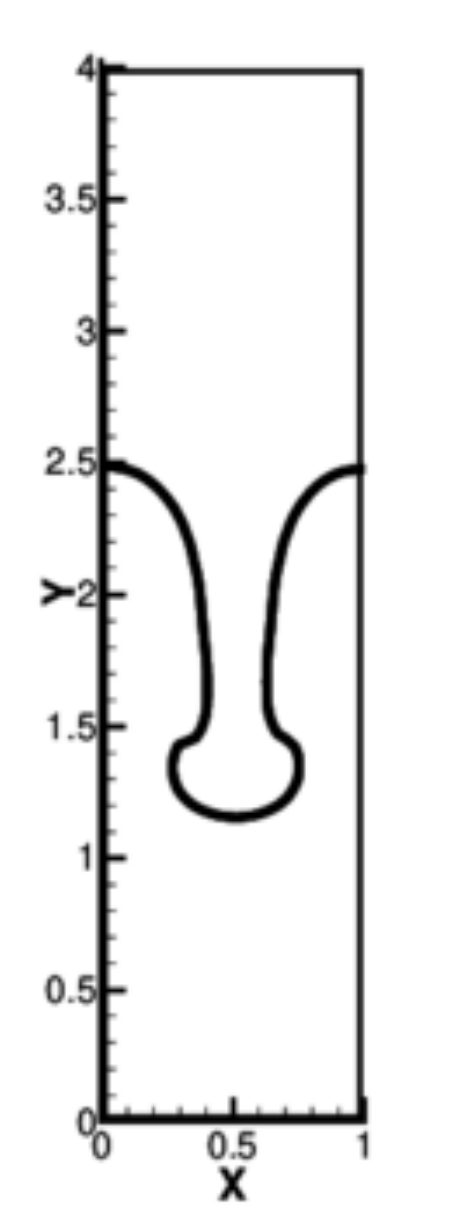}} \hspace{0.8cm}
\subfloat[($t = 0.9s$)]
{\includegraphics[scale=0.6]{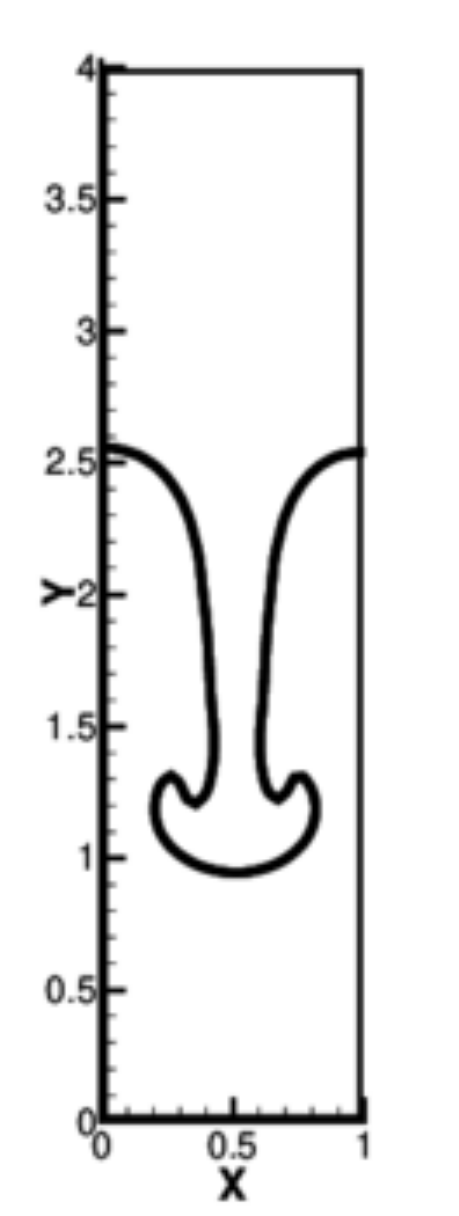}} \hspace{0.8cm}
\subfloat[($t = 1.0s$)]
{\includegraphics[scale=0.6]{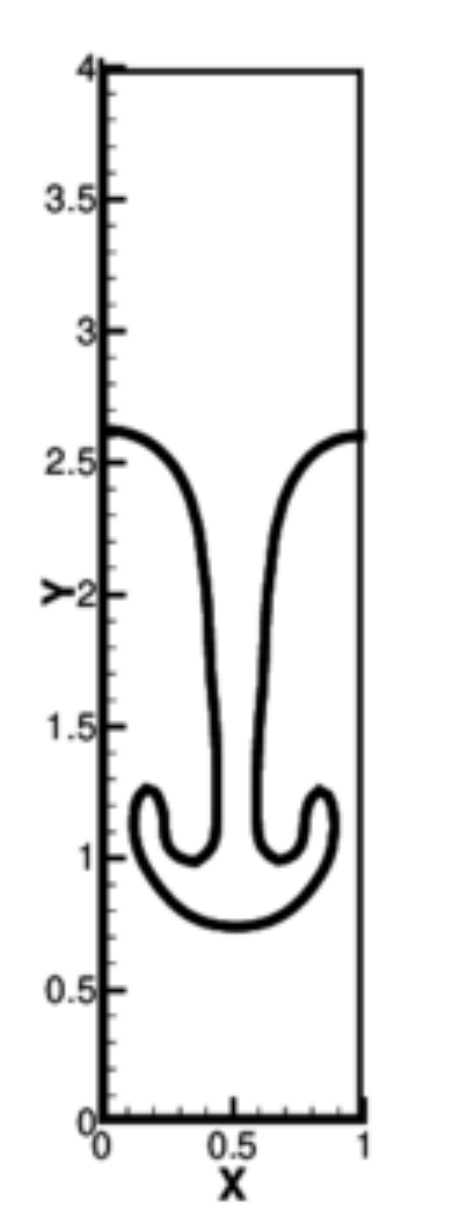}} \hspace{0.8cm}
\subfloat[($t = 1.2s$)]
{\includegraphics[scale=0.6]{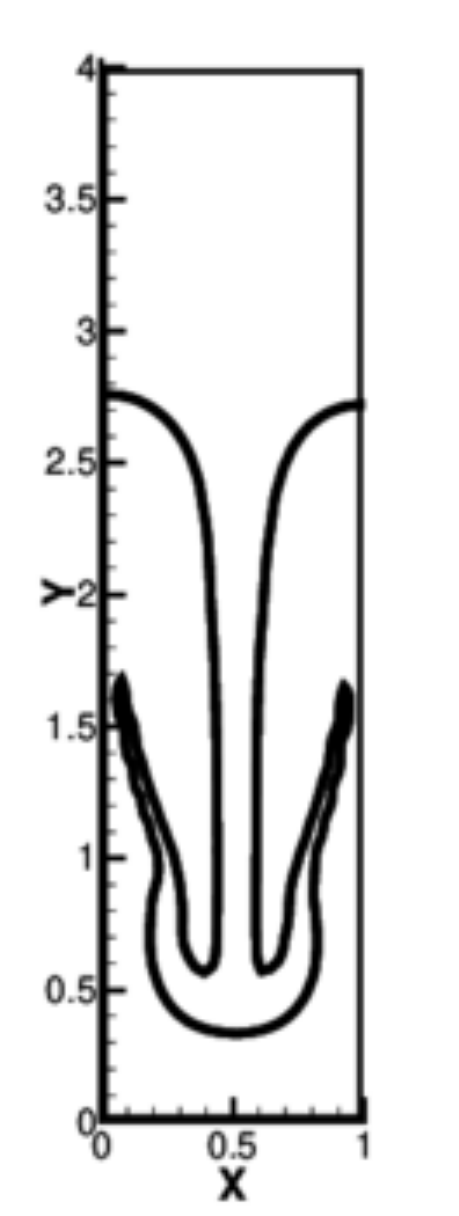}} \hspace{0.8cm}
\subfloat[($t = 1.4s$)]
{\includegraphics[scale=0.6]{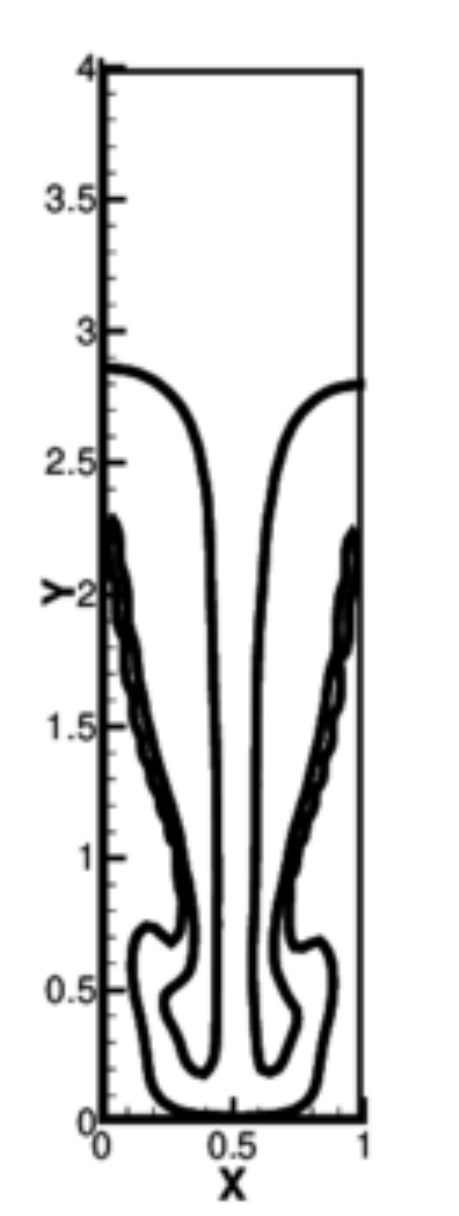}}
\caption{Interface profiles at different time levels, starting from $t
  = 0.8$~s up to $t = 1.4$~s, for the Rayleigh-Taylor instability
  problem. Subfigures from (a) to (e) correspond to the
  CLS-Olsson scheme and from (f) to (j) correspond to the new
  reinitialization scheme.}
\label{fig:RTI_contours}
\end{figure}

\begin{figure}[H]
\centering
\includegraphics[scale=0.5]{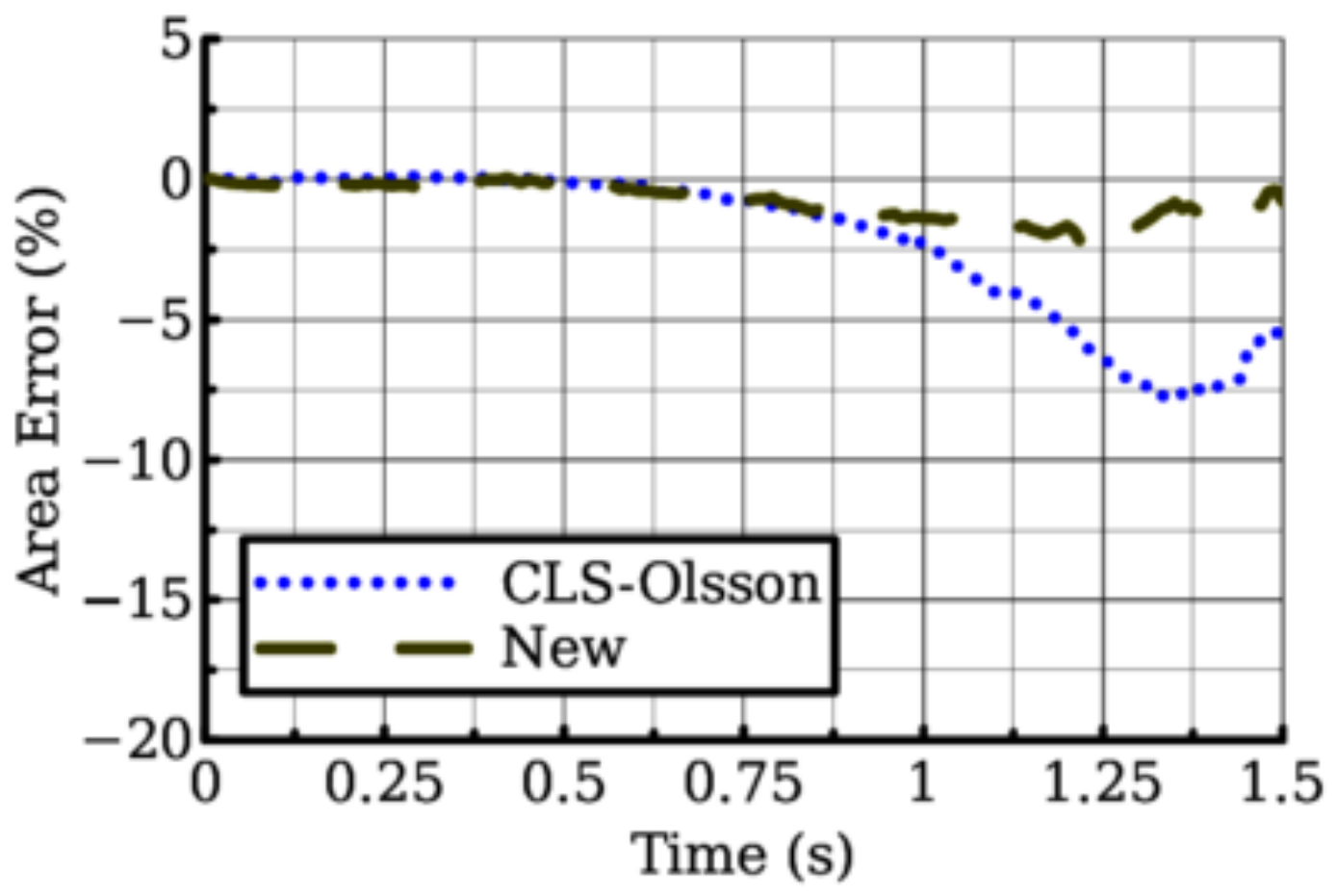}
\caption{The percentage area error plotted with respect to time for
  the Rayleigh-Taylor instability problem.}
\label{fig:RTI_area_error}
\end{figure}

As time progress, the top heavy fluid start to penetrate into the
bottom light fluid resulting formation of an inverted mushroom shaped
structure. Snapshots at different time levels during the evolution of
the fluid-fluid interface for both the existing and new
reinitialiation cases are shown in
Figure~\ref{fig:RTI_contours}. Similar to the previous problem, here
also one can see that the new reinitialization scheme is better in
capturing the thin fluid layer originating from the tips of the
inverted mushroom head. Figure~\ref{fig:RTI_area_error} compares the
percentage area errors computed using equation~(\ref{eq:area_error})
in both the cases. One can clearly see a higher area loss for the
CLS-Olsson case during the later stages of interface evolution.

\subsection{Rising Bubble Problem}
\label{sec:br}
The last problem considered in this section is a rising bubble problem
described in~\cite{Hysing2009}. Unlike previous problems, this one
is more challenging because of the presence of buoyancy, viscosity and
surface tension forces. The problem consists of a circular bubble of
diameter, $d_b = 0.5$~m, placed at the lower half of a rectangular domain
bounded between $0 \leq x \leq 1$~m and $0 \leq y \leq 2$~m, initially
filled with a quiescent liquid. Due to the presence of the buoyant force,
the initial circular bubble will start rising. With the interaction
of the surrounding liquid, the initially circular shape of the bubble
gets deformed. The degree of deformation of the circular bubble
depends upon the Reynolds number ($Re$) and the E\"{o}tv\"{o}s number
($Eo$). The $Re$ and $Eo$ are defined as,
\begin{equation}
Re = \frac{\rho_2 U_g L}{\mu_2}
\end{equation}
and 
\begin{equation}
Eo = \frac{\rho_2 U_g^2 L}{\sigma}
\end{equation}
where, the characteristic length scale, $L = d_b$, and the
characteristic velocity scale, $U_g = \sqrt{g d_b}$. Based on the
level of difficulty, two versions of rising bubble problems are
reported in~\cite{Hysing2009}. The first one (denoted here as
``Case-1'') is relatively simple and the second one (denoted here as
``Case-2'') is more challenging. The physical parameters defining the
two dimensional rising bubble test cases are given in
Table~\ref{tab:phy_parameters}.

\begin{table}[H]
\centering
\caption{Physical parameters defining the two dimensional rising bubble test cases.}
\begin{tabular}{p{1.5cm} p{1.2cm} p{1.2cm} p{1.2cm} p{1.2cm} p{1.2cm} p{1.2cm} p{1.2cm} p{1.2cm} p{1.2cm} p{1.2cm}}
\hline
Test case & $\rho_2$ & $\rho_1$ & $\mu_2$ & $\mu_1$ & $g$ & $\sigma$ & $Re$ & $Eo$ &
$\rho_2/\rho_1$ & $\mu_2/\mu_1$ \\ \hline
Case-1 & 1000 & 100 & 10 & 1   & 0.98 & 24.5 & 35 & 10  & 10   & 10  \\ 
Case-2 & 1000 & 1   & 10 & 0.1 & 0.98 & 1.96 & 35 & 125 & 1000 & 100 \\ \hline
\end{tabular}
\label{tab:phy_parameters}
\end{table}
In both the cases, the initial velocity field is set to zero and the
initial pressure field is set based on gravity. The left and right
boundaries are set to free-slip boundary condition and the top and
bottom walls are set to no-slip boundary condition. The artificial
compressibility parameter, $\beta$, is taken as 10000. The real-time
step is taken as 0.05~s and the pseudo-time step is computed according
to the Courant number 0.9. Numerical simulations are carried out up to
a time level of 4~s. In order to make a quantitative comparison, three
parameters, namely, the rise velocity, the location of centroid and
the circularity of the bubble are reported in~\cite{Hysing2009}. These
parameters are computed as,
\begin{equation}
\text{Rise Velocity, }{v_c} = \frac{\int_{\Omega_b}{\bf u}\cdot{\bf e}_y~\text{d}\Omega_b}{\int_{\Omega_b}\text{d}\Omega_b},
\label{eq:rise_vel}
\end{equation}
\begin{equation}
\text{Centroidal Location, }y_c = \frac{\int_{\Omega_b}{\bf
    x}_b\cdot{\bf e}_y~\text{d}\Omega_b}{\int_{\Omega_b}\text{d}\Omega_b}
\label{eq:centroid}
\end{equation}
and
\begin{equation}
\text{Circularity, }\zeta = \frac{P_a}{P_b} = \frac{\text{Perimeter of
  area-equivalent bubble}}{\text{Perimeter of the bubble}} = \frac{\pi
d_a}{\int_{\Omega}\lVert \nabla \psi \rVert \text{d}\Omega}
\label{eq:circularity}
\end{equation}
where, $\Omega$ is the computational domain, $\Omega_b$ is the region
occupied by the bubble, ${\bf x}_b$ is the position vector inside the
bubble, ${\bf e}_y$ is the unit vector parallel to the $y-$axis and
$d_a$ is the diameter of a circle with area equal to that of the
bubble with circumference $P_b$.

\subsubsection{Case-1}
\label{sec:br1}
\begin{figure}[H]
\centering
\subfloat[CLS-Olsson]
{\includegraphics[scale=0.65]{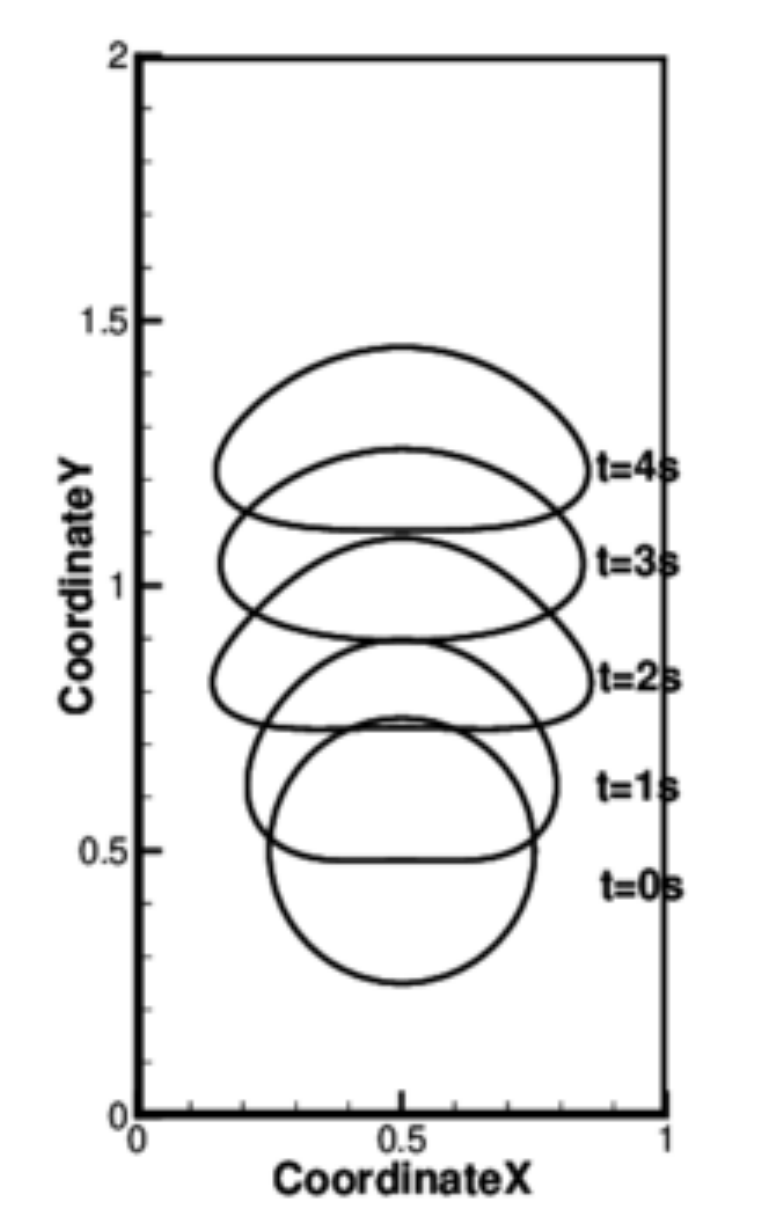}}
\subfloat[New]
{\includegraphics[scale=0.65]{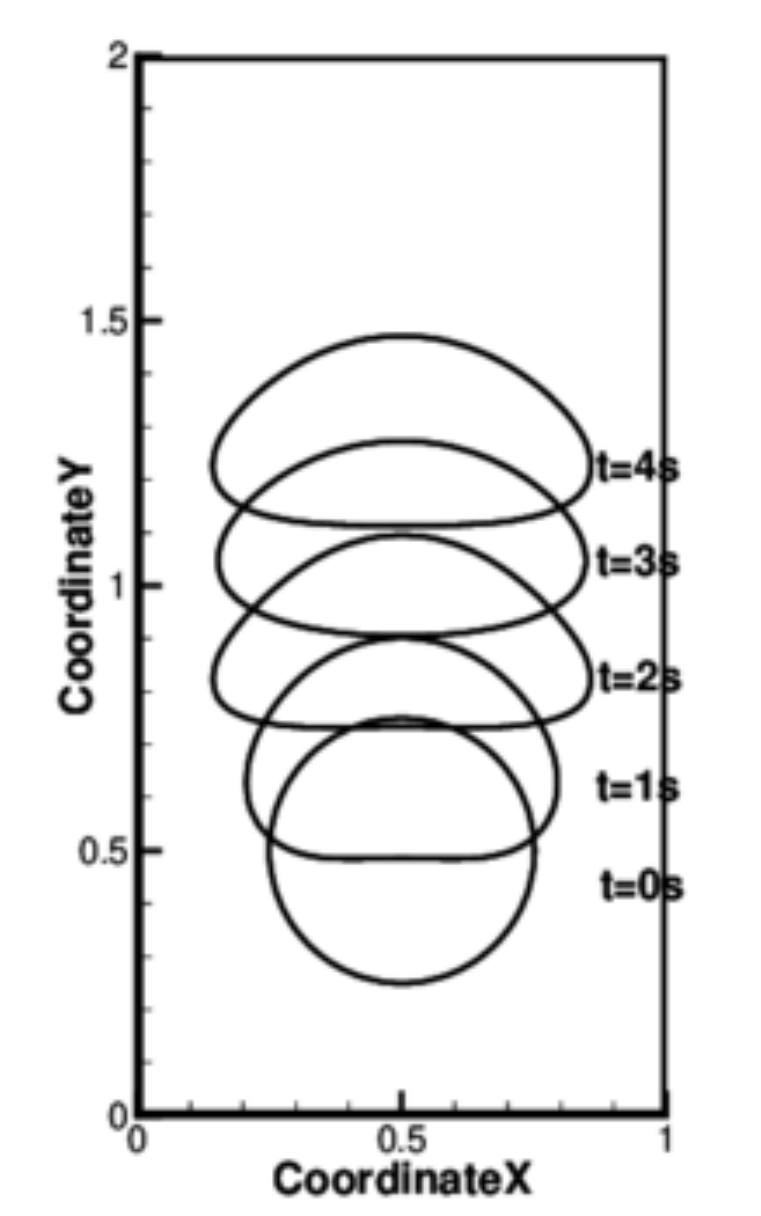}} \hspace{0.8cm}
\caption{Bubble profiles from $t = 0$~s up to $t = 4$~s for the rising bubble problem (Case-1).}
\label{fig:BR_1_bubble_profiles}
\end{figure}

\begin{figure}[H]
\centering
\includegraphics[scale=0.6]{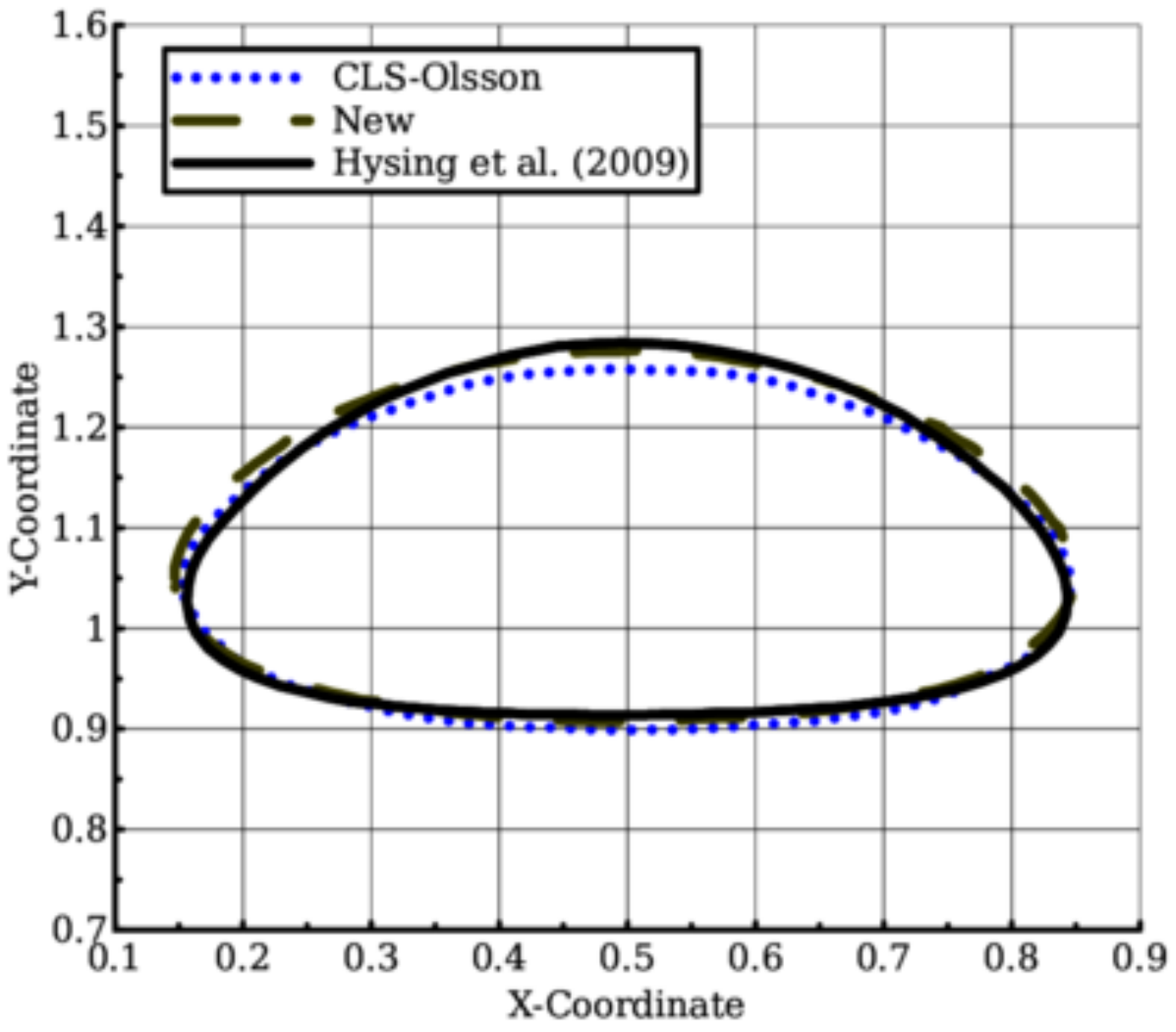}
\caption{Enlarged bubble profiles at $t = 3$~s for the rising bubble problem (Case-1).}
\label{fig:BR_1_bubble_profiles_enl}
\end{figure}

\begin{figure}[H]
\centering
\subfloat[Rise Velocity Vs Time]
{\includegraphics[scale=0.6]{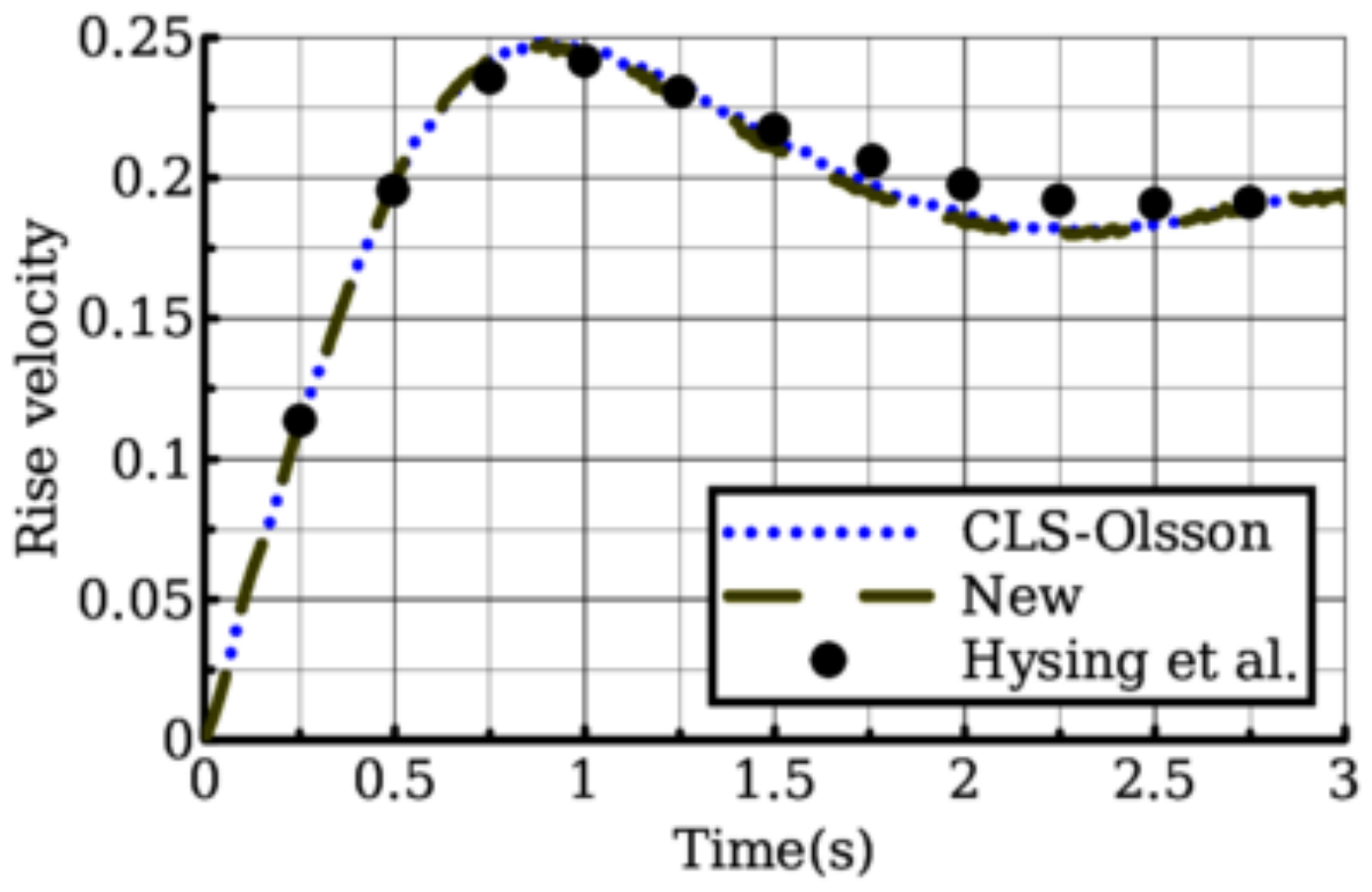}}
\subfloat[Centroid Vs Time]
{\includegraphics[scale=0.6]{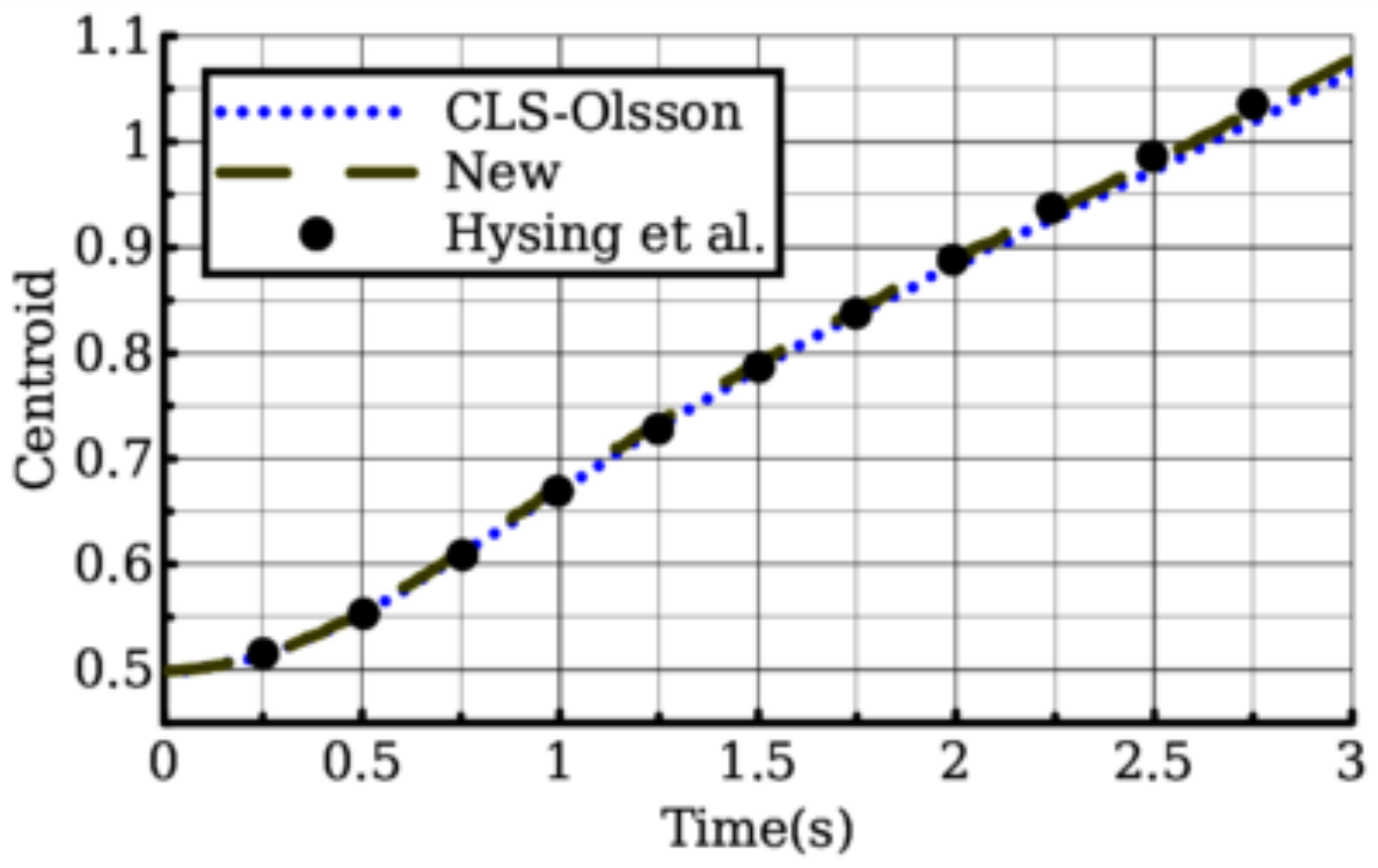}}

\subfloat[Circularity Vs Time]
{\includegraphics[scale=0.6]{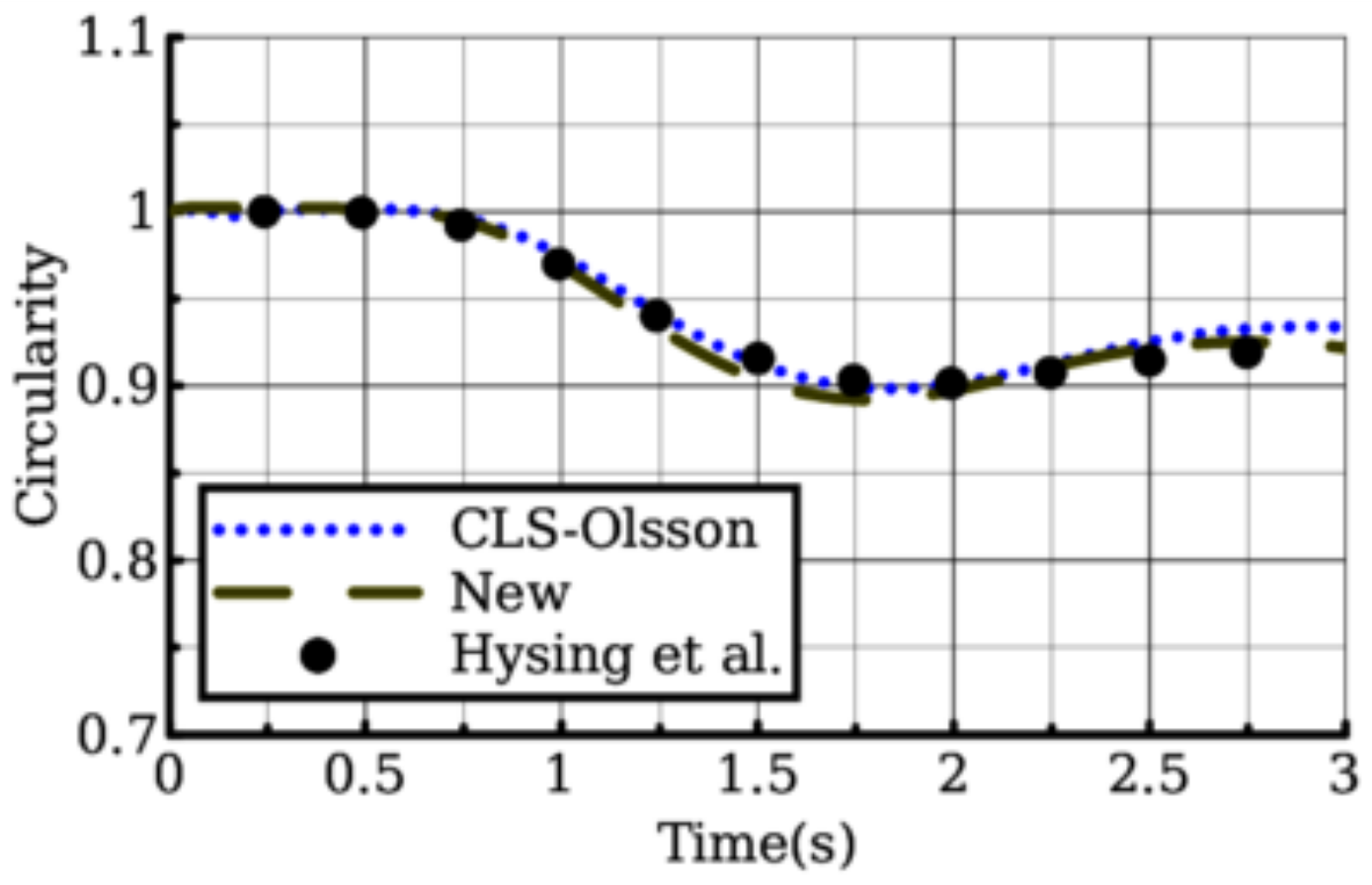}}
\caption{The rise velocity, centroid location and circularity plotted
  with respect to time for the rising bubble problem (Case-1).}
\label{fig:BR_1_parameters_plot}
\end{figure}

\begin{figure}[H]
\centering
\includegraphics[scale=0.6]{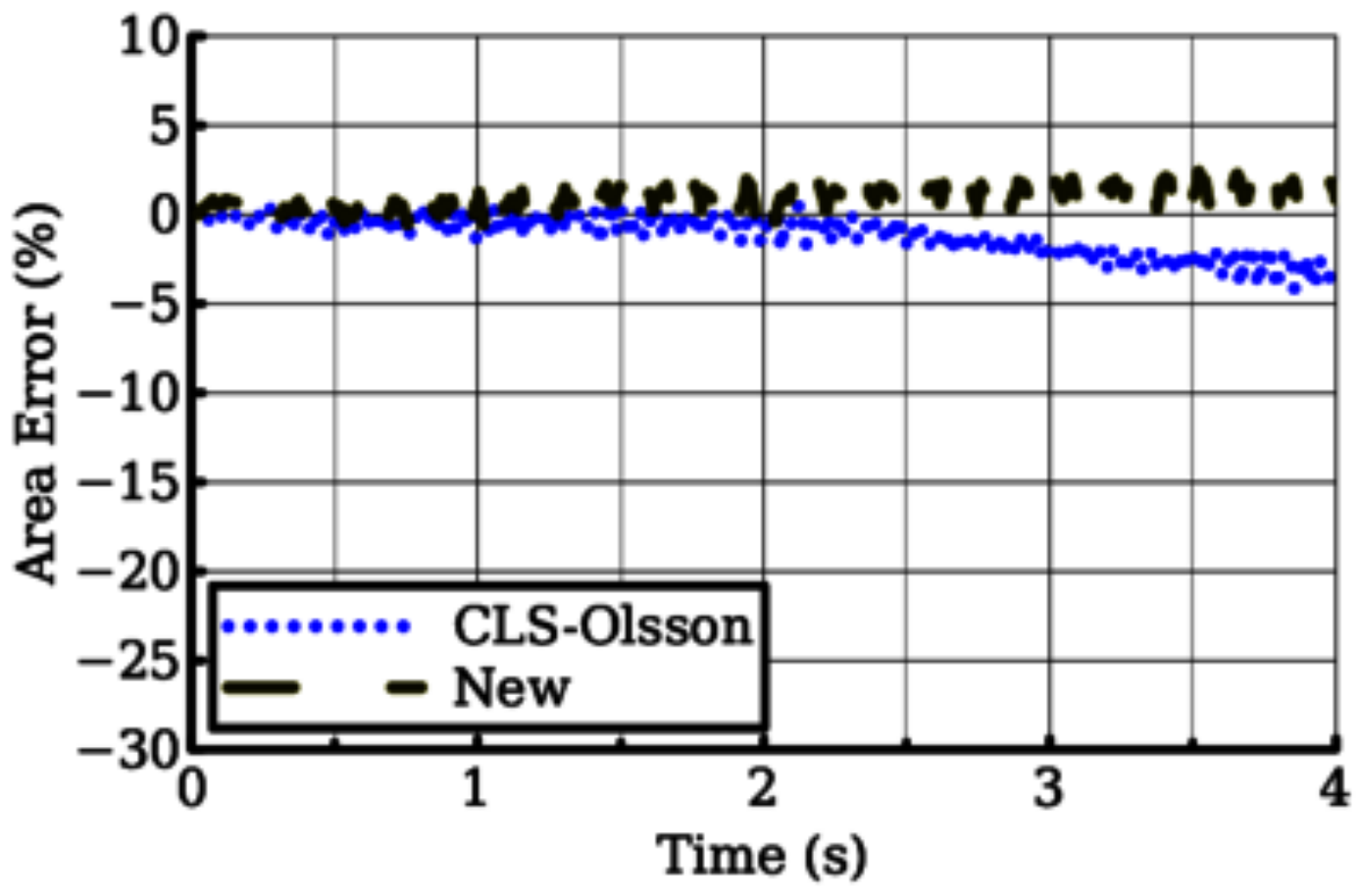}
\caption{The percentage area error plotted with respect to time for
  the rising bubble problem (Case-1).}
\label{fig:BR_1_area_error}
\end{figure}

For the choice of physical parameters of Case-1, the bubble does not
undergo large deformation. The initial circular bubble first stretches
in the horizontal direction and, finally, settles down to an
ellipsoidal profile as it reaches its terminal speed. Numerical
simulations are carried out on a Cartesian mesh of $80 \times 160$
finite volume cells. The bubble profiles at different time levels for
both the CLS-Olsson and the new reinitialization cases are shown in
Figure~\ref{fig:BR_1_bubble_profiles}. One can see that the bubble
profiles for the CLS-Olsson and the new reinitialization cases are
quite similar and match very well with the results reported 
in~\cite{Hysing2009}. In order to make a close comparison, the
terminal shape of the bubbles in both the cases are plotted in
Figure~\ref{fig:BR_1_bubble_profiles_enl} along with the reference
bubble profile of~\cite{Hysing2009}. One can see from
Figure~\ref{fig:BR_1_bubble_profiles_enl} that the bubble profiles of
both the CLS-Olsson and the new reinitialization schemes match very
well with the reference bubble profile. The rise velocity, centroid
location and the circularity of the rising bubble are plotted with
respect to time in Figure~\ref{fig:BR_1_parameters_plot}. Here also,
both the CLS-Olsson and the new reintialization results match
closely with the reference plots. Finally, the percentage area error,
computed using equation~(\ref{eq:area_error}), is plotted with respect
to time in Figure~\ref{fig:BR_1_area_error}. It can be noticed that
the area error is relatively less for the new reinitialization case
compared to that of CLS-Olsson.

\subsubsection{Case-2}
\label{sec:br2}
\begin{figure}[H]
\centering
\subfloat[Cartesian Mesh]
{\includegraphics[scale=0.65]{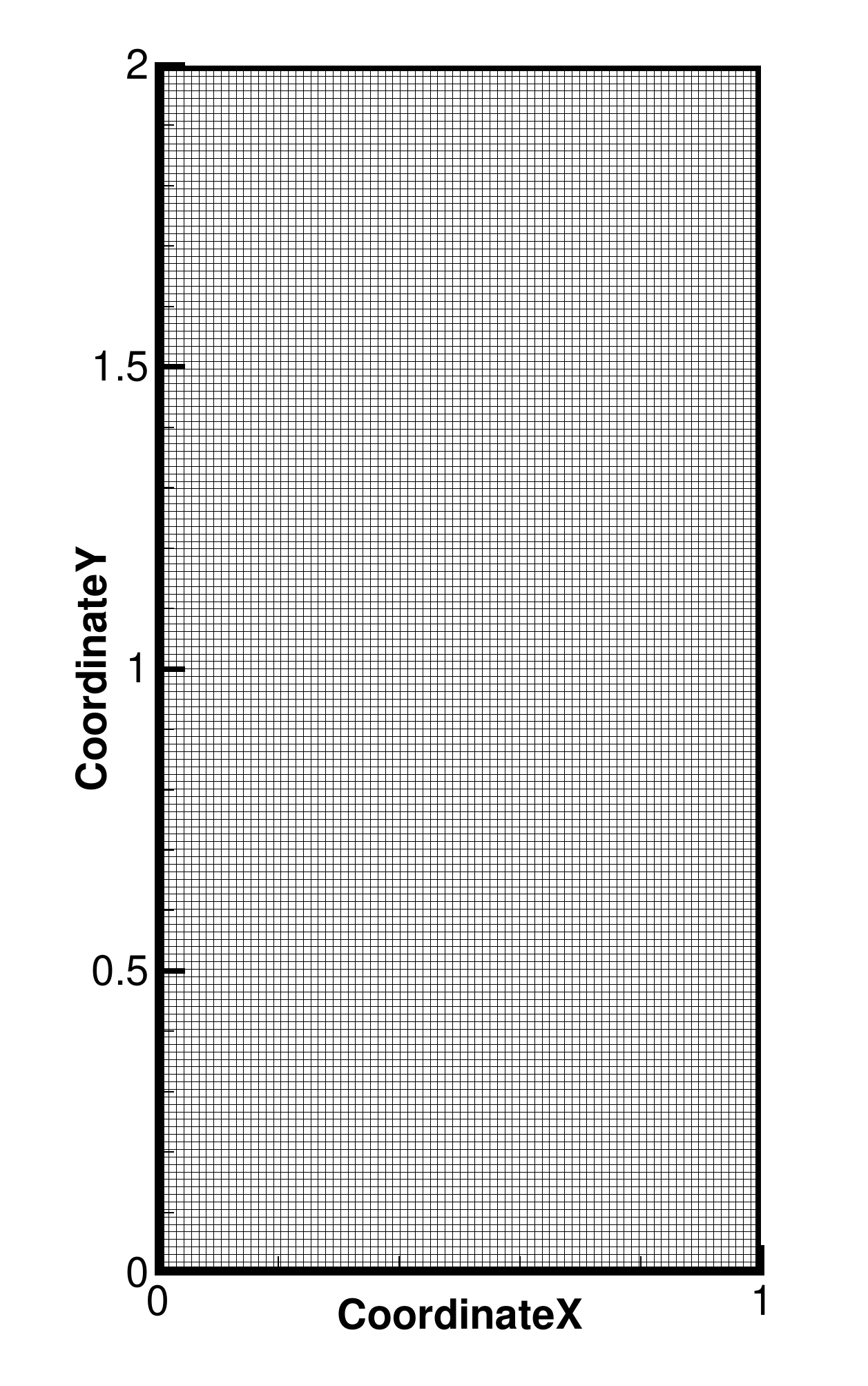}}
\subfloat[Unstructured Mesh]
{\includegraphics[scale=0.65]{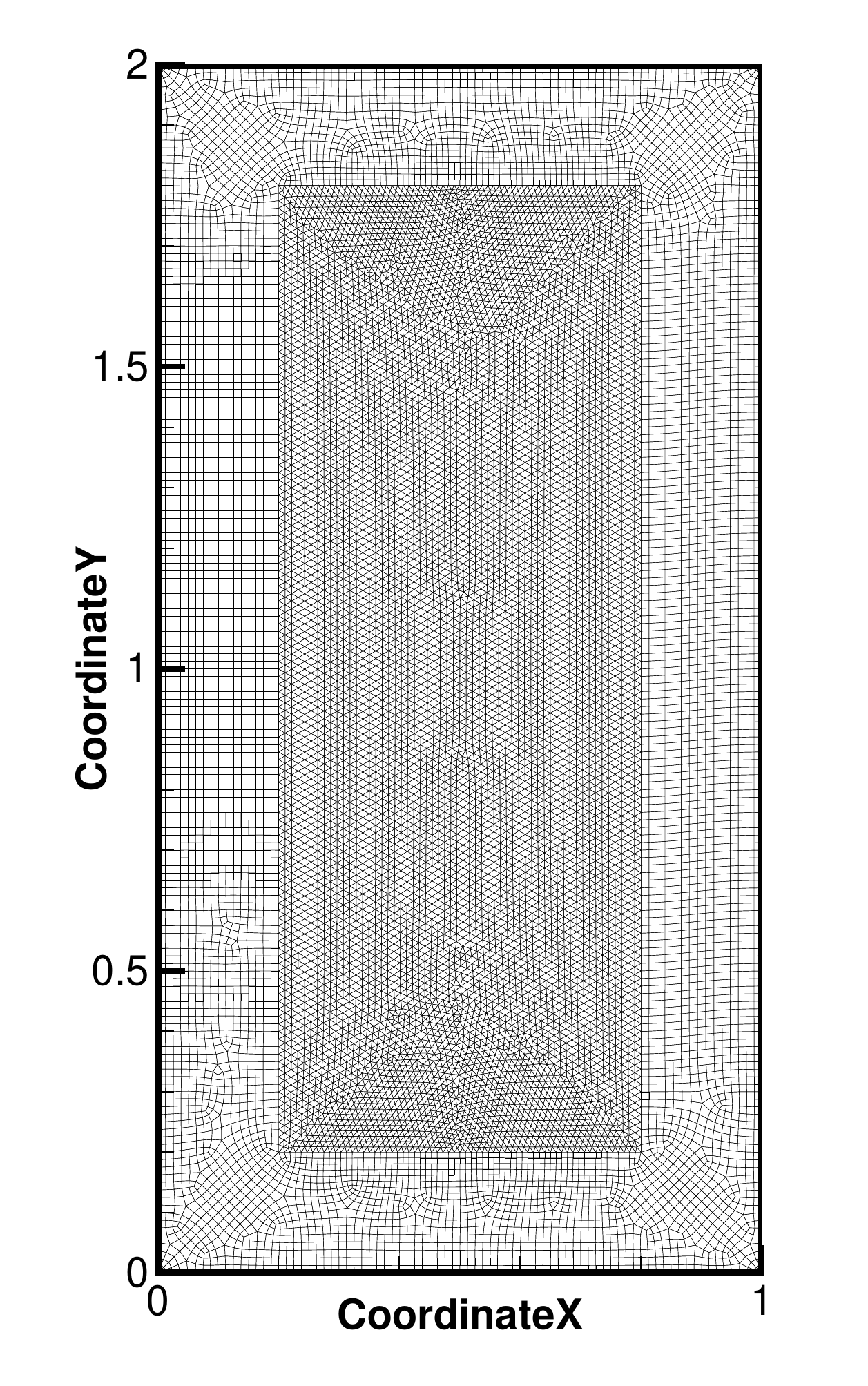}}
\caption{The different meshes used for solving the rising bubble
  problem (Case-2). Subfigures (a) is the $80 \times 160$ Cartesian
  mesh and (b) is the unstructured mesh consisting of 23331 finite
  volume cells of triangular and quadrilateral shape.}
\label{fig:BR_2_meshes}
\end{figure}

\begin{figure}[H]
\centering
\subfloat[($t = 0.0s$)]
{\includegraphics[scale=0.47]{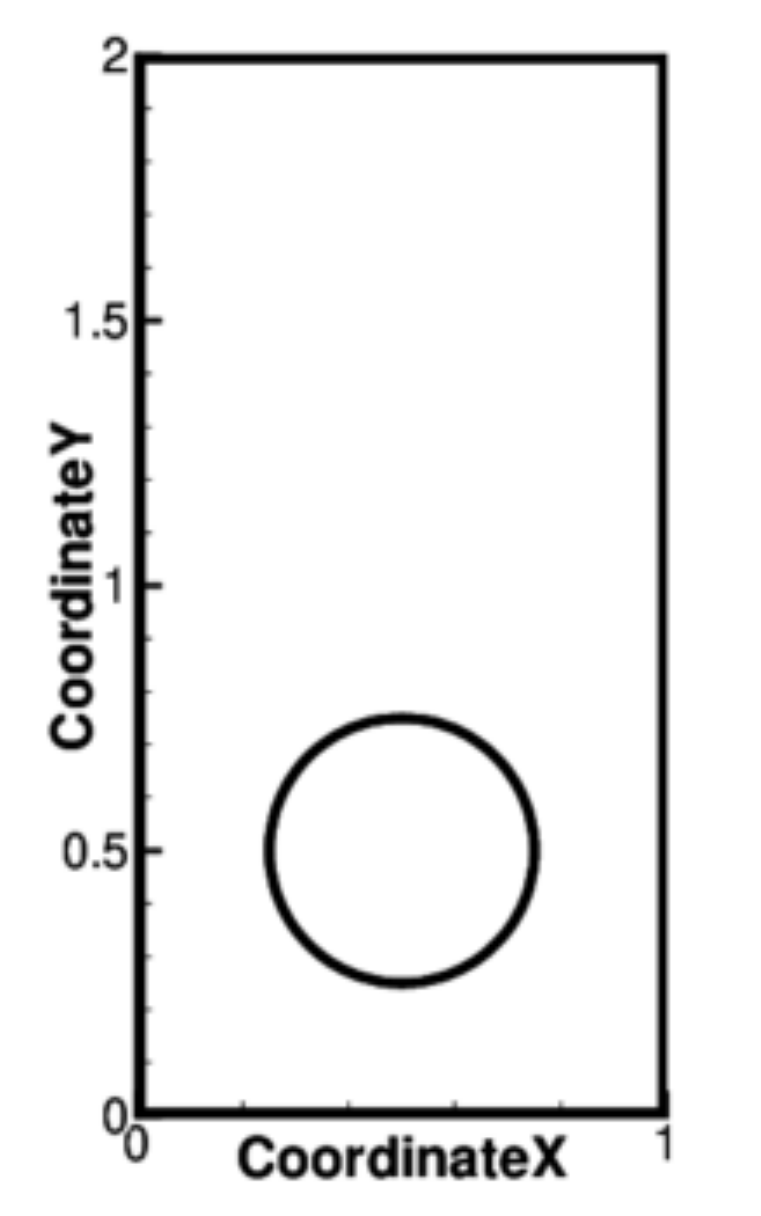}}
\subfloat[($t = 1.0s$)]
{\includegraphics[scale=0.47]{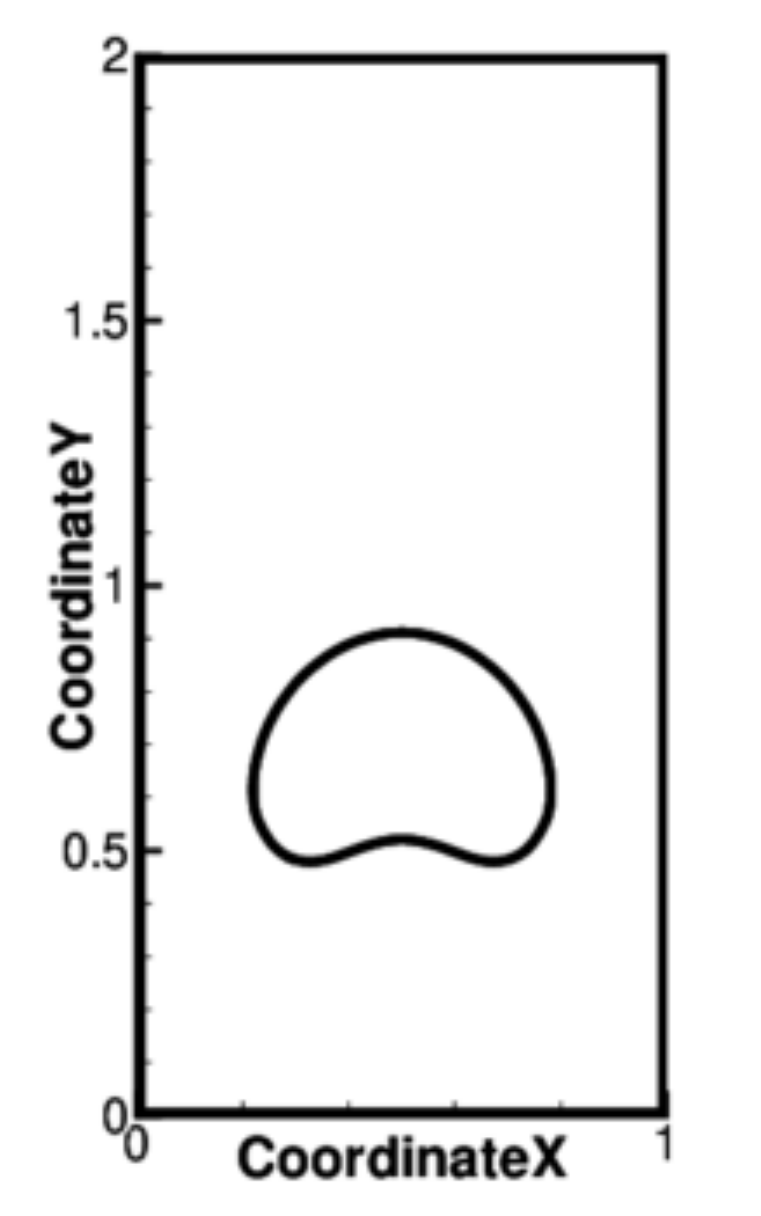}}
\subfloat[($t = 2.0s$)]
{\includegraphics[scale=0.47]{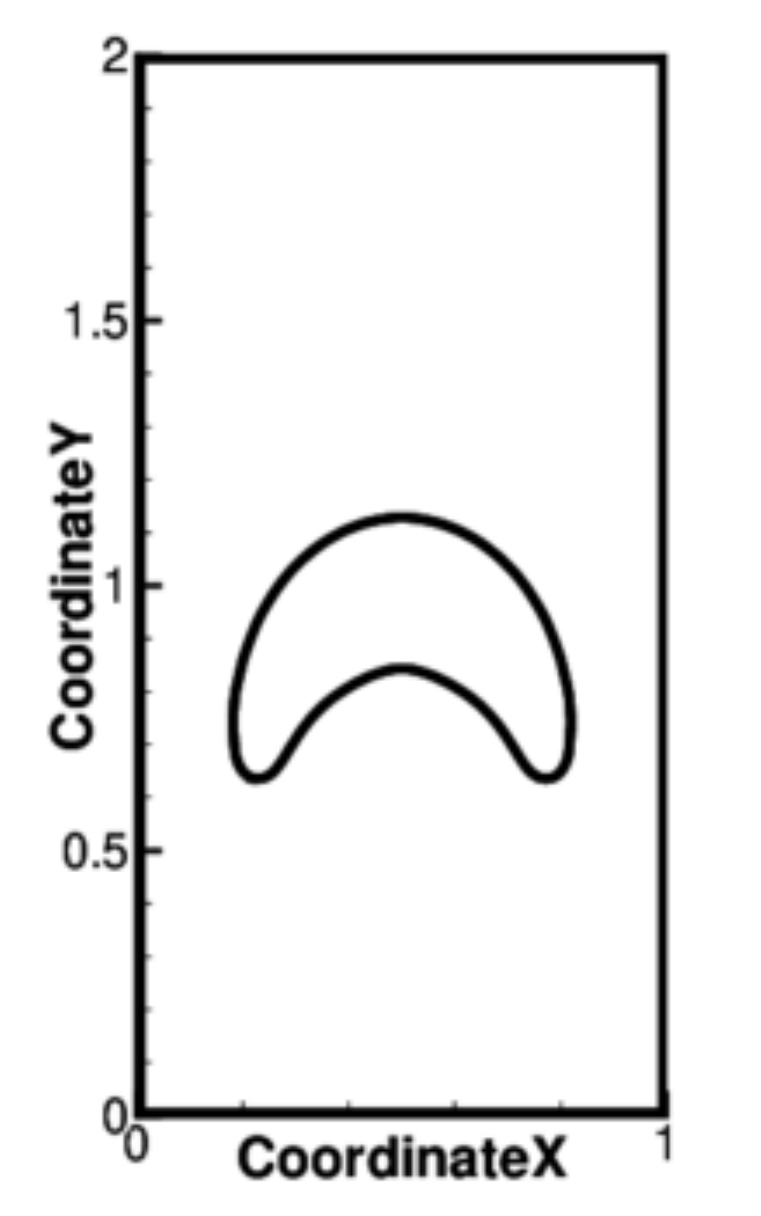}}
\subfloat[($t = 3.0s$)]
{\includegraphics[scale=0.47]{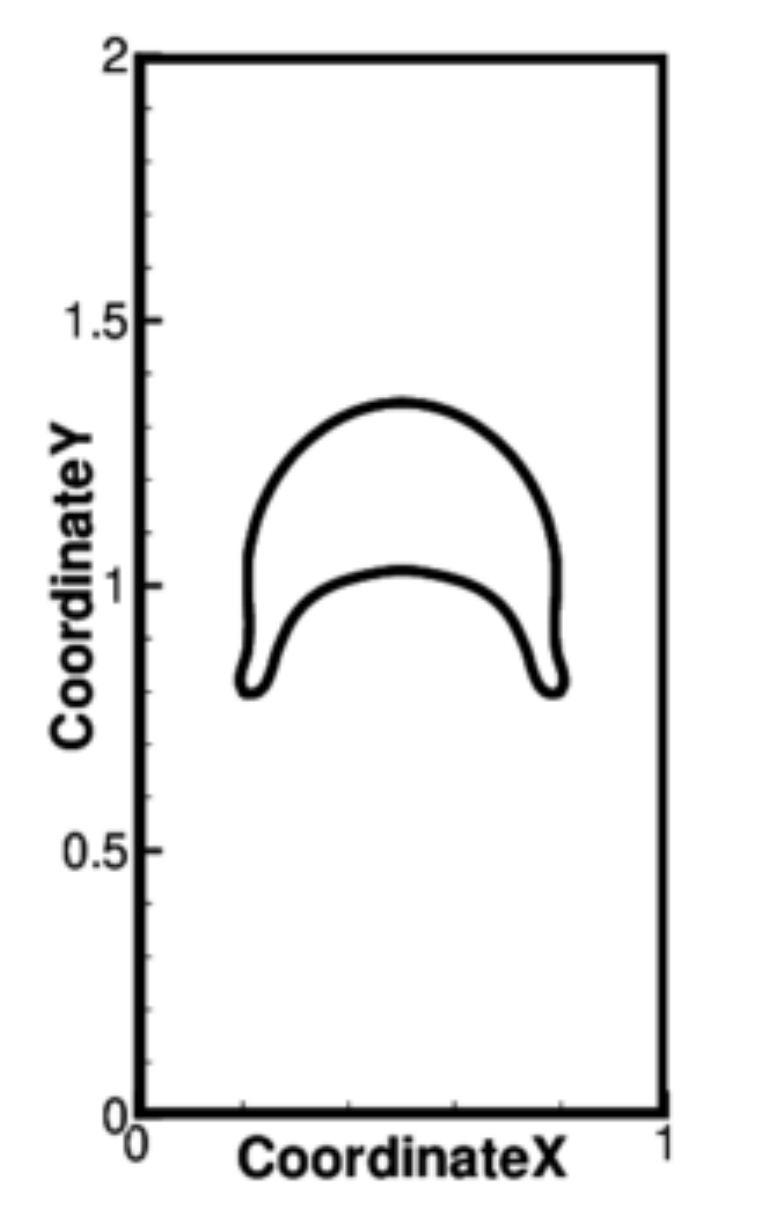}}
\subfloat[($t = 4.0s$)]
{\includegraphics[scale=0.47]{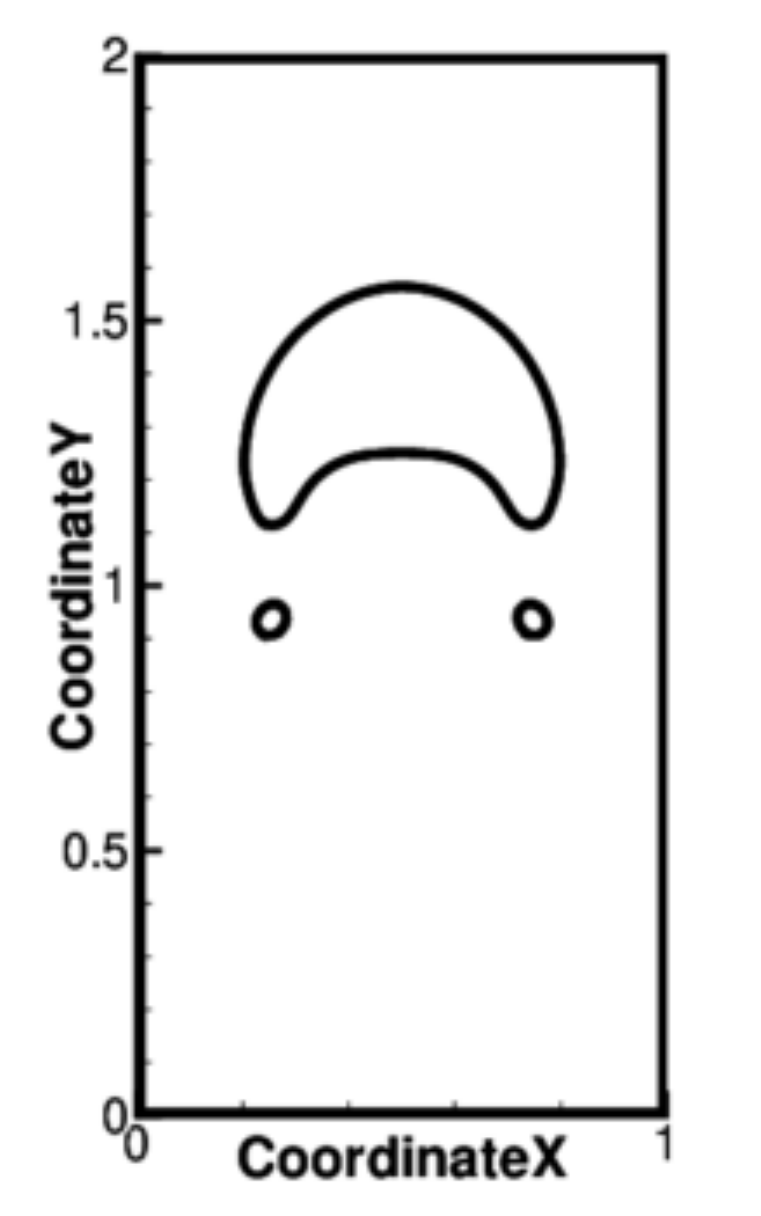}}

\subfloat[($t = 0.0s$)]
{\includegraphics[scale=0.47]{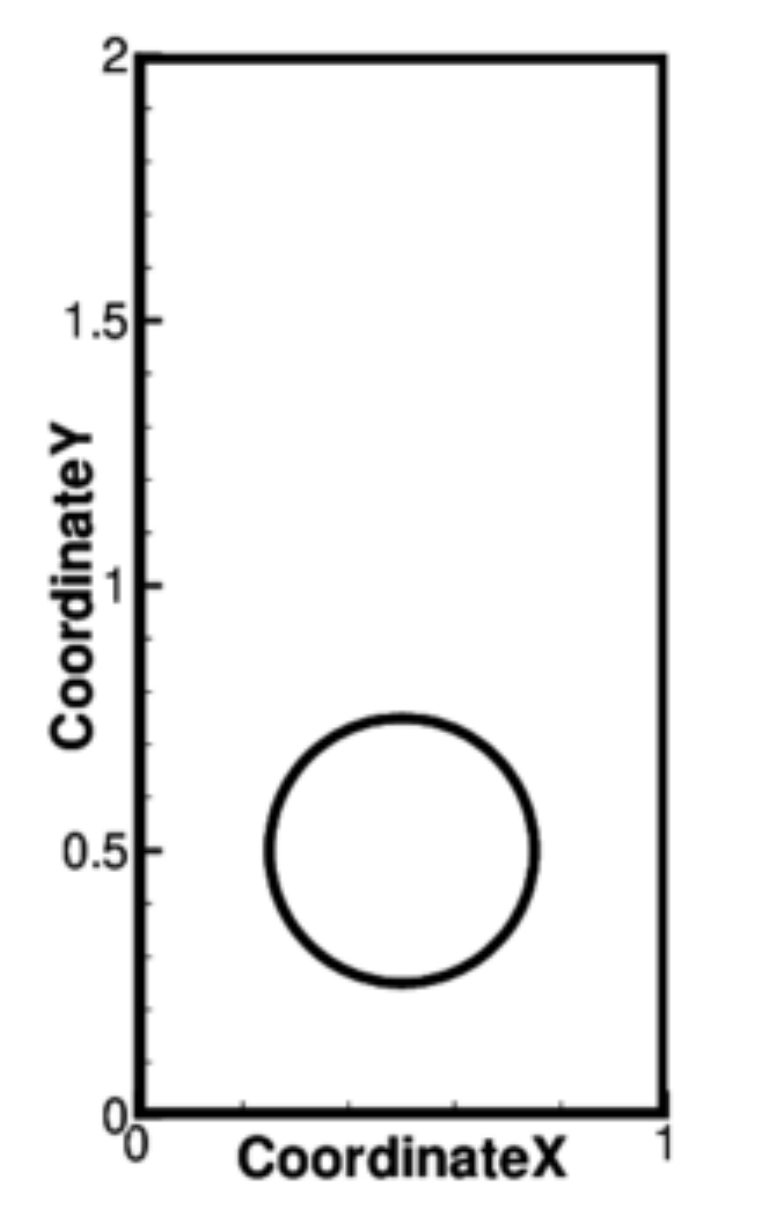}}
\subfloat[($t = 1.0s$)]
{\includegraphics[scale=0.47]{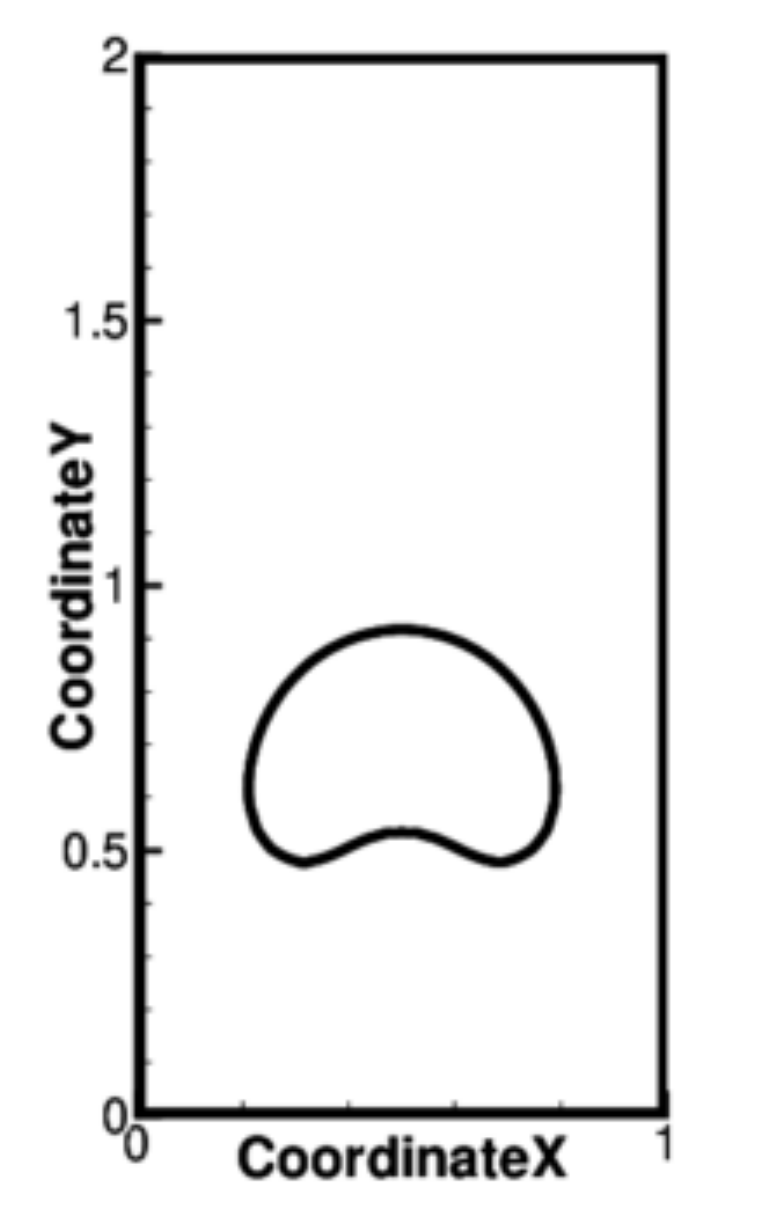}}
\subfloat[($t = 2.0s$)]
{\includegraphics[scale=0.47]{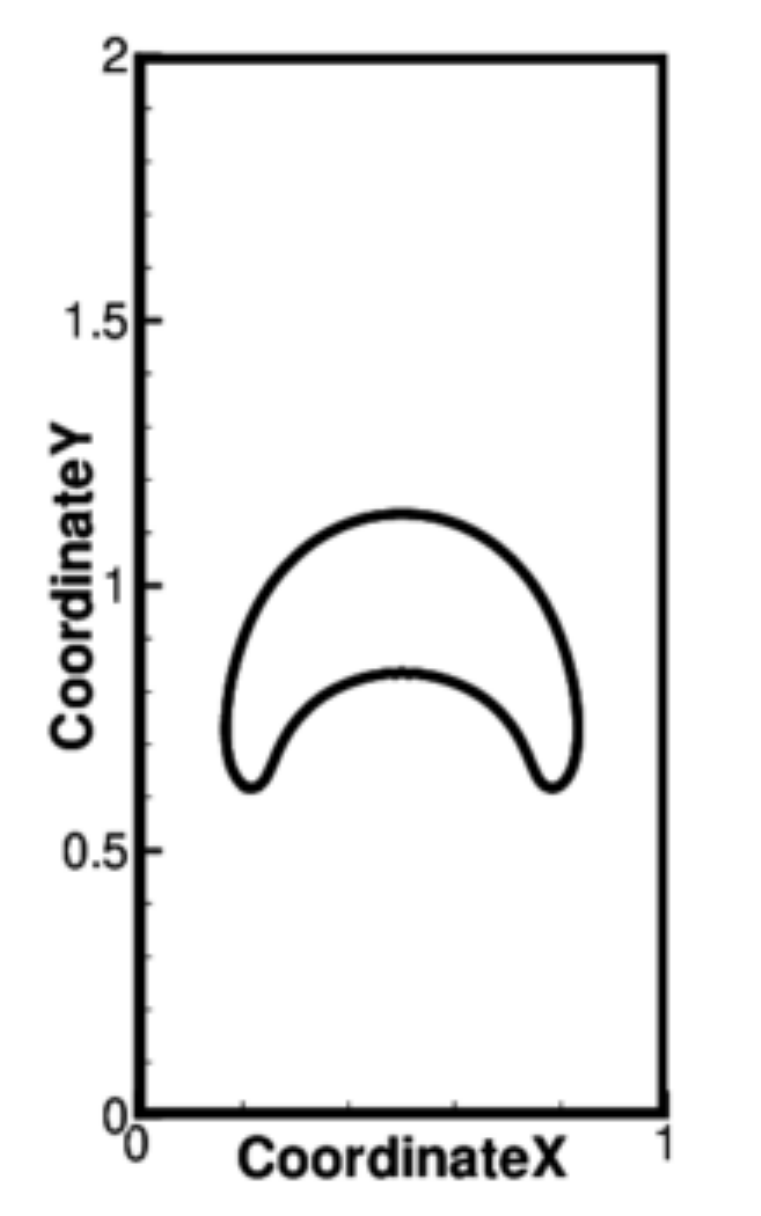}}
\subfloat[($t = 3.0s$)]
{\includegraphics[scale=0.47]{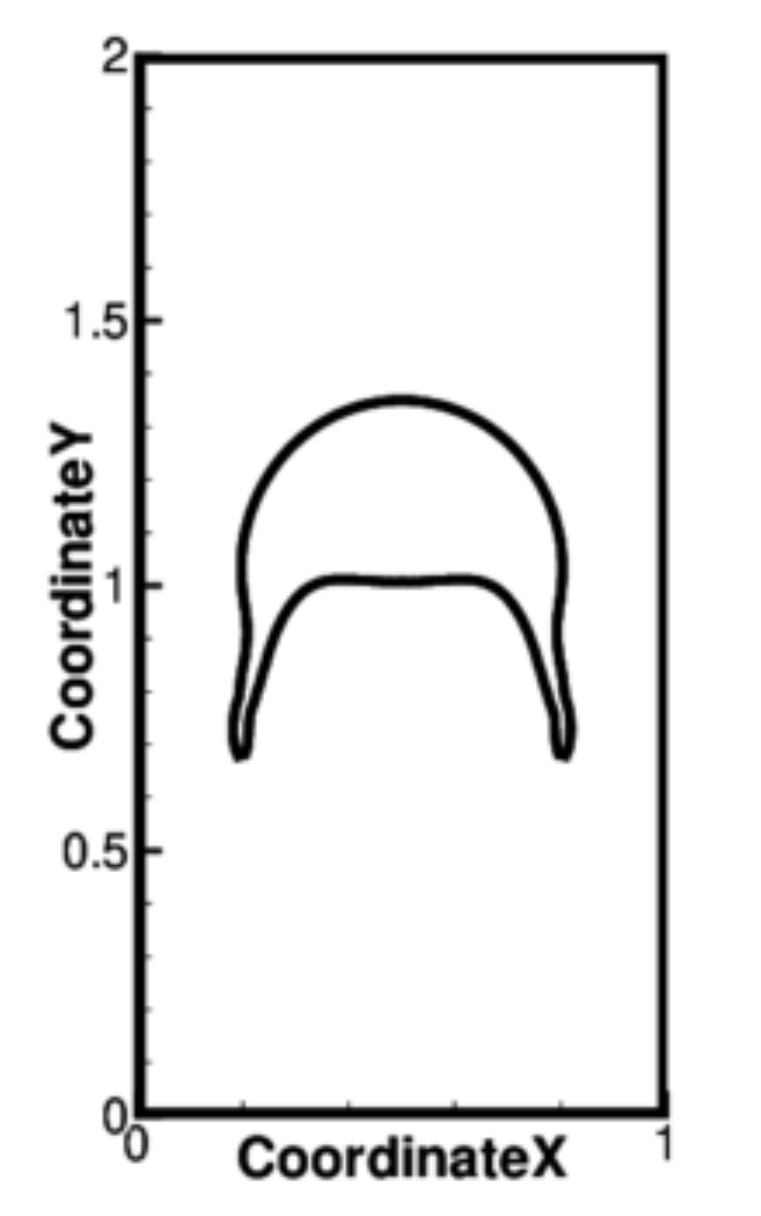}}
\subfloat[($t = 4.0s$)]
{\includegraphics[scale=0.47]{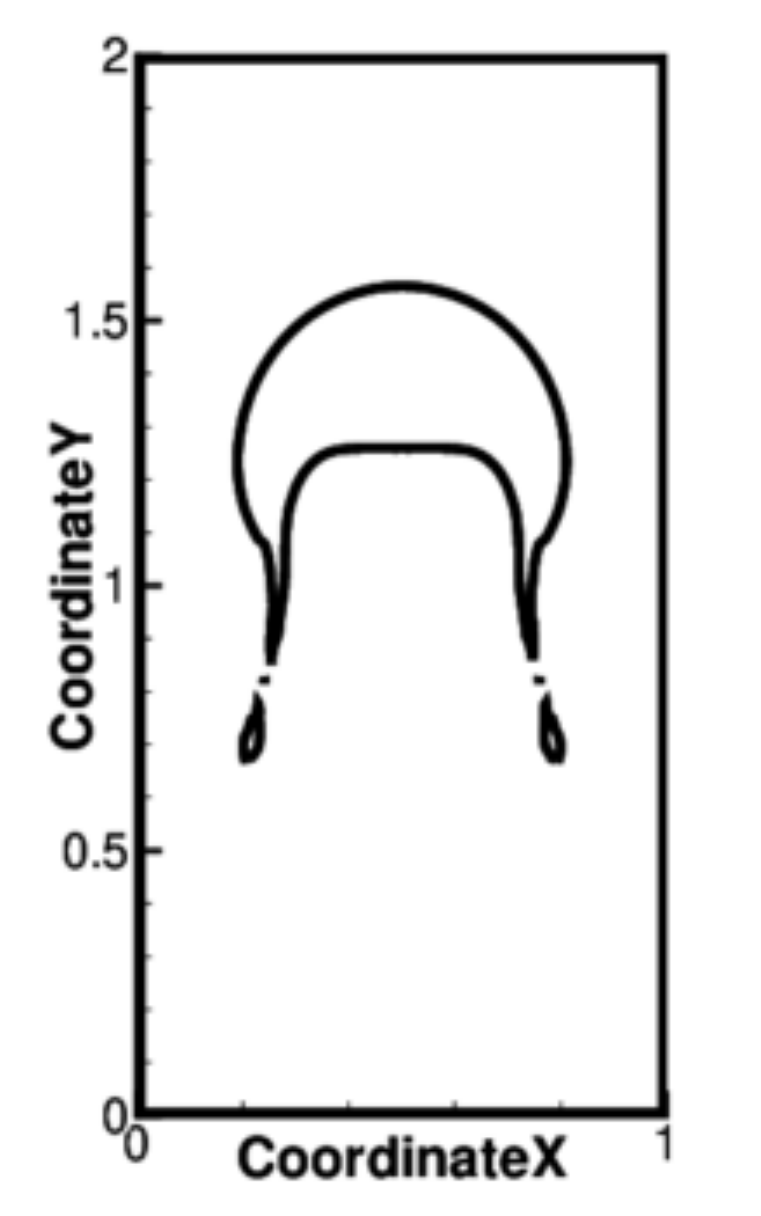}}

\subfloat[($t = 0.0s$)]
{\includegraphics[scale=0.47]{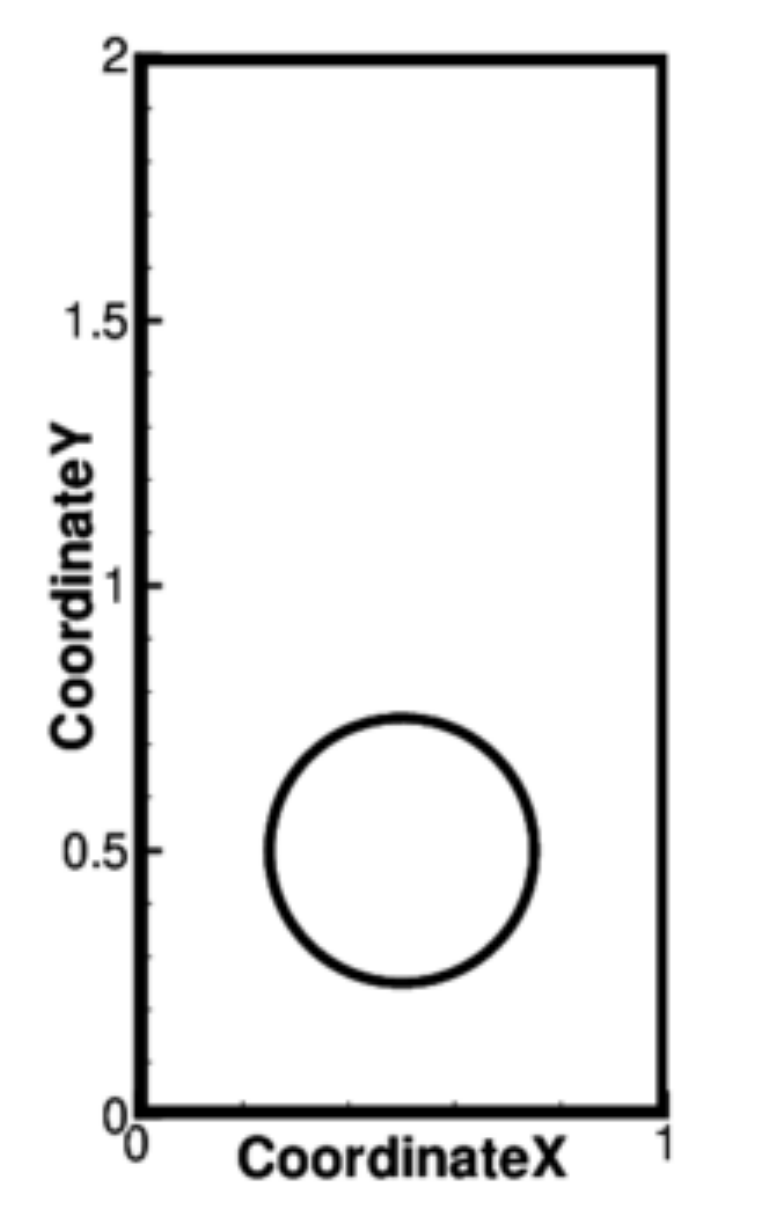}}
\subfloat[($t = 1.0s$)]
{\includegraphics[scale=0.47]{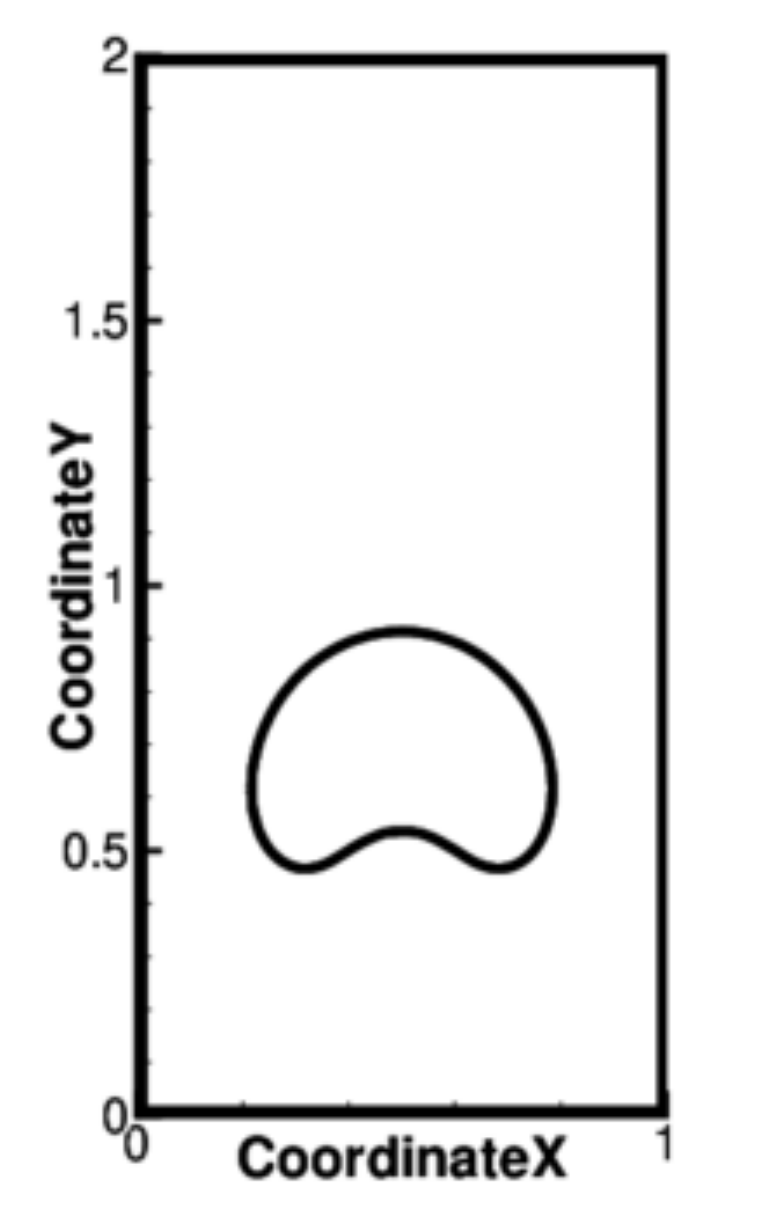}}
\subfloat[($t = 2.0s$)]
{\includegraphics[scale=0.47]{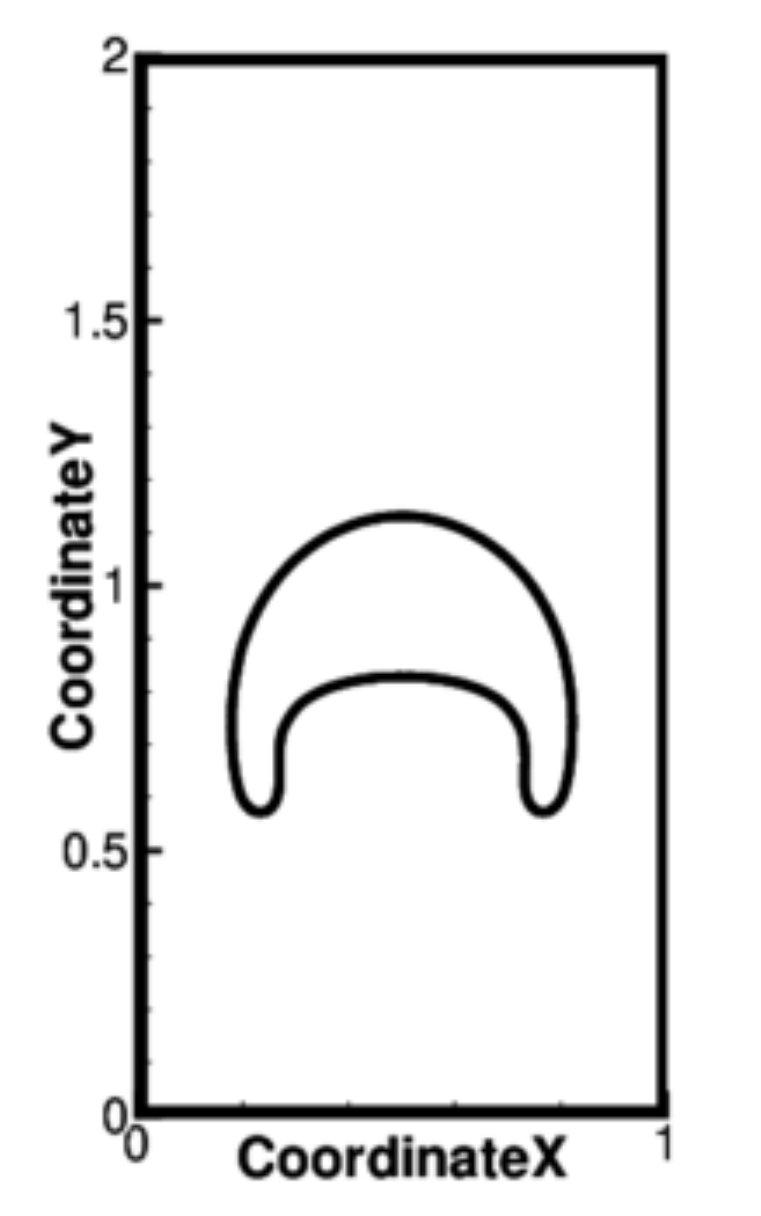}}
\subfloat[($t = 3.0s$)]
{\includegraphics[scale=0.47]{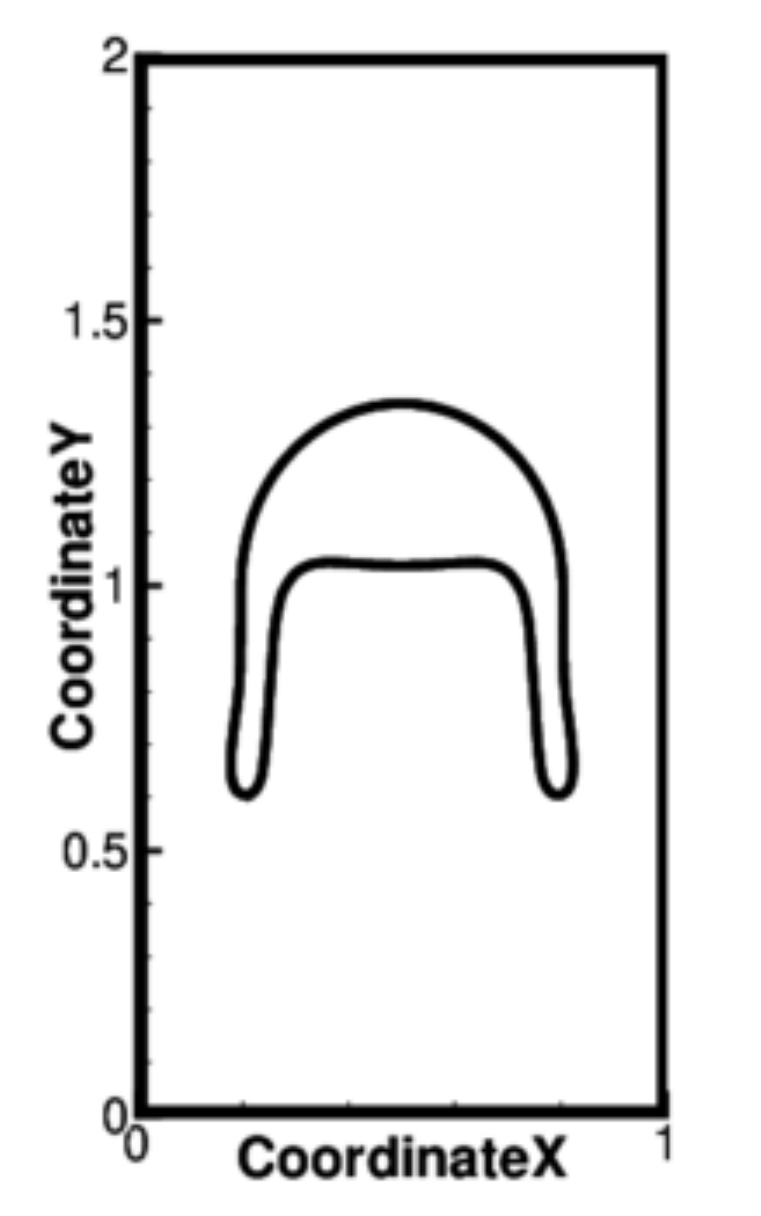}}
\subfloat[($t = 4.0s$)]
{\includegraphics[scale=0.47]{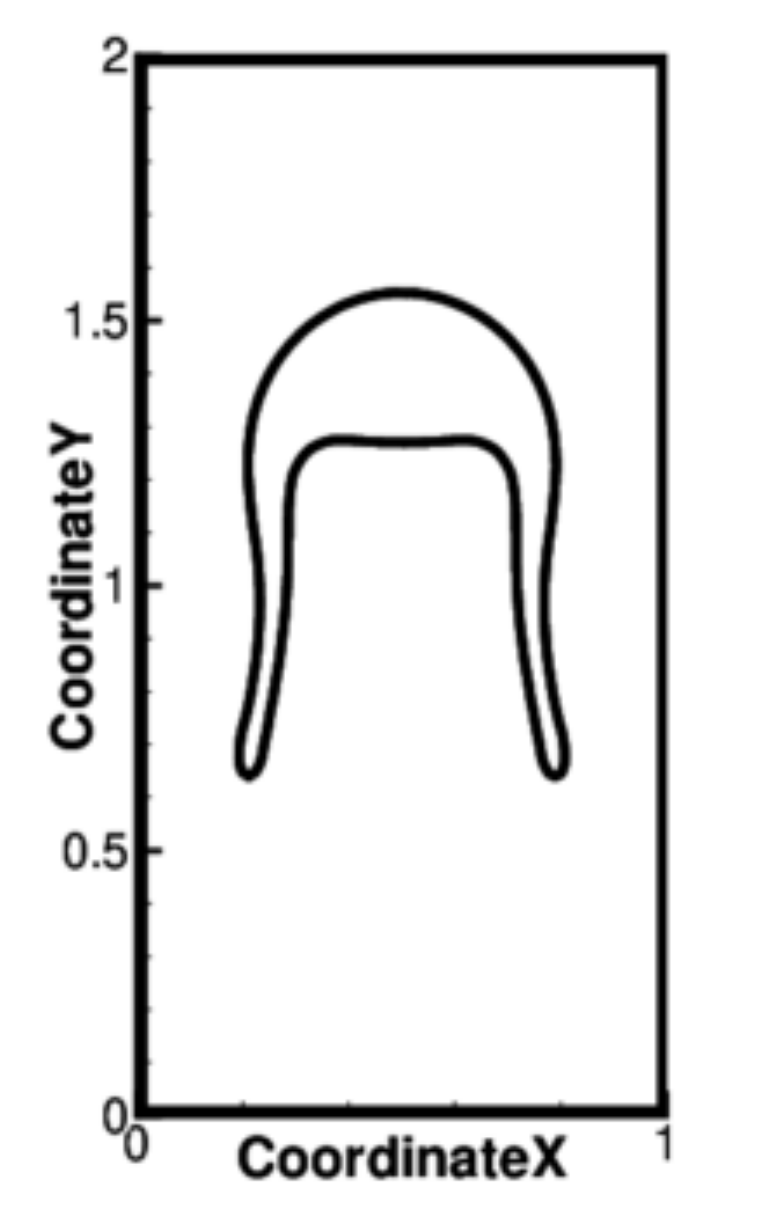}}
\caption{The bubble profiles at different time levels starting from $t
  = 0$~s up to $t = 4$~s for the rising bubble problem (Case-2). The
  subfigures from (a) to (e) correspond to the CLS-Olsson case
  computed on $80 \times 160$ Cartesian mesh,
  from (f) to (j) correspond to the new reinitialization computed on
  $80 \times 160$ Cartesian mesh and from (k) to (o) correspond to
  the new reinitialization computed on the unstructured mesh.}
\label{fig:BR_2_bubble_profiles}
\end{figure}

\begin{figure}[H]
\centering
\subfloat[Rise Velocity Vs Time]
{\includegraphics[scale=0.6]{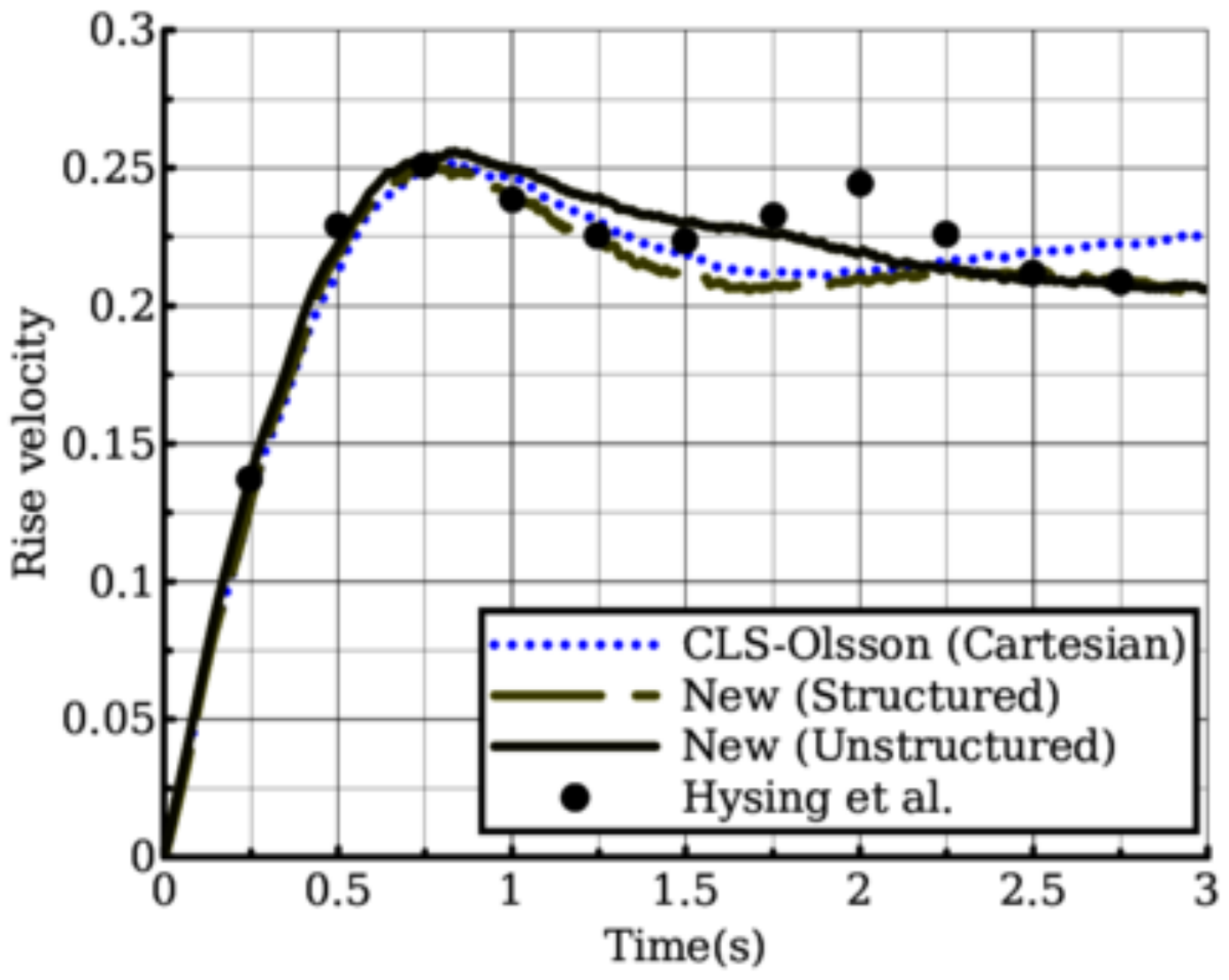}}
\subfloat[Centroid Vs Time]
{\includegraphics[scale=0.6]{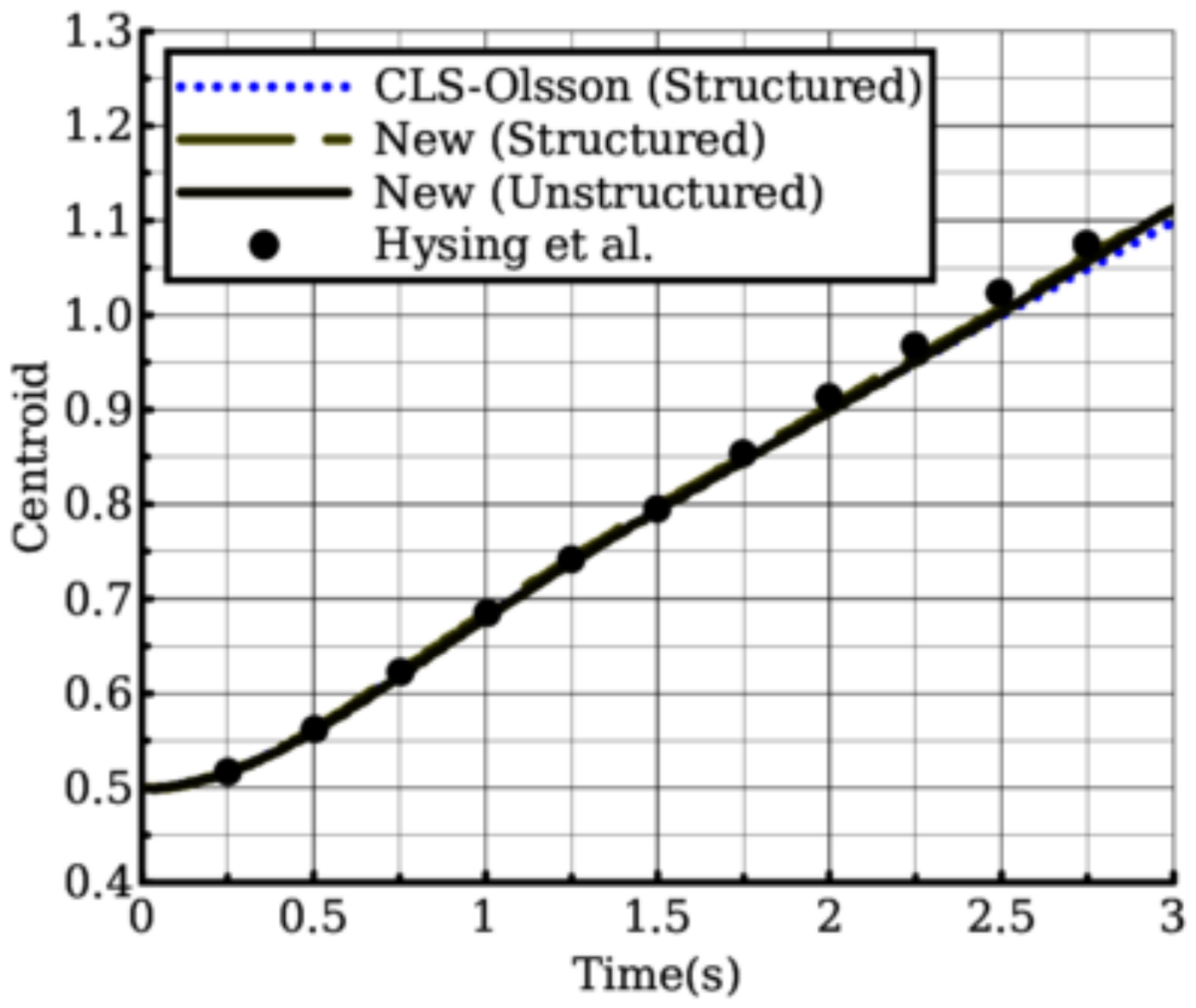}}

\subfloat[Circularity Vs Time]
{\includegraphics[scale=0.6]{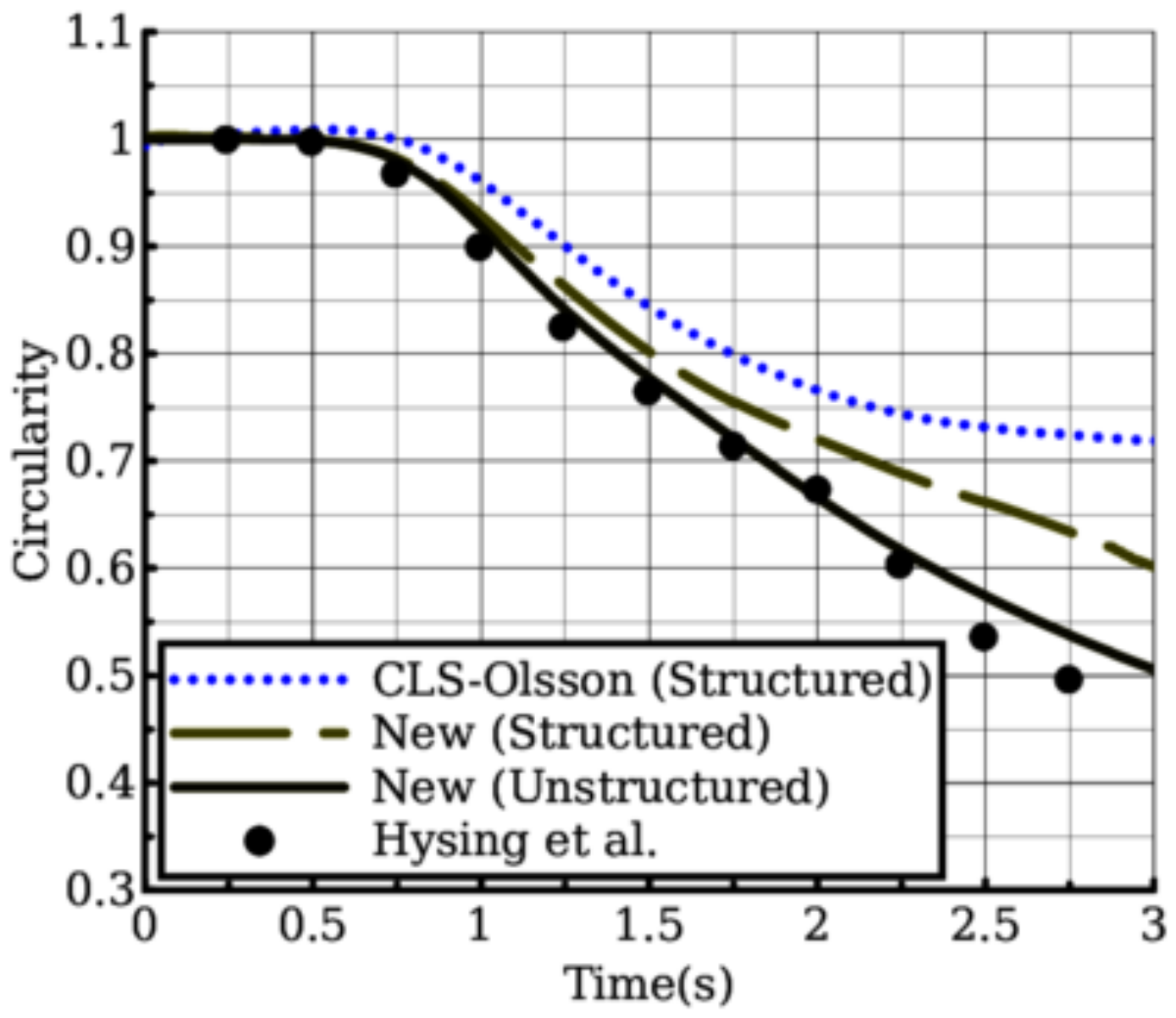}}
\caption{The rise velocity, centroid location and circularity plotted
  with respect to time for the rising bubble problem (Case-2).}
\label{fig:BR_2_parameter_plots}
\end{figure}

\begin{figure}[H]
\centering
\includegraphics[scale=0.6]{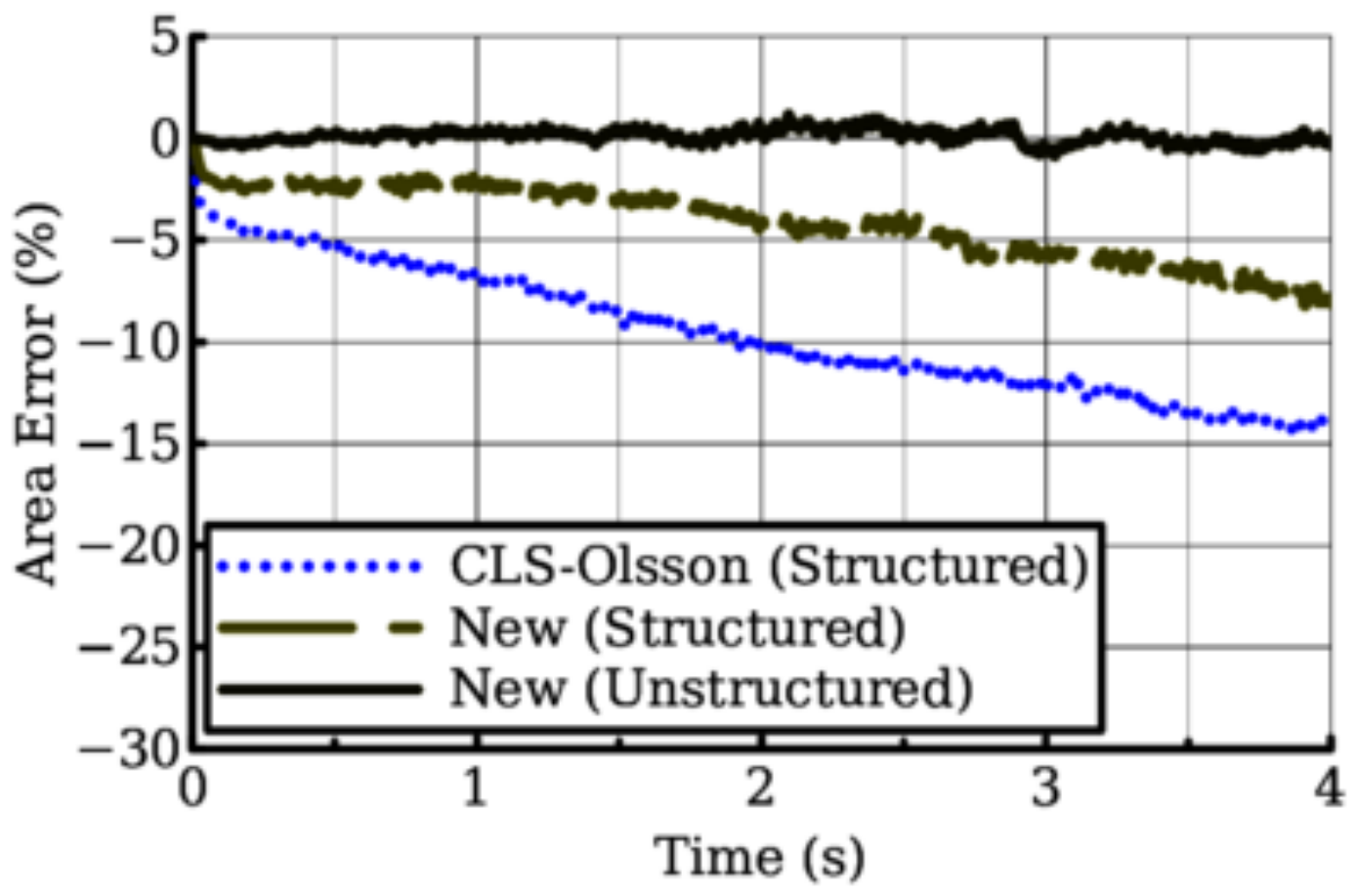}
\caption{The percentage area error plotted with respect to time for
  the rising bubble problem (Case-2).}
\label{fig:BR_2_area_error}
\end{figure}

Unlike the previous case, a large density ratio in this case results
the bubble to deform more and acquire a dimple cap profile with thin
elongated filament like structures originating from both sides. Due to
the complex shape, it is relatively difficult to capture the bubble
profile in Case-2 as compared to the Case-1. Numerical simulations are
carried out on a Cartesian mesh of $80 \times 160$ finite volume
cells. Moreover, in order to demonstrate the ability of the new
reinitialization scheme to deal with complex meshes, the problem is
also solved on an unstructured mesh consisting of 23331 finite volume
cells of triangular and quadrilateral shapes. Due to the clustering of
cells in the bubble path, one may expect improved results in case of
the unstructured mesh case. Figure~\ref{fig:BR_2_meshes} shows the
Cartesian mesh and the unstructured meshes considered in this
problem. The snapshots of bubble profiles at different time levels for
all the three cases are shown in
Figure~\ref{fig:BR_2_bubble_profiles}. One can clearly see from
Figure~\ref{fig:BR_2_bubble_profiles} that the elongated filament
structure is not captured very well in case of the CLS-Olsson
case. Whereas, a better profile of the elongated filament structure is
captured in case of the new reinitialization method solved on the
Cartesian mesh. The bubble profiles captured using the new
reinitialization approach solved on unstructured mesh, however, show a
close resemblance with the fine mesh results reported
in~\cite{Hysing2009}. The rise velocity, centroid location and
circularity of the bubble are plotted with respect to time in
Figure~\ref{fig:BR_2_parameter_plots}. One can see that the results
for the new reinitialization scheme on the unstructured mesh show very
good match with the reference results. Whereas, the results in case of
CLS-Olsson, especially the circularity profile, are far away from the
reference solution. Finally, the percentage area errors, computed
using equation~(\ref{eq:area_error}), are plotted in
Figure~\ref{fig:BR_2_area_error}. One can easily see that, the area
error is highest for the CLS-Olsson case. The new reinitialization
scheme solved on Cartesian mesh shows much less percentage area
error. Whereas, the new reinitialization method solved on unstructured
mesh shows the least area error.

\section{Conclusion}
A new approach to reinitialize the level set function for the CLS
method is formulated in this paper. The two major drawbacks of 
the existing artificial compression based reinitialization procedure,
namely, the unwanted movement of the interface contour and strong
sensitivity towards numerical errors leading to formation of
unphysical fluid patches, are resolved with the new approach. Here,
the existing artificial compression based reinitialization equation is
first examined carefully in order to identify the term responsible for
the movement of the interface contour. After isolating and removing a
curvature dependent velocity term, that is responsible for moving the
interface contour, the reinitialization equation is revised. The
remaining terms in the reinitialization equation are then carefully
replaced with equivalent terms that do not involve contour normal 
vectors. Unlike the compression and diffusion fluxes present in a
typical artificial compression approach, the newly reformulated
approach has a level set sharpening term, responsible for the
narrowing the level set profile, and a balancing term in order to
counteract the effect of sharpening. The combined effect of sharpening
and balancing restores the level set function without causing any
unwanted movement to the interface contour. Due to the absence of
terms involving contour normal vectors, the susceptibility towards
formation of unphysical fluid patches during reinitialization process
has been completely eliminated in the new reinitialization
procedure. As result of the new reinitialization approach, there has been significant improvement in the mass
conservation property. While
solving the new reinitialization equation, one can choose a larger
time step, approximately by a factor of $4/h$, in comparison with the
allowable time step of an artificial compression based
approach. Moreover, the simplified terms also help in significantly
reducing the numerical computation per reinitialization iteration,
aiding an overall reduction in the computational efforts.

In order to evaluate the performance of the new reinitialization
scheme, three types of numerical test cases are carried out. A
set of in-place reinitialization problems demonstrate that the new
reinitialization approach does not unnecessarily move the interface
contour even after large number of reinitialization iterations. The
area and shape errors of the new approach are quantified and compared
against other reinitialization schemes using a set of scalar advection
based test problems. In order to evaluate the performance on more
practical problems, a set of standard incompressible two-phase flow
problems, starting from an inviscid test case to complex test cases
involving viscous and surface tension forces, are solved. Finally, in
order to demonstrate the ability to deal with complex mesh types, an
incompressible two-phase flow problem is also solved on an
unstructured mesh consisting of finite volume cells having triangular
and quadrilateral shapes. The numerical results of the incompressible
two-phase flow problems show superior results as compared to the
existing approach and match very well with the reference solutions
reported in literature. With the enhanced accuracy and improved
ability to deal with complex mesh types, the proposed reinitialization
approach can be efficiently used in solving real life incompressible
two-phase flow problems.

\section{Acknowledgement}
The present work is partially supported by Aeronautics Research \&
Development Board (AR\&DB), with the project Grant number
ARDB/01/1031930/M/I DT. 13.09.2019. We gratefully thank AR\&DB for the
support.

\newpage
\appendix

\section{Central least square estimation of level set gradients}
\label{sec:central_least_square}
\begin{figure}[H]
\centering
\includegraphics[scale=0.75]{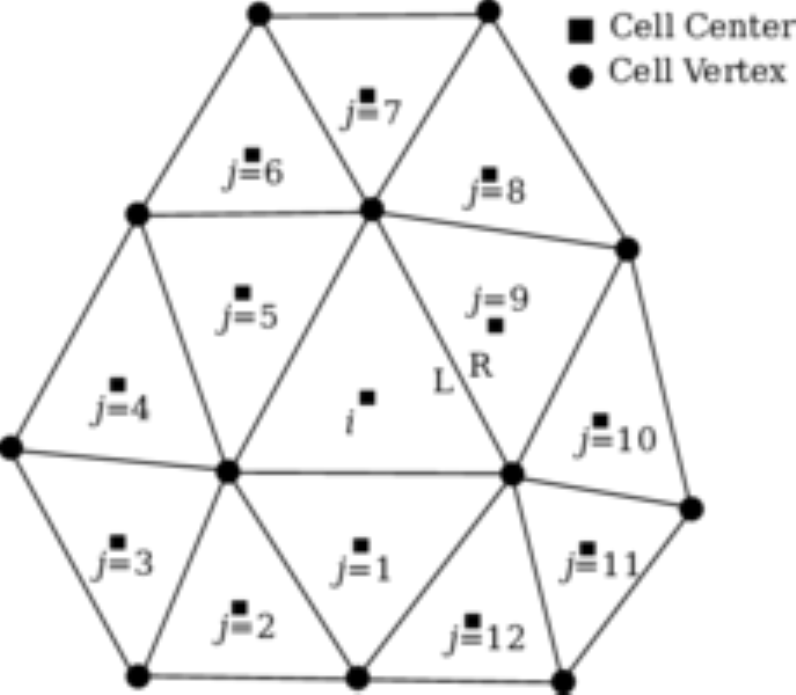}
\caption{A schematic representation of a triangular shaped finite
  volume cell $i$ and its vertex based neighbours, denoted as $j$.}
\label{fig:vertex_based_stencils}
\end{figure}

The second term in equation~(\ref{eq:reinit_new}) involves computation
of level set gradient terms.
\begin{equation}
\lvert \nabla \psi \rvert_i = \sqrt{\left(\frac{\partial \psi}{\partial
      x} \right)_i +
  \left(\frac{\partial \psi}{\partial y} \right)_i}
\end{equation}
These terms are evaluated here using central least square approach.
In order to construct cell center derivatives, a stencil consisting of
vertex based neighbours, as shown in
Figure~\ref{fig:vertex_based_stencils}, is considered. Using Taylor
series expansion, the neighbour cell values, $\psi_j$, of the level
set function are expressed in terms of the value at cell $i$, as, 
\begin{equation}
\psi_j = \psi_i + (x_j - x_i) \left(\frac{\partial
    \psi}{\partial x}\right)_i + (y_j - y_i) \left(\frac{\partial
    \psi}{\partial y}\right)_i \dots
\label{eq:taylor-series}
\end{equation}
where, $(x_i, y_i)$ and $(x_j, y_j)$ are locations of the centroids of
cell $i$ and centroid of the neighbour cell $j$ respectively. 
Upon truncating the higher order terms (after the third order term)
and re-arranging, equations~(\ref{eq:taylor-series}) can be written
as, 
\begin{equation}
\Delta \psi = {\bf S}~\text{d}\psi
\label{eq:taylor-series_matrix}
\end{equation}
where,
\begin{center}
	$\Delta \psi =\left\{\begin{array}{c}
	\psi_1 - \psi_i\\
	\psi_2 - \psi_i\\
	\dots\\
	\dots\\
	\psi_l - \psi_i
	\end{array}\right\}$;\hspace{0.5cm}${\bf S} =
    \left[\begin{array}{cc}
	x_1 - x_i & y_1 - y_i \\
	x_2 - x_i & y_2 - y_i \\
	\dots & \dots \\
	\dots & \dots \\
	x_l - x_i & y_l - y_i \\
	\end{array}\right]$;\hspace{0.5cm}$\text{d}\psi
  =\left\{\begin{array}{c}
	\displaystyle \frac{\partial \psi}{\partial x}\\ \\
	\displaystyle \frac{\partial \psi}{\partial y}\\
	\end{array}\right\}$
\end{center}
The overdetermined system of equation~(\ref{eq:taylor-series_matrix})
can be solved as,
\begin{equation}
\text{d}\psi = \left({\bf S}^{\text{T}}{\bf S}\right)^{-1}
{\bf S}^{\text{T}} \Delta \psi
\label{eq:weighted_normal_method}
\end{equation}
Closed-form expressions for the derivatives can be obtained by
simplifying equation~(\ref{eq:weighted_normal_method}) as,
\begin{subequations}
\begin{align}
\left(\frac{\partial \psi}{\partial x}\right)_i
= \frac{\ell_{22} r_{1} - \ell_{21}
r_{2}}{G}\\
\left(\frac{\partial \psi}{\partial y}\right)_i
= \frac{\ell_{11} r_{2} - \ell_{12} r_{1}}{G}
\end{align}
\label{eq:derivatives}
\end{subequations}
where 
$$
\ell_{11} = \sum_{j=1}^l (x_j - x_i)^2, \hspace{0.5cm} \ell_{22}
=  \sum_{j=1}^l (y_j - y_i)^2, \hspace{0.5cm} \ell_{12} =
\ell_{21}
= \sum_{j=1}^l (x_j - x_i) (y_j - y_i)
$$
$$
r_{1} = \sum_{j=1}^l (x_j - x_i) (\psi_j - \psi_i),
\hspace{0.5cm} r_{2}
= \sum_{j=1}^l (y_j - y_i) (\psi_j - \psi_i), \hspace{0.5cm}
G
= \ell_{11} \ell_{22} - \ell_{12}^2
$$ 

\bibliographystyle{ieeetr}
\bibliography{paper}

\begin{thebibliography}{10}

\bibitem{Olsson2005}
E.~Olsson and G.~Kreiss, ``A conservative level set method for two phase
  flow,'' {\em Journal of Computational Physics}, vol.~210, no.~1, pp.~225 --
  246, 2005.

\bibitem{Olsson2007}
E.~Olsson, G.~Kreiss, and S.~Zahedi, ``A conservative level set method for two
  phase flow {\uppercase{ii}},'' {\em Journal of Computational Physics},
  vol.~225, no.~1, pp.~785 -- 807, 2007.

\bibitem{Shukla2010}
R.~K. Shukla, C.~Pantano, and J.~B. Freund, ``An interface capturing method for
  the simulation of multi-phase compressible flows,'' {\em Journal of
  Computational Physics}, vol.~229, no.~19, pp.~7411 -- 7439, 2010.

\bibitem{Mccaslin2014}
J.~O. McCaslin and O.~Desjardins, ``A localized re-initialization equation for
  the conservative level set method,'' {\em Journal of Computational Physics},
  vol.~262, pp.~408 -- 426, 2014.

\bibitem{Desjardins2008}
O.~Desjardins, V.~Moureau, and H.~Pitsch, ``An accurate conservative level
  set/ghost fluid method for simulating turbulent atomization,'' {\em Journal
  of Computational Physics}, vol.~227, no.~18, pp.~8395 -- 8416, 2008.

\bibitem{Yohei2012}
Y.~Sato and B.~Ničeno, ``A conservative local interface sharpening scheme for
  the constrained interpolation profile method,'' {\em International Journal
  for Numerical Methods in Fluids}, vol.~70, no.~4, pp.~441--467, 2012.

\bibitem{Zhao2014}
L.~Zhao, J.~Mao, X.~Bai, X.~Liu, T.~Li, and J.~Williams, ``Finite element
  implementation of an improved conservative level set method for two-phase
  flow,'' {\em Computers \& Fluids}, vol.~100, pp.~138 -- 154, 2014.

\bibitem{Waclawczyk2015}
T.~Wac{\l}awczyk, ``A consistent solution of the re-initialization equation in
  the conservative level-set method,'' {\em Journal of Computational Physics},
  vol.~299, no.~Supplement C, pp.~487 -- 525, 2015.

\bibitem{Chiodi2017}
R.~Chiodi and O.~Desjardins, ``A reformulation of the conservative level set
  reinitialization equation for accurate and robust simulation of complex
  multiphase flows,'' {\em Journal of Computational Physics}, vol.~343, pp.~186
  -- 200, 2017.

\bibitem{Tabar2018}
N.~Shervani-Tabar and O.~V. Vasilyev, ``Stabilized conservative level set
  method,'' {\em Journal of Computational Physics}, vol.~375, pp.~1033 -- 1044,
  2018.

\bibitem{Parameswaran2020}
S.~Parameswaran and J.~C. Mandal, ``A novel reinitialization scheme for
  conservative level set method,'' {\em Manuscript submitted for publication},
  vol.~00, pp.~000 -- 000, 2020.

\bibitem{Brackbill1992}
J.~Brackbill, D.~Kothe, and C.~Zemach, ``A continuum method for modeling
  surface tension,'' {\em Journal of Computational Physics}, vol.~100, no.~2,
  pp.~335 -- 354, 1992.

\bibitem{Chorin1967}
A.~J. Chorin, ``A numerical method for solving incompressible viscous flow
  problems,'' {\em Journal of Computational Physics}, vol.~2, pp.~12 -- 26,
  1967.

\bibitem{Parameswaran2019-2}
S.~Parameswaran and J.~C. Mandal, ``A conservative finite volume method for
  incompressible two-phase flows on unstructured meshes,'' {\em Manuscript
  submitted for publication}, 2020.

\bibitem{Parameswaran2019}
S.~Parameswaran and J.~C. Mandal, ``A novel {R}oe solver for incompressible
  two-phase flow problems,'' {\em Journal of Computational Physics}, vol.~390,
  pp.~405 -- 424, 2019.

\bibitem{Gottlieb2005}
S.~Gottlieb, ``{On high order strong stability preserving runge-kutta and multi
  step time discretizations},'' {\em Journal of Scientific Computing}, vol.~25,
  pp.~105--128, Oct. 2005.

\bibitem{Gaitonde1998}
A.~L. Gaitonde, ``A dual-time method for two-dimensional unsteady
  incompressible flow calculations,'' {\em International Journal for Numerical
  Methods in Engineering}, vol.~41, no.~6, pp.~1153--1166, 1998.

\bibitem{Yang2014}
Y.~Aiming, C.~Sukun, Y.~Liu, and Y.~Xiaoquan, ``An upwind finite volume method
  for incompressible inviscid free surface flows,'' {\em Computers \& Fluids},
  vol.~101, pp.~170 -- 182, 2014.

\bibitem{Martin1952}
J.~C. Martin and W.~J. Moyce, ``Part {IV}. an experimental study of the
  collapse of liquid columns on a rigid horizontal plane,'' {\em Philosophical
  Transactions of the Royal Society of London A: Mathematical, Physical and
  Engineering Sciences}, vol.~244, no.~882, pp.~312--324, 1952.

\bibitem{Puckett1997}
E.~G. Puckett, A.~S. Almgren, J.~B. Bell, D.~L. Marcus, and W.~J. Rider, ``A
  high-order projection method for tracking fluid interfaces in variable
  density incompressible flows,'' {\em Journal of Computational Physics},
  vol.~130, no.~2, pp.~269 -- 282, 1997.

\bibitem{Hysing2009}
S.~Hysing, S.~Turek, D.~Kuzmin, N.~Parolini, E.~Burman, S.~Ganesan, and
  L.~Tobiska, ``Quantitative benchmark computations of two-dimensional bubble
  dynamics,'' {\em International Journal for Numerical Methods in Fluids},
  vol.~60, no.~11, pp.~1259--1288, 2009.

\end{thebibliography}

\end{document}